# Human-Centered Artificial Intelligence (HCAI): Foundations and Approaches

Wei Xu

HCAI Labs, California, USA

https://orcid.org/0000-0001-8913-2672

## Abstract:

Artificial Intelligence (AI) is a transformative yet double-edged technology that can advance human welfare while also posing risks to humans and society. In response, the Human-Centered Artificial Intelligence (HCAI) approach has emerged as both a design philosophy and a methodological complement to prevailing technology-centered AI paradigms. Placing humans at the core, HCAI seeks to ensure that AI systems serve, augment, and empower humans rather than harm or replace them. This chapter establishes the conceptual and methodological foundations of HCAI by tracing its evolution and recent advancements. It introduces key HCAI concepts, frameworks, guiding principles, methodologies, and practical strategies that bridge philosophical HCAI principles with operational implementation. Through an analytical review of the emerging characteristics and challenges of AI technologies, the chapter positions HCAI as a holistic paradigm for aligning AI innovation with human values, societal well-being, and sustainable progress. Finally, this chapter outlines the structure and contributions of the *Handbook of Human-Centered Artificial Intelligence*. The purpose of this chapter is to provide an integrated foundation that connects HCAI conceptual frameworks, principles, methodology, and practices for this handbook, thereby paving the way for the content of subsequent chapters. Ultimately, the success of AI will be measured not solely by its technical sophistication but by its enduring capacity to enhance human performance, safeguard human dignity, and strengthen societal benefits and trust.

## 1. Introduction

Artificial Intelligence (AI) technology has become a transformative force across nearly all aspects of modern life, reshaping people's daily work and life. Notably, large language models (LLMs), such as OpenAI's GPT-5 and Google's Gemini 3, exemplify the powerful capabilities of contemporary AI systems to generate fluent, context-aware language, supporting applications from virtual assistants to automated content creation. These models are central to the rise of generative AI, a broader class of technologies that can support writing, summarization, image composition, video generation, music composition, coding, and translation from learned data patterns, representing significant milestones in AI development for creativity and productivity. While AI brings immense benefits to society, it simultaneously amplifies risks, including bias, opacity, misinformation, and the loss of human agency (Paeth & McGregor, 2025; Raunak & Kuhn, 2024, 2025; Ridley, 2025; Sison et al., 2024; Endsley, 2023). For example, reported AI incidents rose to 233 in 2024, a 56.4% year-over-year increase (Stanford HAI, 2025), and the AI Incident Database continued to log dozens more in mid-2025 (AIID, 2025), while security telemetry shows AI-driven automation fueling 36,000 scans/second and rising credential abuse (Fortinet, 2025). Global initiatives, from the *EU Artificial Intelligence Act* to the *NIST AI Risk Management Framework,* reflect an accelerating effort to align AI innovation with human values, accountability, and safety (EU, 2024; NIST, 2023).

Within this context, *Human-Centered Artificial Intelligence (HCAI)* has emerged as both a research paradigm and practical design approach emphasizing that AI must be developed *for* and *with* humans, ensuring AI technology amplifies, augments, enhances, and empowers humans rather than harms or replaces humans (Shneiderman, 2020; Xu, 2019). Historically, a similar pattern was seen during the 1980s' personal computing (PC) boom. At the time, computer systems were primarily designed for experts, often neglecting human needs and user experience. Accordingly, a human-centered design (HCD) approach was proposed to address the challenges (Norman, 1986). The rise of HCD practices spurred the emergence and development of new fields, including user experience and human-computer interaction (HCI) (Shneiderman, 1987; Xu, 2003a). Today, as society vigorously promotes and advances new AI technology, humans once again face the question of whether we should continue to adhere to the principle of being human-centered. More importantly, neglecting human-centered principles in the advancement and adoption of AI technology could lead to far more severe consequences than in the past.

This chapter situates HCAI within this broader transformation, offering an integrated foundation that connects HCAI conceptual frameworks, principles, methodology, and practices. It first examines the emerging characteristics and challenges of AI technology, including potential negative impacts on humans if not appropriately developed and used. It then introduces leading HCAI frameworks along with emerging concepts shaping the field. It then elaborates on the HCAI methodology for operationalizing HCAI, providing methodological guidance. Finally, it highlights the strategies for HCAI practice. In doing so, this chapter provides the conceptual and methodological foundation for the *Handbook of Human-Centered Artificial Intelligence*, which aims to bridge the gap between technical innovation and human well-being, placing humans, rather than algorithms, at the center of AI's future. This chapter introduces the overarching themes and theoretical foundations that underpin the entire handbook.

## 2. Emerging characteristics of AI technology

### 2.1 Transitioning to Human-AI Interaction

Human-computer interaction (HCI) is a cross-disciplinary field that emerged in the early days of the PC era. It primarily studies the interaction between humans and non-AI computing systems (Shneiderman, 1987; Xu, 2003b). Today, people interact with a variety of AI systems daily, transitioning to human-AI interaction. The evolution from traditional human-computer interaction to human-AI interaction marks a paradigm shift in how humans engage with technology (Xu et al., 2023). While HCI, rooted in the PC era, focused primarily on the user experience and functionality of non-AI computing systems, human-AI interaction introduces systems with varying degrees of automation and cognitive capabilities. Enabled by advances in big data, deep learning, and computational power, AI systems can now perceive, learn, adapt, and in some contexts operate independently. These AI systems can operate in unpredictable environments in certain operational scenarios, interact with humans collaboratively rather than merely as tools, and complete tasks beyond the scope of earlier automation technologies (Rahwan et al., 2019; Kaber, 2018; Shneiderman, 2020; Xu, 2020).

Table 1 presents a comparative analysis between traditional non-AI-based computing systems and AI-based systems. The comparison underscores that the transition from *human–computer interaction* to *human–AI interaction* introduces not only enhanced technical capabilities but also fundamentally new paradigms of human–machine interaction.

Table 1: A comparative analysis between non-AI-based computing systems and AI systems

| Characteristics | Non-AI Automated Systems | AI-Based Systems |
|---|---|---|
| Examples | Office software, washing machines, elevators, and automated manufacturing lines | Chat GPT, intelligent decision-support systems, automated vehicles, service robots |
| Machine behavior | Fixed algorithms and deterministic logic, no capacity for adaptation | Learning, adaptation, and self-execution; behaviors may evolve, be uncertain, or biased |
| Machine's role | Tools assisting human tasks | Possible collaborators with humans, varying by automation level |
| Machine output | Deterministic and predictable | Probabilistic, uncertain, context-dependent |
| Human-machine relationship | Unidirectional interaction (human to machine) | Bidirectional (two-way) collaborative interaction |
| Complementarity | No complementarity, static functional allocation between humans and machines | Dynamic complementarity between human intelligence and machine intelligence (hybrid intelligence) |
| Human-like sensing ability | Limited or none | Present (via multimodal sensing technologies) |
| Human-like cognitive abilities | Absent | Present to varying degrees (e.g., pattern recognition, learning, reasoning) |

| Self-execution | Requires manual activation | Independent execution in specific contexts |
|---|---|---|
| Machine's adaptation to unpredictable environments | Not possible | Adaptive behaviors are possible depending on the design |
| Initiating abilities | Only humans possess the capacity to actively initiate actions, while machines function passively by executing instructions provided by humans | Both humans and machines can now actively initiate actions; AI systems can generate actions proactively based on implicit collaborative interaction cues, such as human behavior, intentions, emotions, and operational context. |
| Directionality | A one-way, human-directed model of trust and control, in which machines lack reciprocal situation awareness, trust, and task-sharing with humans. | A two-way relationship characterized by shared trust, mutual situation awareness, and collaborative interaction and task execution between humans and AI systems, while preserving human decision-making authority |
| Human intervention | Always required, humans are the final decision makers | Human oversight and authority remain crucial, while systems may act proactively |

As shown in Table 1, non-AI systems primarily function as passive tools, executing predefined rules and relying heavily on continuous human input. In contrast, AI systems enable a shift toward *bidirectional (two-way) collaborative interaction,* beyond the traditional *unidirectional (one-way) interaction (human-to-machine*). These AI systems can proactively sense and interpret human states (e.g., physiological, cognitive, and emotional), while providing humans with richer situational awareness through multimodal interfaces. Depending on the degree of autonomy of the AI system, an AI system (including single or multiple AI agents) can have cognitive abilities similar to humans (perception, learning, adaptation, independent operation, etc.), can independently complete some specific tasks in specific scenarios, and has a certain adaptability to some unpredictable environments, and can independently complete tasks that previous automation technology could not complete (Kaber, 2018; Rahwan et al., 2019; Xu, 2020). Although current AI technology cannot yet realize all these new features, with continued development and further improvements in machine intelligence, AI systems will gradually acquire them.

It is evident that there are significant differences between these two types of interactions. These new features of AI systems also introduce a range of challenges distinct from traditional human-computer interaction, necessitating novel approaches to optimize interaction between humans and AI systems.

## 2.2 Emerging human-machine relationship

A human–machine relationship shapes how humans and technology work together. A good relationship fosters trust and balance, helping individuals remain in control while leveraging technology's strengths. A poor one can lead to dependence or a loss of control. Understanding is key to creating technology that truly serves humans.

Technological advancements have historically propelled the development of human–machine relationships, simultaneously fostering growth in human factors science fields such as human factors engineering, human–computer interaction, and user experience (Xu & Ge, 2020). As illustrated in Figure 1, the relationship initially followed a pre–World War II paradigm of *human adaptation to machines (i.e., a machine-centered approach), in which* optimization relied primarily on the human side (e.g., operator skill and training). After the war, it shifted to a *machine adaptation to human* paradigm (i.e., a human-centered approach), in which system optimization was achieved through the design of human–machine interfaces in mechanical systems, marking a transition from a machine-centered to a human-centered perspective. Since then, a series of human-centered approaches and disciplines have emerged and evolved, reflecting advances in human knowledge. As computers became more prevalent, this relationship evolved into human–computer interaction, in which system optimization involves both designing effective user interfaces and providing user training. In this stage, non-AI computing systems primarily functioned as auxiliary tools to support human operations.

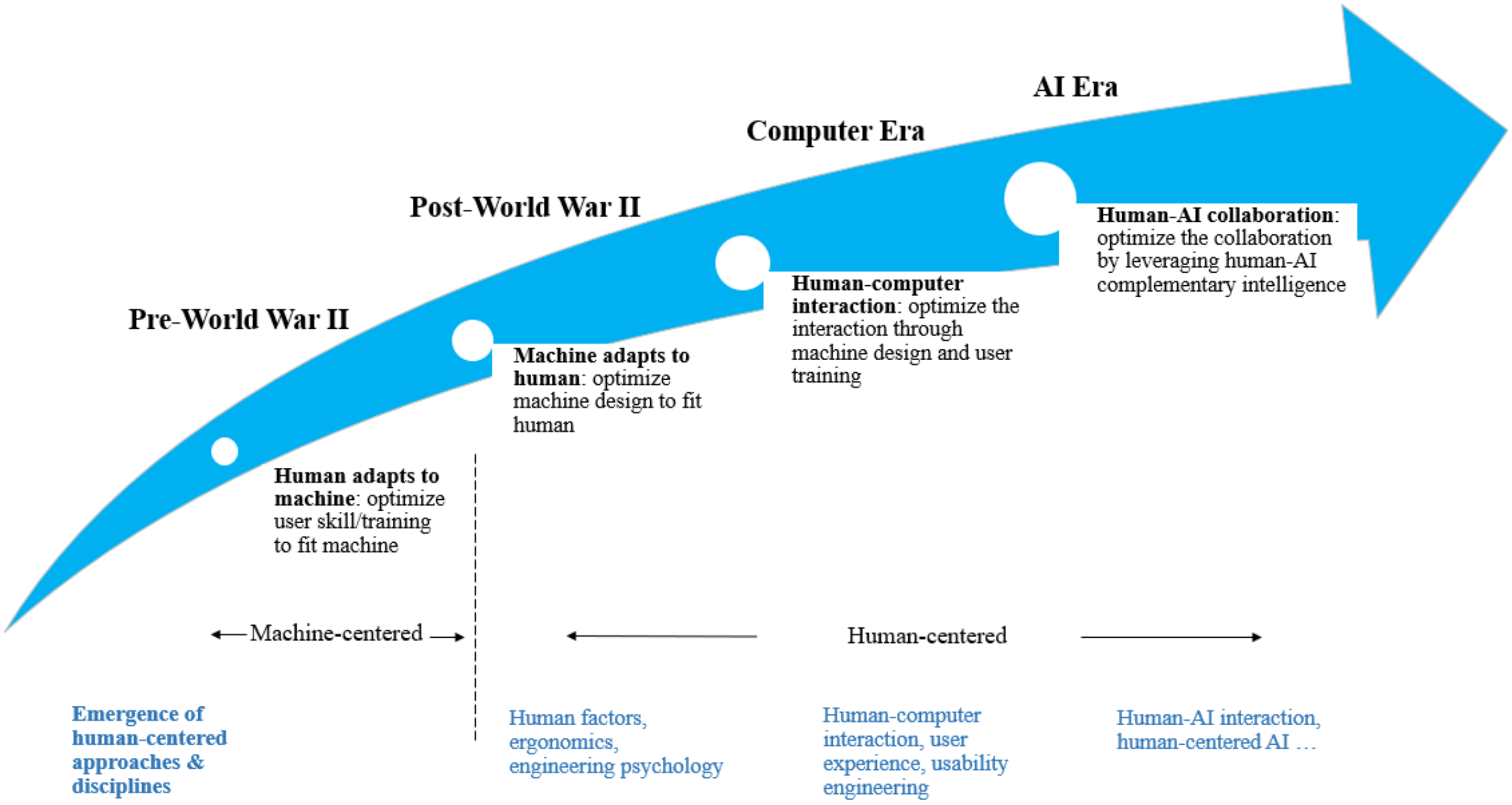

Figure 1 Evolution of the human-machine relationship across eras

The transition to human–AI interaction has also given rise to a new form of human–machine relationship. In the era of AI, rapid advances are profoundly reshaping the traditional human–machine relationship. AI agents with human-like cognitive capabilities (Table 1) are evolving beyond their earlier role as auxiliary tools that merely support human operations. They are increasingly interacting with humans in two-way, collaborative ways. This transformation gives rise to a new form of relationship, *human–AI collaboration*, in which AI systems simultaneously function as both tools and possible collaborators, embodying a dual role of *"tool + collaborator"* (Brill, Cummings et al., 2018; Shively, Lachter et al., 2018; Xu & Ge, 2020).

The human-AI collaboration can be further represented by a conceptual framework of human-AI joint cognitive systems (HAJCS) (Xu & Gao, 2024). As illustrated in Figure 2, unlike traditional non-AI computing systems, this framework treats an AI system (with one or more AI agents) as a cognitive agent capable of performing some human-like cognitive tasks. Therefore, a human-AI system can be represented as a human-AI joint cognitive system in which two cognitive agents interact collaboratively. Depending on the level of technical advancement, an AI system can perceive, recognize, learn, and reason about the user's state, the environmental context, and other relevant factors, and subsequently carry out more automated actions (Kaber, 2018). Thus, AI technologies enable the shift toward bidirectional, collaborative interaction.

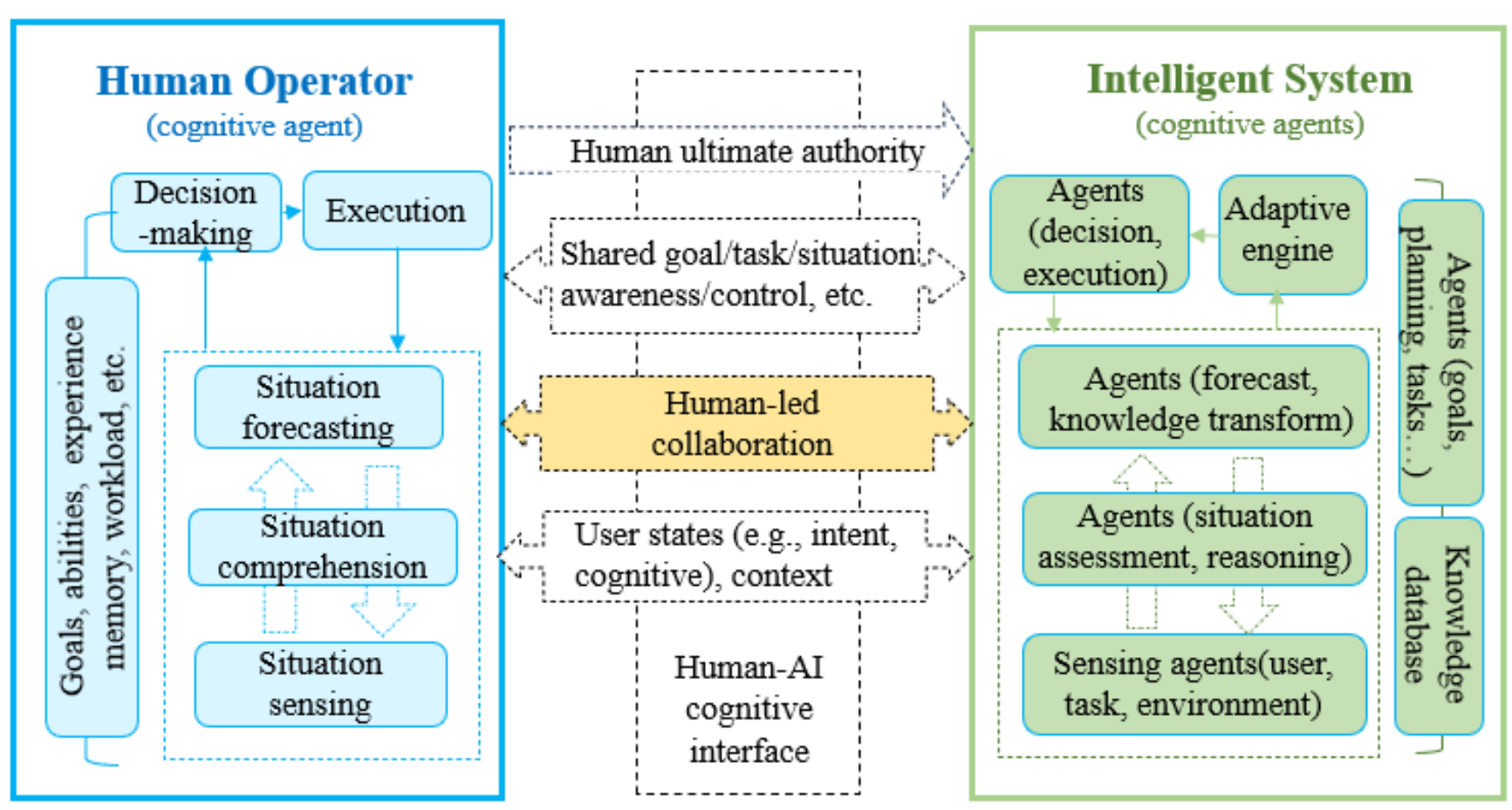

Figure 2 A conceptual framework of human-AI joint cognitive systems (HAJCS)
[adapted from (Xu & Gao, 2024)]

This framework uses Endsley's situation awareness theory to represent the information-processing mechanisms of human and machine cognitive agents (Endsley, 1995), specifically the cognitive processes of human operators in perceiving, understanding the current environmental state, and predicting future states. As shown in Figure 2, this framework employs a similar cognitive mechanism to model the information processing of the machine cognitive agent. Notably, the framework emphasizes human leadership and ultimate authority in human-AI collaboration. In addition, it models AI agents by reflecting how humans process information, thereby providing a conceptual reference for the development of AI systems. It views humans and AI as a unified system, where overall performance depends on their synergy. This perspective emphasizes the importance of defining interdependent roles and strengthening mutual collaboration to improve joint performance.

Nevertheless, the debate persists over whether AI should function as a teammate or remain a tool in its interactions with humans (NAS, 2021; Naikar et al., 2025). Critics, who have valid concerns, argue that adopting the "teaming" paradigm may contradict HCAI principles, risking a loss of human oversight and control (Shneiderman, 2021, 2022). From the HCAI perspective, as emphasized by the human-AI joint cognitive systems framework, any development of AI systems inspired by the human–AI teaming paradigm must retain human leadership roles and ultimate authority over AI systems, ensuring that AI technology remains under human control to serve humans rather than a fully autonomous system to harm them (Xu & Gao, 2024; Gao, Q., Xu, et al., 2025).

Thus, the emerging human–machine relationship, *human–AI collaboration*, is neither purely instrumental nor tool-like, nor competitive or replacement-oriented. Instead, it represents a complementary, possible collaborative partnership between humans and AI systems, while emphasizing the human leadership role. Importantly, the emerging human-machine relationship has design implications for AI systems. AI systems can complement human capabilities and achieve synergistic performance gains that neither humans nor AI alone can attain through approaches such as human–AI hybrid intelligence (NAS, 2021; Xu & Gao, 2024; Pileggi, 2024). The performance of a human-AI system depends not only on the capability of individual components but also on the complementarity and collaboration between humans and AI systems. Provided that human leadership and ultimate authority are maintained in collaborations, optimizing this collaboration between humans and AI in design can facilitate the development of human-centered AI systems (Gao, Q., Xu, et al., 2025).

### 2.3 The double-edged sword effect of AI

While AI technologies continue to advance rapidly, they also present a *double-edged sword effect*. This effect highlights the dual nature of AI: when developed and applied responsibly, it can significantly enhance human welfare and societal progress; however, when misused, poorly designed, or inadequately governed, it may introduce significant ethical, social, and security risks (Raunak & Kuhn, 2024, 2025). For example, early research by Yampolskiy et al. (2019) found that deploying unsuitable AI systems has led to incidents that harm fairness, justice, and safety. Another example is that generative AI poses important challenges because its fluent, human-like outputs can strongly influence how people think, decide, and act. Core issues include overreliance and automation bias, poor user understanding of probabilistic generation and hallucinations, erosion of human skills through cognitive offloading, and reduced meaningful human control as AI shifts from a tool to a perceived collaborator. Generative AI may also amplify bias, obscure accountability, and shape human choices through subtle framing effects (Kang et al., 2025; Fields et al., 2025). Several projects and initiatives use these public data to demonstrate the utility and importance of AI incident reporting for various AI safety stakeholders, such as Organisation for Economic Co-operation's (OECD) automated AI Incident Monitor (AIM) (OECD, 2025), the Database of AI Litigation (DAIL, 2025), the AI Incident Database (AIID, 2025), the AI, Algorithmic, Automation Incidents and Controversies (AIAAIC) repository (AIAAIC, 2025), and the Political Deepfakes Incident Database (PDID) (Walker et al., 2024). Specifically, the AI Incident Database has documented more than 1,000 AI-related accidents (AIID, 2025), including cases such as self-driving vehicles fatally striking pedestrians, trading algorithms triggering market "flash crashes," and facial recognition errors leading to wrongful arrests. Similarly, according to the AIAAIC database, the number of newly reported AI-related incidents and controversies was approximately twenty-six times greater than in 2012 (Stanford Institute of HAI, 2025). Research has highlighted several main limitations of AI technologies (NAS, 2021; Xu Wei, 2019, 2020; Endsley, 2023; Ozmen Garibay et al., 2023).

- *Vulnerability*: AI performs well within the scope covered by its programming or training data, but when it encounters new behavior categories or different data statistical distributions that are different from previous learning, it may perform poorly, leading to performance defects during the learning cycle.
- *Perception Limitations*: If the information input is incorrect, it may disrupt AI's learning of higher cognitive processes.
- *Potential Bias*: Limited training datasets or biases inherent in the data itself can lead to potential output biases in intelligent systems.
- *Uninterpretability*: The "black box effect" and opacity issues of machine learning (ML) algorithms make the outputs of intelligent systems difficult to interpret and understand.
- *Lack of Causal Models*: ML technology is based on simple pattern recognition and lacks causal patterns, leading to an inability to understand causal relationships, predict future events, simulate the impact of potential actions, reflect on past behavior, or learn.
- *Hallucination in generative AI*: The phenomenon where AI systems produce outputs (e.g., text, images, or audio) that are plausible-sounding but factually incorrect, nonsensical, or entirely fabricated. This is especially prominent in large language models, which generate information based on statistical patterns rather than verified truth and undermine user trust, accountability, and safety
- *Development Bottleneck Effect*: It is difficult for machine intelligence to simulate human higher cognitive abilities, leading to a bottleneck effect in the isolated approach of developing machine intelligence.
- *Autonomy Effect*: The autonomy feature of AI may lead to human factors problems like those of automation technology, affecting the performance of operators, including automation confusion and irony, reduced situation awareness, "out-of-the-loop" control issue, human decision bias, degradation of manual skills, etc.
- *Ethical Issues*: AI systems can generate a series of issues, including data privacy, user justice, and fairness.
- *Independent Operability*: In the foreseeable future, AI alone will not be able to handle many complex or unfamiliar situations; human involvement will remain necessary to achieve desired results. In many application scenarios, AI cannot replace humans, and humans must retain the final decision-making power.

Similarly, nuclear and biochemical technologies also have a double-edged sword effect. The rational use of nuclear and biochemical technologies can benefit humanity, but the irrational use of nuclear and biochemical technologies to manufacture and use nuclear and biochemical weapons could potentially have disastrous consequences for humanity and society. However, the development of nuclear and biochemical technologies is highly centralized, and only a few countries possess this expertise. The development and use of AI technologies are decentralized and global in nature, with relatively low barriers to entry, making effective regulation particularly challenging. Moreover, the autonomous characteristics of AI systems further complicate oversight. Consequently, the responsible and rational development and use of AI technologies are especially critical.

### 2.4 Emerging characteristics of the third wave of AI

From a broader perspective, the evolution of AI technology can be divided into three waves, each characterized by distinct technological capabilities and conceptual paradigms. Developing a thorough understanding of the trends and defining characteristics of the current wave of AI is crucial for formulating practical approaches to advance AI innovation responsibly.

As highlighted in Table 2, the third wave of AI emerged from significant technological advances that broadened the range and impact of AI applications. These advancements have shifted AI from an experimental research domain to a mature, application-driven ecosystem capable of reshaping industries and societies. Most notably, this movement represents a fundamental shift by integrating technological advances, practical design, and a design philosophy focused on *human needs* (Xu, 2019). This stage redefines AI from a purely technical pursuit into a cross-disciplinary systems engineering endeavor that extends beyond the traditional boundaries of AI and computer science. For example, AI research and development have begun to center on a set of human factors outlined below.

Table 2 The Three Waves of AI and Their Key Characteristics (adapted from Xu, 2019)

| | **First Wave** (1950s-1980s) | **Second Wave** (1990s-2010s) | **Third Wave** (2010s – present) |
|---|---|---|---|
| **Main characteristics of technologies** | Early "symbolism and connectionism" models, initial expert systems | Statistical models, artificial neural networks in pattern recognition, and expert systems | Breakthrough in deep machine learning, big data, and computing power |
| **Human needs** | Not satisfied | Not satisfied | Beginning to deliver practical AI solutions that address real-world problems in daily work and life, with an emphasis on ethically aligned design, human needs, explainable AI, intelligent human–AI interfaces, and enhanced user experiences, etc. |
| **Focus** | Technical solutions | Technical solutions | Beginning to explore integrated solutions that address applications, practical usages, technological advancements, human and societal benefits of AI, and AI risk management and governance. |
| **Characteristics** | Academia-driven | Academia-driven | Technological enhancement + application-oriented solutions + human-centered approach |

*User Needs and AI Usage.* AI is increasingly addressing genuine user needs within specific application contexts. People identify practical scenarios where AI-based solutions deliver tangible value to individuals, organizations, and society. Consequently, more user-facing AI applications are emerging to solve real-world problems, accompanied by sustainable business models built around these solutions. This marks a fundamental departure from the first two waves of AI.

*Human–Machine Hybrid Enhanced Intelligence.* AI research and applications have placed growing emphasis on integrating human roles into AI systems, promoting a paradigmatic shift toward viewing humans and machines as components of a unified human–machine system. Scholars argue that, regardless of how advanced machines become, they cannot fully replicate higher-order human cognitive abilities. Developing machine intelligence in isolation has thus become a bottleneck constraining the sustainable evolution of AI (Zheng et al., 2017). To address these limitations, researchers advocate for human–machine hybrid enhanced intelligence, such as human-in-the-loop and human-on-the-loop systems, in which humans serve as cognitive or decision-making nodes within the intelligent loop, enabling more robust hybrid intelligence.

*"Data + Knowledge" Dual-Driven AI.* Earlier AI technologies primarily relied on knowledge bases and inference engines to simulate human reasoning (e.g., IBM's Deep Blue). While interpretable, such systems faced persistent challenges in knowledge collection, representation, modeling, algorithmic conversion, and execution. The third wave of AI, driven by big data and deep learning, has achieved broad generalization capabilities, exemplified by large language models (LLMs), yet faces new bottlenecks, including algorithmic fragility, data- and computation-dependence, low robustness, and a lack of interpretability and reliability. The generative AI (e.g., LLM) thus requires a *dual drive* of data and knowledge, leveraging the four essential elements of human knowledge, data, algorithms, and computing power to overcome the limitations of prior approaches (Nie et al., 2023; Wang & Wang, 2024).

*AI Interpretability.* People increasingly recognize the severity of the "black box" problem. AI systems must generate transparent, interpretable results that allow users to understand underlying reasoning processes and make informed decisions, thereby fostering trust in AI technology (Gunning et al., 2019).

*Ethical AI.* As ethical questions about AI become more prominent, both researchers and policymakers are examining its social and moral impacts. Many governments and companies have set up ethical AI frameworks focused on human values, fairness, data protection, and privacy to address ethical issues (Zhou & Chen, 2023).

*Human-Controllable AI.* As AI systems become increasingly autonomous, maintaining meaningful human oversight in critical decision-making is essential. Researchers have proposed concepts such as *meaningful human control* to ensure human accountability and prevent unintended consequences arising from AI-based decision-making (de Sio & den Hoven, 2018).

Collectively, these *human factors* encompass human needs, characteristics, values, roles, knowledge, and abilities, which are absent from the first two waves of AI. This represents a paradigmatic shift from a *technology-centered* to a *human-centered* mindset. Although current AI technologies have not yet achieved these human-centered goals, these human factors will shape the direction of AI technology. This shift not only complements existing AI development approaches but also addresses their limitations, ensuring that AI evolves in a manner that is sustainable, trustworthy, and aligned with human well-being.

Thus, the third wave of AI can be characterized by the integration of *technological enhancements, application-oriented solutions, and a human-centered approach,* a triad that defines the future direction of AI development (see Table 1) (Xu, 2019). It calls for an integrative approach rooted in systems thinking to address emerging challenges related to human factors, ethics, and sociotechnical integration (Xu, Dainoff, Ge, & Gao, 2023). The third wave thus emphasizes not only *what* AI can do, but also *how* it should be developed and applied to advance human values and societal well-being responsibly. Humans still hold an irreplaceable role in the development and use of AI systems.

In conclusion, advancements in human-AI interaction have led to an innovative relationship between humans and AI systems. Although these advancements bring notable advantages, AI technologies also introduce potential risks. It is essential to implement strategies that optimize human benefits while mitigating the risks.

## 3. Human-Centered AI (HCAI)

### 3.1 HCAI Concepts and Frameworks

In response to the challenges posed by advancing AI, researchers have begun exploring a *human-centered AI (HCAI)* approach. In 2018, Stanford University established the first "Human-Centered AI (HAI)" research institute (Li, 2018; 2023). Earlier researchers, including Shneiderman (2020), Xu (2019), Riedl (2019), Dignum and Dignum (2020), AI-HLEG (2020), and Auemhammer (2020), have introduced various HCAI frameworks. This section highlights key concepts and frameworks in HCAI, presenting a comprehensive, collected overview across scholars and a variety of perspectives.

#### 3.1.1 Ben Shneiderman's frameworks

1) The Two-Dimensional HCAI Framework

Shneiderman (2020a) introduced a two-dimensional framework for HCAI, characterized by the *degree of human control* and the *degree of AI autonomy* (see Figure 3**)**. This framework emphasizes that greater autonomy does not necessarily imply less human control; instead, it encourages the design of AI systems that remain controllable by humans.

Figure 3 The two-dimensional framework with the goal of reliable, safe, and trustworthy AI (Shneiderman, 2020a)

Shneiderman's key insight is to decouple "levels of automation" from "levels of human control." Rather than accepting the traditional one-dimensional trade-off, where greater automation implies reduced human control (Parasuraman et al., 2000), he proposes that designers should often pursue both high automation and high human control simultaneously. This combination leads to systems that are not only reliable, safe, and trustworthy but also enhance human performance. As illustrated in Figure 3, the quadrant framework also helps reason about system failures: over-automation (high automation, low human control) and over-control (high human control, low automation), each introduce distinct risks. The main point is about context: when quick, automated actions are needed, where human learning matters most, and how systems, user interfaces, and governance can improve both control and automation simultaneously (Shneiderman, 2020a).

2) HCAI Design Metaphors

To move designers beyond the assumption of "AI as an autonomous teammate by default," Shneiderman (2022) introduced four metaphors that foreground system design (Figure 4). *Supertools* highlight powerful user interfaces that amplify human intent, characterized by information-rich displays, direct manipulation, and reversible actions. *Tele-bots* maintain human authority over remote effectors through teleoperation with continuous, high-fidelity feedback. *Active appliances* encapsulate autonomy within clearly defined, well-understood task boundaries, such as those found in modern washing machines and elevators. Finally, *control centers* orchestrate automated systems through human oversight, auditability, and responsive intervention mechanisms. The unifying principle across these metaphors is clarity of responsibility, achieved through effective human-AI interaction and user interfaces of AI systems.

**Design Metaphors**

| | | |
|---|---|---|
| **Intelligent Agents**<br>Manifests Cognitive & Linguistic Abilities, Appears to Think & Be Knowledgeable | ❯ | **Supertools**<br>Augments Human Abilities, Empowers Users, Enhances Human Performance |
| **Teammates**<br>Acts as a Partner or Collaborator, Interacts using Natural Language | ❯ | **Tele-bots**<br>Boosts Human Perceptual & Motor Skills, Enables Remote Operations |
| **Assured Autonomy**<br>Sets Goals, Makes Independent Decisions, Improves Behavior | ❯ | **Control Centers**<br>Supports Human Control & Situation Awareness, Enables Preventive Actions |
| **Social Robots**<br>Anthropomorphic, Humanoid, Bionic, Bio-inspired, Emotionally Intelligent | ❯ | **Active Appliances**<br>Consumer-oriented, Widely Used, Low Cost, Easy to Use, Reliable |

Figure 4 HCAI Design Metaphors: Supertools, Tele-bots, Active Appliances, and Control Center (Shneiderman, 2022)

This reframing does not deny the use of social robots or teammate metaphors, provided that humans remain the final authority for AI systems in decision-making and controllability. In practice, "supertools" plus "control centers" map neatly to the HCAI quadrant's target region (high control and high automation) as illustrated in Figure 4, while "active appliances" live where tasks are standardized, and "tele-bots" keep human-in-the-loop in the high-stakes environments (Shneiderman, 2022).

3) AI's Two Grand Goals and Compromised Design Strategy

Furthermore, Shneiderman (2020c) defined the aims of AI into two overarching goals: human emulation and useful application (Figure 5). *Emulation* seeks to replicate human cognition, perception, and motor abilities to build AI systems that perform tasks as well as or better than humans. *Application* aims to develop widely used, practical AI products and services that enhance human capabilities while preserving human oversight. Shneiderman argues that confusion between these goals, such as treating computers as autonomous teammates or humanoid robots as necessary design forms, creates potential risks and unrealistic expectations. He proposes reframing AI design toward HCAI systems that amplify human performance, emphasizing supervisory control, teleoperation, and mechanoid appliances rather than autonomous systems.

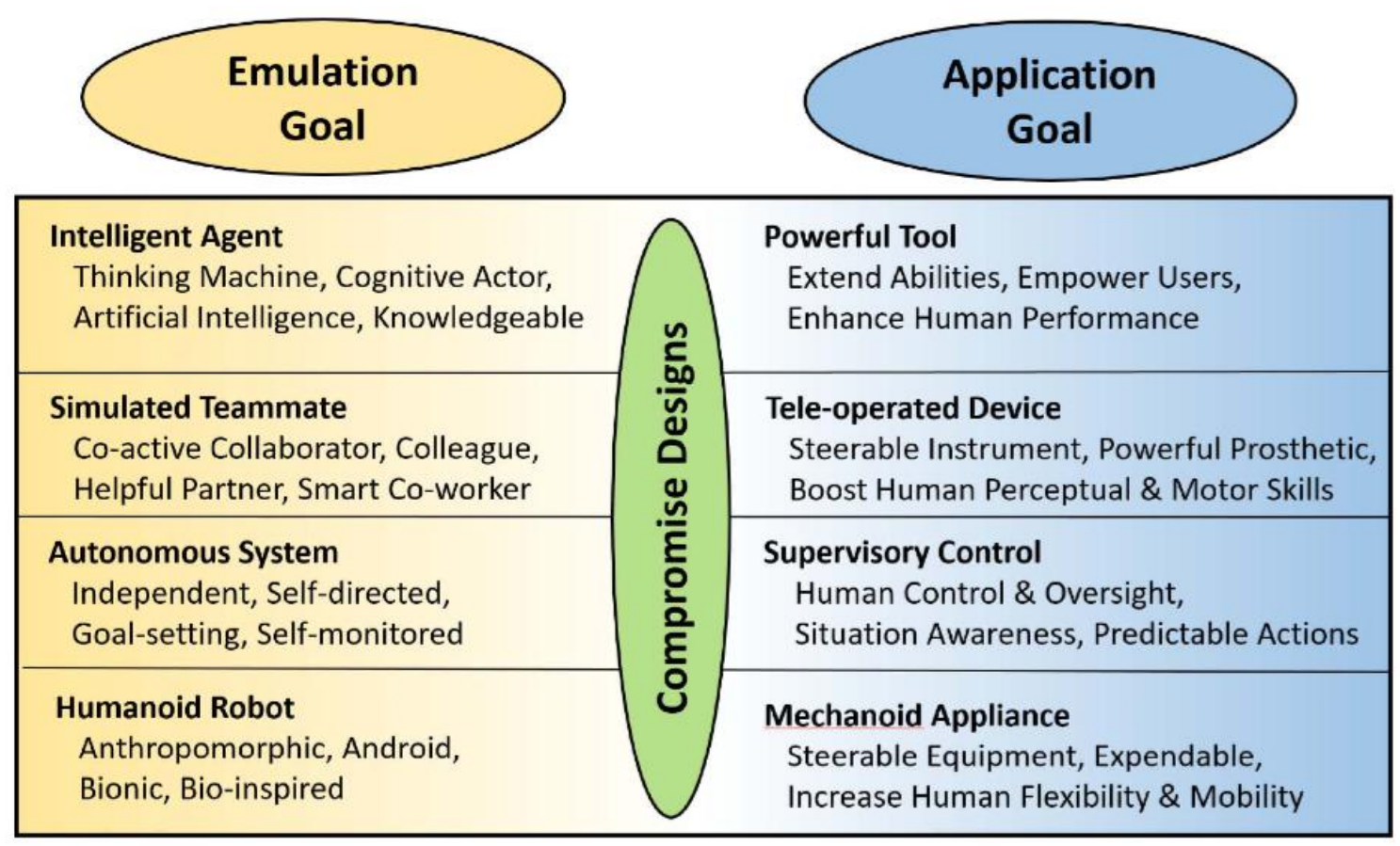

Figure 5 Two grand AI goals and compromise strategies for HCAI systems (Shneiderman, 2020c)

In outlining the two grand goals of AI, Shneiderman identifies four recurring design mismatches: intelligent agent vs. powerful tool, simulated teammate vs. teleoperated device, autonomous system vs. supervisory control, and humanoid robot vs. mechanoid appliance (Shneiderman, 2020c). He suggests *compromise strategies* integrating AI with human-computer interaction (HCI) methods. He stresses that trustworthy AI requires reliability, safety, and user control achieved through comprehensible, predictable, and controllable user interfaces. By combining AI's computational power with HCI principles (e.g., direct manipulation and feedback loops), designers can create systems that strengthen human creativity and accountability. Shneiderman calls for human-centered AI applications that deliver real societal benefits across domains such as healthcare, education, and environmental sustainability, shifting the focus from machines that think like humans to technologies that empower humans.

4) Governance Structures for reliable, safe, and trustworthy AI

To adopt and promote HCAI, Shneiderman (2020b) proposes a governance structure to achieve HCAI goals: reliable, safe, and trustworthy (RST) AI (see Figure 6). The governance structures include four levels: (1) Team: reliable systems based on software engineering practices; (2) organization: a well-developed safety culture based on sound management strategies; (3) industry: trustworthy certification by external review; and (4) government regulation.

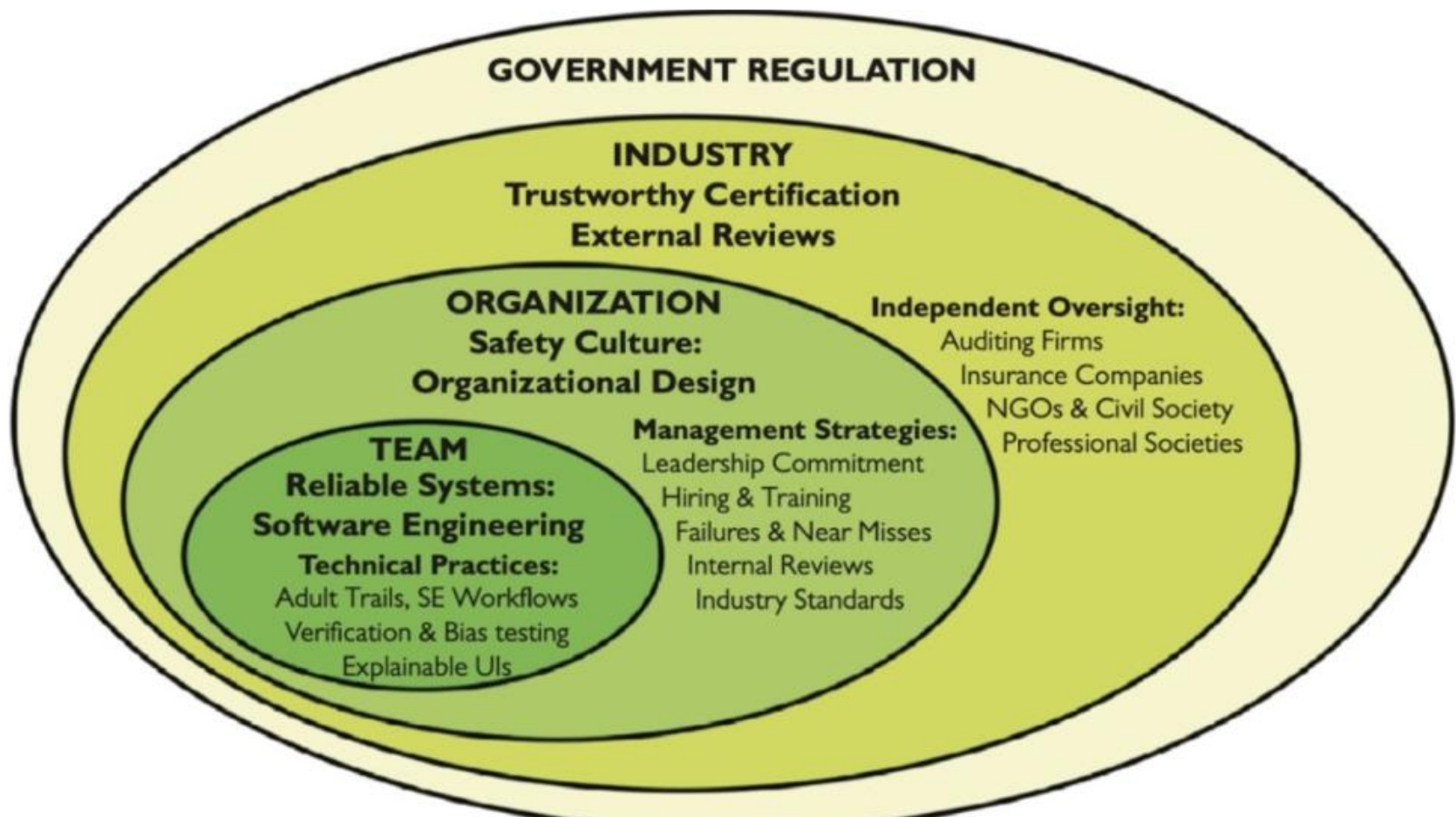

Figure 6 Governance Structures for HCAI (Shneiderman, 2020b)

Within this governance structure, Shneiderman advances beyond abstract principles to operational governance through a set of structured mechanisms for HCAI. He delineates three complementary pillars: (1) *reliability*, achieved through rigorous engineering practices such as validation, bias testing, and audit trails; (2) *safety culture*, fostered through organizational management strategies including leadership commitment, near-miss reporting, and internal review boards; and (3) *trustworthy certification*, ensured through independent oversight by auditors, standards bodies, insurers, and regulators. Across these pillars, he articulates fifteen actionable recommendations spanning the team, organizational, and industry/regulatory levels to translate ethical values into consistent operational practice. The central argument is that HCAI cannot be realized through algorithms alone; it demands robust processes, verifiable evidence, and enduring accountability.

In summary, Ben Shneiderman's HCAI frameworks integrate conceptual, design, governance, and practical dimensions into a unified vision: AI should achieve both high automation and high human control to deliver systems that are reliable, safe, and trustworthy. His two-dimensional framework breaks the false trade-off between autonomy and control. In contrast, the design metaphors (i.e., supertools, tele-bots, active appliances, and control centers) reframe AI as empowering tools rather than autonomous teammates. Complementing this, his focus on augmenting and empowering humans emphasizes applications over emulation. To operationalize ethics, he advances multi-level governance structures spanning teams, organizations, industry, and government, ensuring accountability and certification. Together, these contributions provide a pragmatic, human-centered roadmap for building AI that amplifies human performance, strengthens responsibility, and enhances societal benefit.

### 3.1.2 Wei Xu's frameworks

1) The Technology–Human Factors–Ethics (THE) Triangle Framework

As one of the earliest HCAI frameworks, Xu (2019) proposed the *Technology–Human Factors–Ethics (THE) Triangle*, which integrates technological innovation, human factors principles, and ethical alignment (see Figure 7 and Table 3). This framework emphasizes that the design and deployment of AI must simultaneously *integrate* the three perspectives. This HCAI framework represents a critical early attempt to promote the HCAI approach. It emphasizes a systematic design approach that fosters the synergy and complementarity among these perspectives. It therefore advocates a design mindset grounded in the systematic, balanced integration of technology, human factors, and ethical principles throughout the AI lifecycle, from conception to deployment and monitoring.

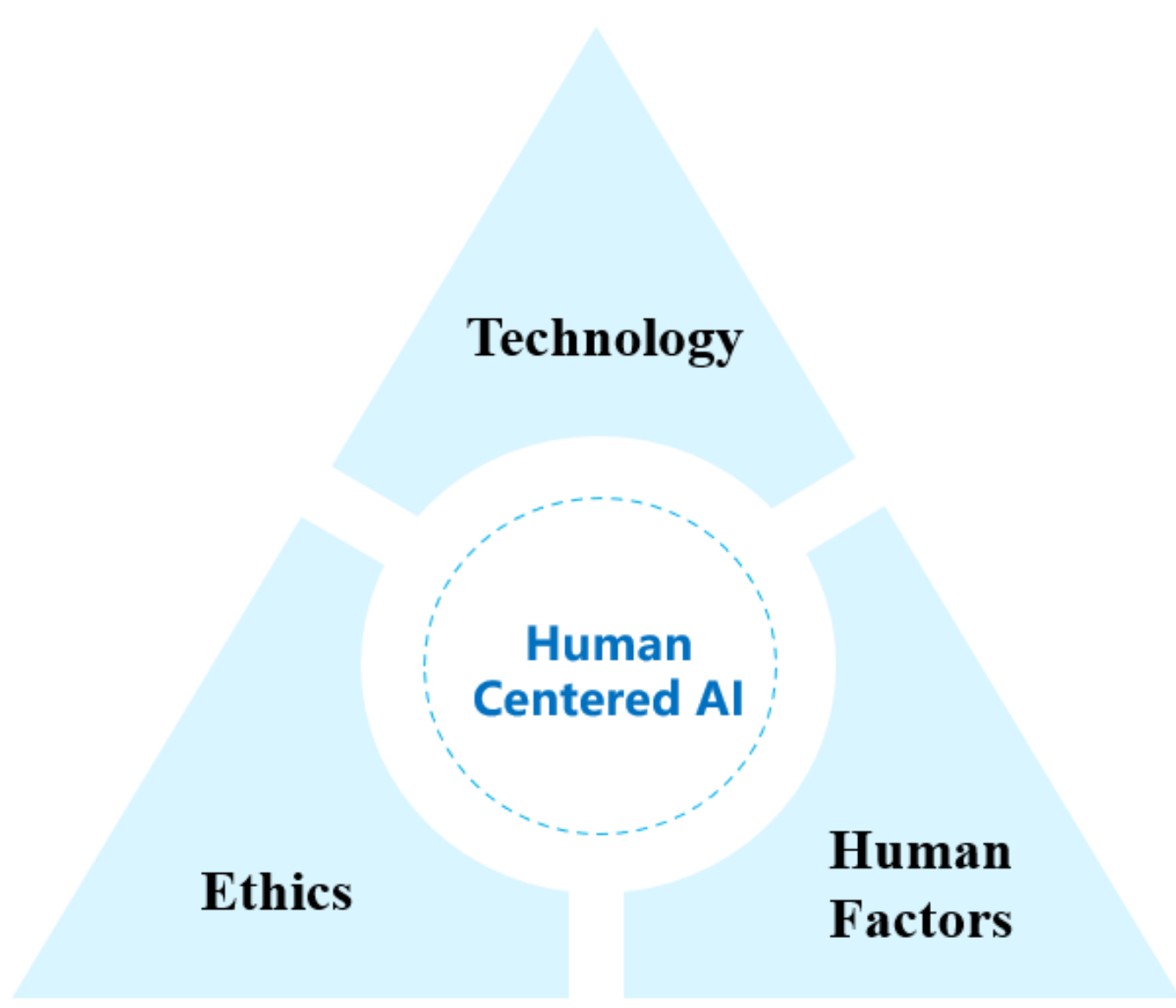

Figure 7 The Technology–Human Factors–Ethics (THE) Triangle HCAI Framework
and the optimal integration zone for human-centered AI solutions [adapted from (Xu, 2019)]

Table 3  The three perspectives and key factors of the THA Triangle HCAI framework

| Perspectives | Key Factors | Implications |
| --- | --- | --- |
| **Technology** | Model/algorithm, interaction technology, robustness & reliability, security & resilience, explainability mechanism, data quality & representation, scalability & efficiency, system integration, scalability and efficiency | • Defines *what* can be built and how safely and effectively it performs<br>• Provides *what:* the actual capability and feasibility<br>• Enables human-centered/ethical practices |
| **Human Factors** | User experience, accessibility, trust & interpretability, cognitive compatibility & mental model, emotional intelligence, human controllability, human reskilling, training & education | • Guide to *how* humans and AI should interact<br>• Guide the design of AI solutions<br>• Provides the principles for design |
| **Ethics** | Fairness & non-discrimination, transparency, accountability & auditability, culture, privacy & data protection, sustainability, inclusivity & justice, regulatory compliance | • Guides to *what* should be done<br>• Motivates and constrains AI use<br>• Provides *why:* the normative justifications and societal values |

Specifically, if AI development focuses solely on technological considerations while neglecting ethical considerations, it risks harming humanity; if it neglects human factors, AI adoption will falter, and economic returns will be limited. Conversely, if it emphasizes human needs and ethics but neglects technological advancements, the result may lack innovation and scalability. Thus, a viable HCAI solution resides in the overlapping middle area, as illustrated in Figure 7, which represents the *optimal integration zone*. Its goal is to ensure that AI serves and empowers humans, enhancing their abilities and well-being rather than replacing or harming them. To achieve its goals, the HCAI framework underscores the importance of interdisciplinary collaboration to ensure a holistic approach to HCAI practice.

2) Hierarchical Human-Centered AI (hHCAI)

Sociotechnical systems (STS) theory, originating from the mid-20th century, emphasized the joint optimization of social and technical subsystems (Trist & Bamforth, 1951; Clegg, 2000). Xu and Gao (2025) extend this tradition into the era of AI by proposing an Intelligent Sociotechnical Systems (iSTS) framework. The central insight is that the classical joint optimization is insufficient in the face of AI technologies; instead, human-AI joint optimization must be prioritized**.** This shift ensures that advances in AI technology serve human values and organizational goals without marginalizing human well-being. iSTS further conceptualizes sociotechnical system design for AI across multiple levels, beginning with individual AI systems, extending to interactions among multiple systems within organizational or ecosystem contexts, and ultimately scaling to a societal level. The iSTS framework thus offers a multi-layered approach to embedding AI within human-centered sociotechnical environments, positioning it as a foundational concept for the design of large-scale AI systems.

Inspired by the iSTS and other frameworks (Xu & Gao, 2025; Herrmann & Pfeiffer, 2023), Xu and Gao (2025) further proposed a conceptual framework of *Hierarchical Human-Centered AI (hHCAI)* (see Figure 8). As shown in Figure 8, hHCAI extends the scope of HCAI practice; it frames HCAI as a multi-level paradigm shift: individual → organizational → ecosystem → macrosocial, encompassing individual human-AI systems (keeping human-in/on/the loop), organizational environments (keeping organization-in-the-loop), cross-system ecosystems (keeping ecosystem-in-the-loop), and macrosocial contexts (keeping society-in-the-loop). In practice, it provides organizations and policymakers with a structured foundation for aligning AI development with human and societal needs, ensuring that HCAI advances are both theoretically grounded and operationally impactful. hHCAI highlights that sustainable and responsible AI adoption requires coherence across all levels simultaneously. Improving algorithms or focusing only on specific human-AI systems does not suffice; without alignment at the organizational, ecosystem, and societal levels, AI may fail due to broader systemic issues. This hierarchical lens provides both researchers and practitioners with a scalable roadmap for implementing HCAI at multiple levels. For further elaboration, refer to Section 4.3 of this chapter.

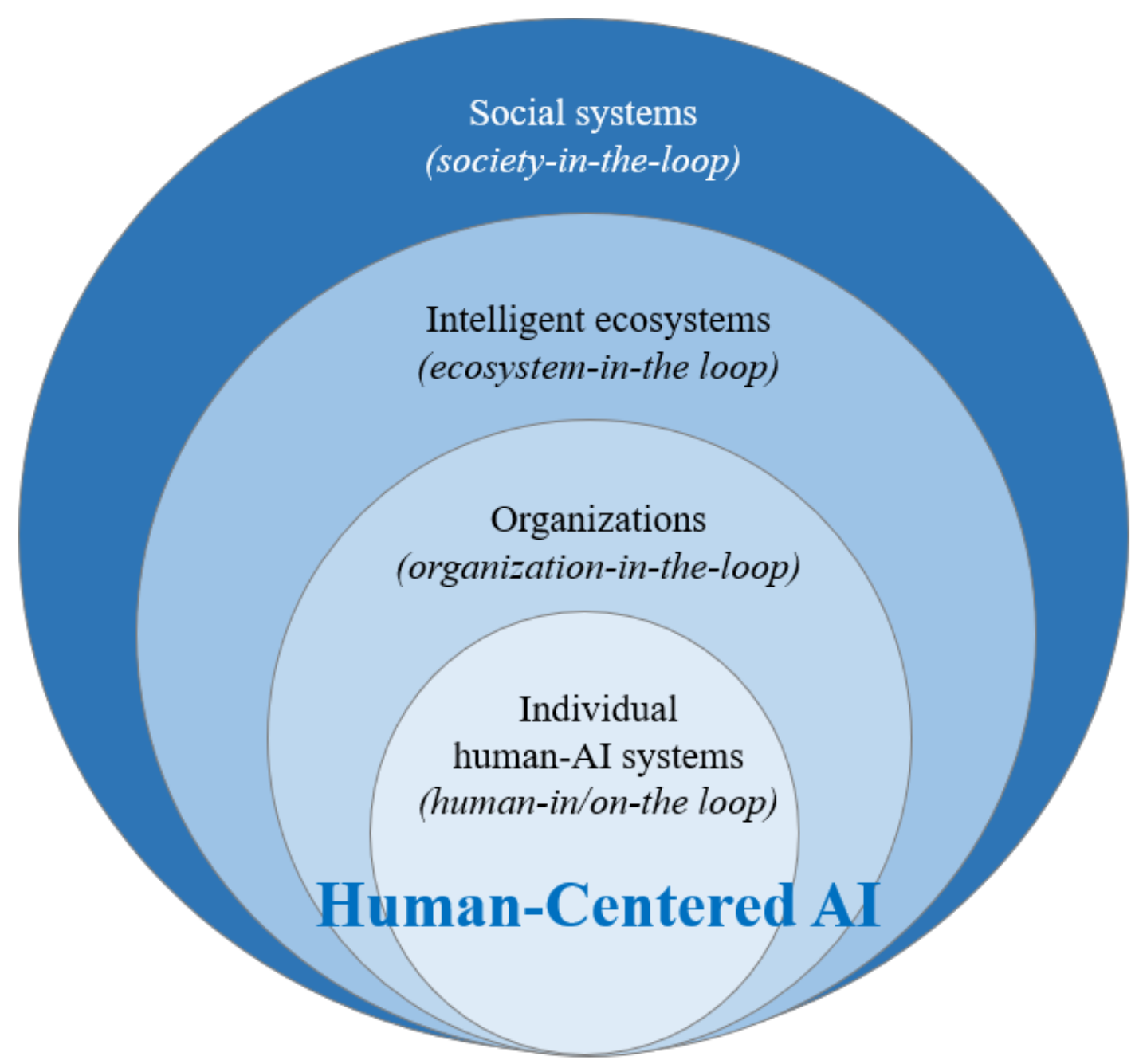

Figure 8 Hierarchical human-centered AI (hHCAI) framework [adapted from (Xu & Gao, 2025)]

3) HCAI Methodological Framework (HCAI-MF)

Despite its strong conceptual foundation, implementing HCAI remains challenging. A key obstacle is the lack of comprehensive methodologies to bridge the gap between HCAI design philosophy and practical execution (Hartikainen et al., 2022; Bingley et al, 2023; Capel et al., 2023; Mazarakis et al., 2023). To address these gaps and advance HCAI adoption, Xu, Gao, and Dainoff (2025) introduced a comprehensive HCAI methodological framework (HCAI-MF) that provides structured, practical guidance for HCAI implementation.

As shown in Figure 9, HCAI-MF includes five components: HCAI requirement hierarchy (i.e., HCAI guiding principles, HCAI design guidelines, and HCAI product requirements), HCAI method taxonomy (e.g., HCAI strategy, human-centered computing, human-AI interaction technology and design, human controllability, human-centered AI risk management and governance), HCAI process, HCAI interdisciplinary collaboration approach, and HCAI multi-level design paradigms. HCAI-MF aims to transform HCAI from a design philosophy into a systematic, actionable practice methodology, providing structured guidance that bridges the gap between conceptual principles and real-world implementation. For further elaboration, refer to Section 4 of this chapter.

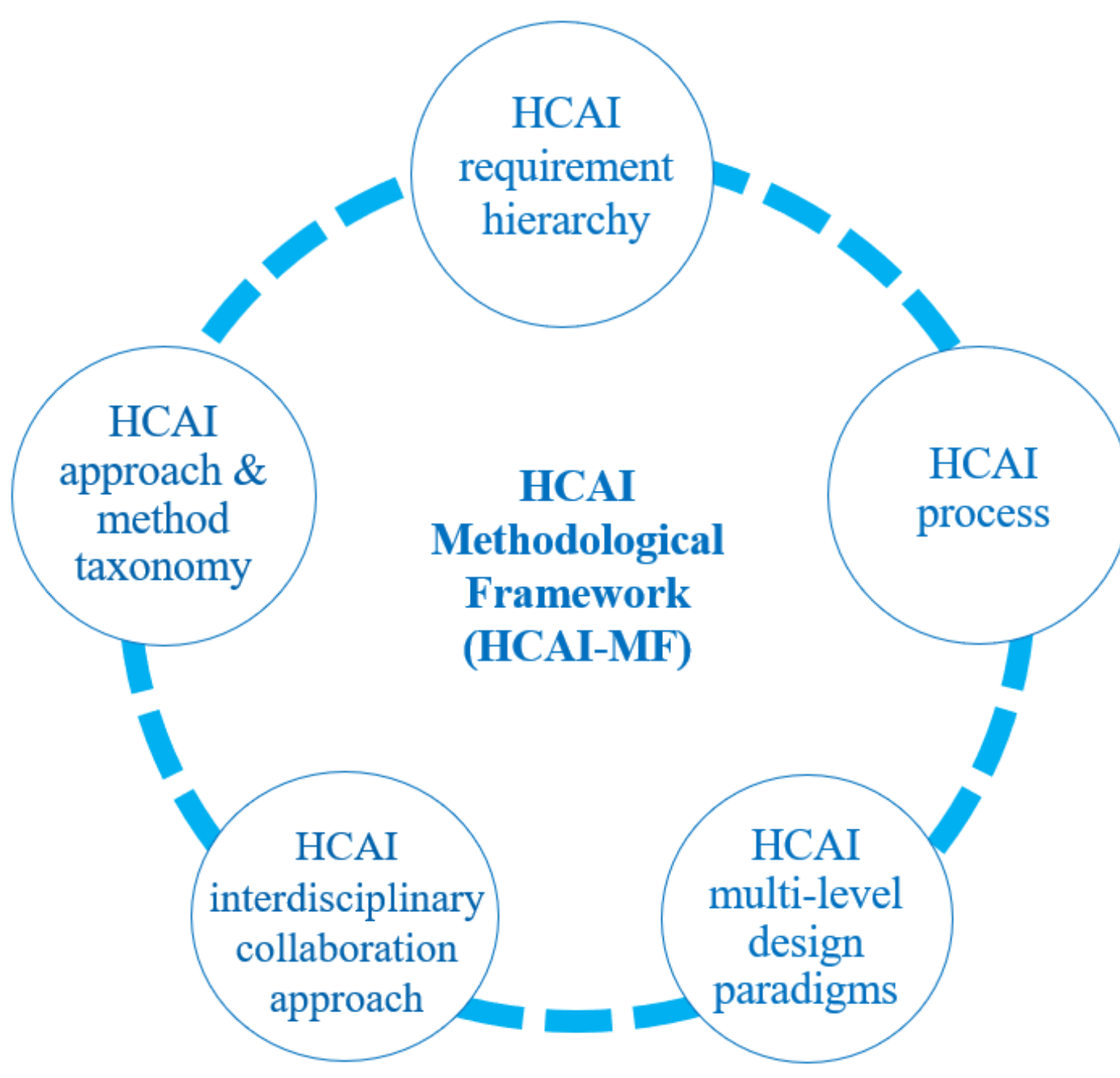

Figure 9 An HCAI methodological framework (HCAI-MF) (Xu, Gao, & Dainoff, 2025)

4) HCAI Maturity Model (HCAI-MM)

An HCAI maturity model provides organizations with a structured, progressive roadmap for operationalizing HCAI (Hartikainen et al., 2023). To facilitate the adoption of HCAI in organizations, Winby and Xu (2025) proposed a Human-Centered AI Maturity Model (HCAI-MM). As illustrated in Figure 10, the HCAI-MF offers a staged framework that organizations can use to assess and enhance their HCAI integration, progressing from ad hoc efforts to fully optimized, continuously improving practices. Each stage reflects greater institutionalization of principles, processes, and metrics for HCAI practice.

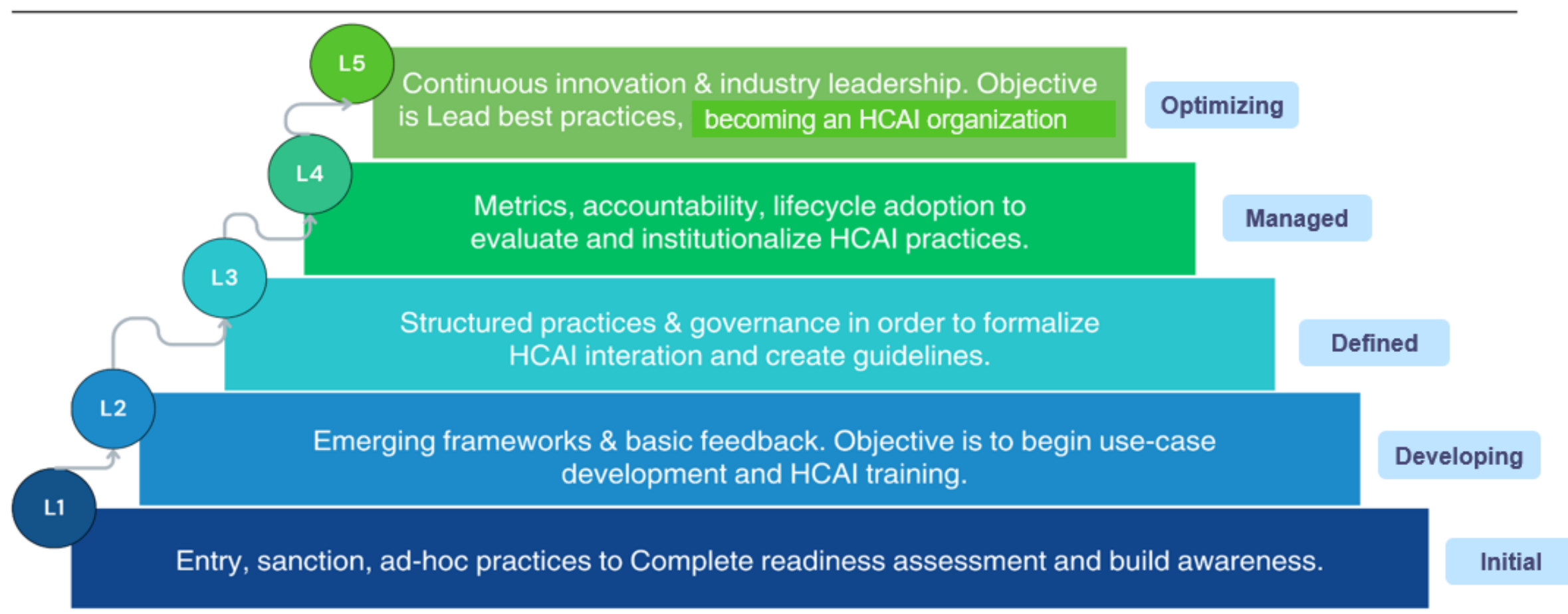

Figure 10: High-level description and objectives of HCAI Maturity Model Stages (Winby & Xu, 2025)

More specifically, HCAI-MM defines five progressive levels of organizational adoption. At *Level 1 (Ad Hoc)***,** HCAI practices are unsystematic and reliant on individual efforts. *Level 2 (Repeatable)* introduces basic, partially applied HCAI guidelines in response to external pressures. At *Level 3 (Defined)*, practices become standardized across projects, with formalized processes and cross-disciplinary collaboration. *Level 4 (Managed)* integrates continuous monitoring and measurable HCAI indicators such as fairness audits, usability metrics, and incident reporting pipelines. Finally, *Level 5 (Optimizing)* represents HCAI organizations that institutionalize HCAI as part of their culture and strategy, achieving continuous learning, predictive governance, and alignment with societal ecosystems.

Recently, Xu and his team have continued to advance HCAI frameworks in the emerging areas, such as *human-centered human-AI collaboration (HC-HAC)* (Gao, Xu et al., 2025), *human-centered human-AI interaction (HC-HAII)* (Xu, 2025), *human-centered privacy (HCP) approach to AI* (Sun, Xu, & Gao, 2025), and *human-centered artificial social intelligence (HC-ASI)* (Pan, Xu, & Gao, 2025).

In sum, Wei Xu's HCAI frameworks collectively advance a multidimensional, practice-oriented vision for HCAI. His early *Technology–Human Factors –Ethics (THE) Triangle* laid the groundwork by integrating technical robustness, human factors design, and ethical safeguards as inseparable pillars of HCAI. Building on the intelligent sociotechnical systems (iSTS) framework, the *Hierarchical HCAI (hHCAI)* framework extends this vision into a multi-layered model, spanning individual human–AI systems, organizations, ecosystems, and societal levels. To bridge HCAI design philosophy and practical execution, the *HCAI Methodology Framework (HCAI-MF)* provides structured guidance through requirements, methods, processes, and design paradigms. Complementing this, the *HCAI Maturity Model (HCAI-MM)* provides organizations with a staged pathway for assessing progress across key dimensions. These frameworks evolve from conceptual foundations → methodological guidance → organizational practice → scalable ecosystems and sociotechnical integration. Together, they form a comprehensive multi-level roadmap for HCAI, enabling its design, evaluation, and institutionalization in a scalable, systemic, and sustainable manner.

#### 3.1.3. Six HCAI grand challenges

In 2022, led by Ben Shneiderman and Gavriel Salvendy, an international consortium of twenty-six experts from academia, industry, and government reached a consensus on the foundational directions for HCAI (Ozmen Garibay, Winslow ... & Xu, 2023). The resulting *Six HCAI Grand Challenges* collectively define a vision for developing AI systems that benefit humanity, ensuring that technological advancement in AI aligns with human values, societal well-being, and sustainable progress (see Figure 11). These challenges include: (1) *Human well-being*, ensuring AI development enhances quality of life and supports both hedonic and eudaimonic aspects of flourishing; (2) *Responsible AI,* advocating for accountability and ethical design throughout the AI lifecycle; (3) *Privacy*, ensuring

AI systems respect data ownership, transparency, and individual rights; (4) *Human-centered design and evaluation*, emphasizing human-centered participatory and iterative methods to align AI systems with user needs and values; (5) *Governance and oversight*, calling for regulatory frameworks that safeguard human and ecological well-being; and (6) *Human–AI interaction*, focusing on developing AI that complements human cognition and promotes mutual understanding in decision-making contexts.

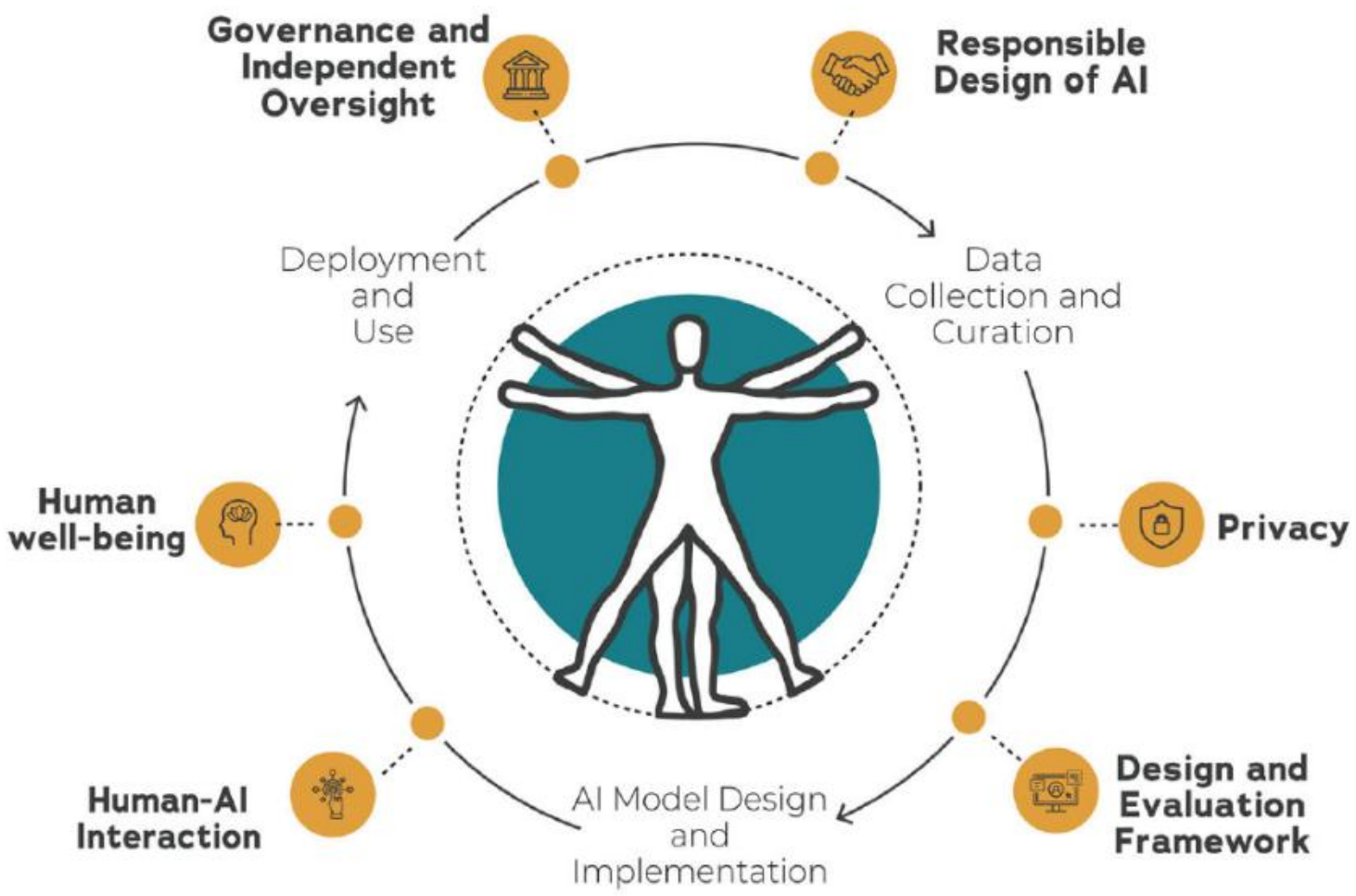

Figure 11 Six HCAI grand challenges (Ozmen Garibay et al., 2023)

To address these challenges, Ozmen Garibay et al. (2023) argue that HCAI approaches should integrate sociotechnical design, ethics-by-design, and participatory methodologies across the entire AI development lifecycle. For human well-being, this means embedding well-being metrics and user experience measures into system objectives. Responsible AI requires transparent algorithms, auditability, and continuous stakeholder engagement. Privacy is managed through privacy-preserving computing methods and clear data governance rules. Human-centered design fosters collaboration between engineers, psychologists, and ethicists to co-create systems aligned with human goals. Governance relies on multi-level frameworks integrating existing human and ecological welfare indicators. Finally, advancing human–AI interaction involves developing inclusive design paradigms, enhancing shared situation awareness, and promoting trust, and shared control in hybrid human–AI systems. The work also offers a set of actionable recommendations to engage stakeholders from diverse domains and roles in collaboratively promoting the implementation of HCAI (Ozmen Garibay et al., 2023).

### 3.1.3 Other Emerging HCAI Concepts and Frameworks

In addition to the work reviewed above, other scholars have explored HCAI from different perspectives, including Humanistic AI Design (Auernhammer, 2020), Human-Centered Explainable AI (Ehsan et al., 2020), and Human-Centered Machine Learning (Kaluarachchi et al., 2021). Recent developments in HCAI research and applications reveal an expanding diversity of domains, methodologies, and socio-technical challenges (Régis et al., 2025; Germanakos et al., 2025). For example, across sectors, HCAI has been explored in medicine (Coskun, 2026), the workplace (Coursaris et al., 2025), and beyond; on the technology front, research on human-centered large language models (LLMs) (Di Fede et al., 2026), and human-centered social computing (Ye & Wang, 2026). Collectively, these developments illustrate how HCAI is evolving into a multidimensional paradigm that bridges human values, ethical principles, and technological innovation across domains and platforms.

Scholars' continued exploration of HCAI concepts and approaches has deepened the understanding of HCAI and its practical implications. For example, to clarify the meaning and scope of HCAI, Capel and Brereton (2023) analyzed over 250 studies and identified four main themes: explainable AI, human-centered design and evaluation, human-AI teaming, and ethical AI with an emerging focus on human-AI interaction. They found that HCAI research varies widely, ranging from algorithmic modeling of humans to participatory design and ethics-driven frameworks. Explainability research seeks transparency in AI decision-making; design studies use HCI methods to align AI with user needs; human-AI teaming studies explore collaboration between humans and AI; and ethical work focuses on fairness, accountability, and human rights. The authors mapped HCAI along two axes: human values–led vs. AI-led and designed vs. in-use, showing tensions between technical and ethical approaches. They concluded that HCAI is a diverse, interdisciplinary field requiring collaboration to ensure AI enhances human agency and societal well-being.

Schmager, Pappas, and Vassilakopoulou (2025) offer a systematic review of HCAI literature and provide a comprehensive definition that captures the richness of the scholarly understanding of HCAI by focusing on the purposes, human values, and desired AI properties in the creation of AI systems through the application of human-centered design practices. They argue that HCAI seeks to augment human capabilities while maintaining human control over AI systems by considering the necessity, context, and ethical and legal conditions of the AI system, and by promoting individual and societal well-being. They offer a detailed description of the purposes (augmentation, AI-Autonomy, and automation), values (ethical, protection, and performance), and properties (oversight, comprehension, and integrity). Their review also covers HCAI processes, methods, and tools from a human-centered design perspective, emphasizing user involvement and iterative development.

Inspired by a literature survey and analysis, Desolda et al. (2025) define HCAI systems as: "AI systems that are designed, developed, and evaluated by involving users in the process, with the goal of increasing performances and satisfaction of humans in specified tasks. HCAI systems aim to be useful and usable for specified users, who might be described through a formal model, to reach their specified goals in their context of use, while being reliable, safe to use, and trustworthy." They advocate interdisciplinary collaboration in HCAI practice across fields including HCI, AI, software engineering, and ethics. Furthermore, they specify HCAI design solutions and evaluation solutions across these fields. These detailed discussions and recommendations aim to identify more pertinent approaches to creating HCAI systems.

#### 3.1.4 Summary

Despite the diverse perspectives and approaches scholars have taken to advance HCAI concepts and frameworks, they share a common understanding of HCAI as both a guiding design principle and a methodological counterbalance to technology-centered approaches to AI development. HCAI places humans at the core, emphasizing their needs, values, ethics, controllability, capabilities, and experiences throughout the development, design, deployment, and use of AI systems. Its goal is to ensure that AI serves and empowers humans, enhancing their abilities and well-being rather than functioning as fully autonomous systems that replace or harm them.

The collaborative contributions underscore that HCAI is not a fixed paradigm, but rather an evolving journey. Over time, HCAI has evolved from an early focus on ethical and technical integration to encompass multi-level organizational, ecosystem, and sociotechnical perspectives. It is now expanding from fundamental concepts and design philosophy toward practical methodologies, governance, and interdisciplinary collaboration. As more scholars advance frameworks, principles, and empirical studies, HCAI continues to move forward as a shared endeavor, fostering HCAI systems across individual, organizational, ecosystem, and societal levels.

### 3.2 HCAI Guiding Principles

HCAI concepts and frameworks provide a foundation for understanding and implementing HCAI's situatedness across humans, organizations, and sociotechnical systems. They define the "big picture" and articulate why human-centeredness matters, and how to conceptualize and operationalize it across different scales and perspectives. On the other hand, HCAI's guiding principles serve as core values or standards that set the overall direction for HCAI, influencing the creation and application of knowledge, practices, and decision-making processes. They articulate the core priorities, values, and boundaries of the HCAI field, ensuring that actions remain consistent with shared purposes and standards. These guiding principles also provide a shared language across disciplines, enabling engineers, human factors experts, ethicists, policymakers, and users to collaborate more effectively to shape AI systems.

The overall landscape of HCAI guiding principles reflects a growing convergence across academia, industry, and policy around ensuring AI systems remain aligned with human values, needs, and societal well-being (Jobin et al., 2019). Early frameworks provided conceptual foundations (e.g., Shneiderman, 2020; Xu, 2019), while global initiatives such as the EU Trustworthy AI Guidelines (AI-HLEG, 2019) and the NIST AI Risk Management Framework (NIST, 2023) translated these principles into governance and compliance structures. Across these efforts, common themes consistently emerge, such as transparency and explainability, accountability and responsibility, fairness and inclusivity, safety and reliability, privacy and data protection, and sustainability (e.g., Ozmen Garibay et al., 2023; Google PAIR, 2019; Amershi et al., 2019; Floridi et al., 2018; Richards et al., 2023; Rong et al., 2023; Chatila et al., 2019; Stefan et al., 2020; Wickramasinghe et al., 2020; Tahaei et al., 2023; Wu et al., 2022). For example, generative AI presents both powerful opportunities and critical design challenges. These systems, such as LLMs and image generators, can augment human creativity, support communication, and personalize digital experiences. However, their effectiveness hinges on ensuring that they remain accountable, transparent, and aligned with human values. Recent human-centered perspectives expand the scope by emphasizing user experience, human augmentation, human controllability, and human-AI collaboration (Xu, Gao, & Dainoff, 2025).

However, challenges of HCAI guiding principles lie in moving from aspirational values to practical implementation (Akbarighatar et al., 2024; Mäntymäki et al., 2025). While there is broad agreement on core themes such as fairness, transparency, accountability, and safety, translating these abstract principles into operational requirements, technical standards, and governance practices remains challenging (Floridi et al., 2018; Jobin et al., 2019). Principles often conflict or involve trade-offs, such as privacy versus personalization or explainability versus performance, requiring context-sensitive balancing (Sanderson et al., 2023). Cultural and contextual differences complicate efforts to create guidelines that are both globally applicable and locally relevant (Costea & Kamaldeen, 2025). Most HCAI design guidelines focus on abstract concepts such as human values, morals, and privacy, which are difficult to operationalize in practice. Although many HCAI guiding principles, such as explainable AI (Rong et al., 2024), ethical AI (Chatila et al., 2019), fair AI (Stefan et al., 2020), trustworthy AI (Wickramasinghe et al., 2020), and responsible AI (Tahaei et al., 2023), have been proposed, they often overlap conceptually. For instance, ethical, fair, and responsible AI share significant similarities.

Table 4 summarizes HCAI guiding principles into nine categories to address these challenges. These principles constitute the conceptual foundation of the HCAI design philosophy and guide efforts to realize HCAI design goals. As this area grows and changes, these key principles will likely develop further to reflect new findings and real-world experiences.

| **HCAI Guiding Principles** | **Definition and HCAI Goals** | **Illustrative examples and selected chapters of this handbook** |
|---|---|---|
| Human augmentation | Design AI to amplify, empower, and enhance human abilities and potential. This ensures AI serves as a supportive tool rather than replacing human judgment | (Chapter 4) Hybrid intelligence<br>(Chapter 20) AI-Augmented Computational Modeling of Human Behavior |
| Human controllability | Allow humans to understand, influence, oversee, and override AI systems when necessary; Ensure AI systems must remain under human authority and controllability | (Chapter 24) From Automation to Autonomy through AI: Enabling and Retaining Human Controllability<br>(Chapter 26) Human-in/on-the-loop design for human controllability |
| Ethical alignment | Develop AI in alignment with ethical and societal norms to preserve human values, foster trust, and minimize harm. | (Chapter 2) Human value alignment in AI<br>(Chapter 37) Ethical AI standards and governance: A perspective of human-centered AI |
| User Experience | Create interactions that are engaging, intuitive, accessible, and aligned with user expectations. | (Chapter 14) Design thinking and AI: Facilitating HCAI solutions<br>(Chapter 16) Designing human-centered AI experiences |
| Human-led collaboration | Ensure that humans direct and oversee collaboration with AI, maintaining control and responsibility while leveraging AI to enhance human capabilities. | (Chapter 9) Designing Artificial Social Intelligence for Human-AI Team Effectiveness<br>(Chapter 17) Human-AI co-creation: A New Interaction Paradigm for Human-AI Interaction |
| Transparency and explainability | Provide explainable and understandable AI output for humans to enhance human trust and empower informed decisions. | (Chapter 15) Human-centered AI design principles for generative AI<br>(Chapter 46) LLM explainability |
| Accountability and responsibility | Ensure the implementation of responsible AI mechanisms that establish clear accountability for humans (e.g., operators, developers) in relation to AI actions | (Chapter 25) Human-controllable AI: Meaningful Human Control<br>(Chapter 45) Trustworthy and responsible LLM |
| Safety and reliability | Prioritize human safety, maintain reliability in diverse scenarios to ensure resilience and reduce potential risks to humans. | (Chapter 30) AI risk, safety, and incident reporting<br>(Chapter 31) AI risk management frameworks |
| Sustainability | Develop AI to support environmental, social, and economic well-being to prioritize human well-being while aligning with sustainability. | (Chapter 54) Sustainable AI: An environmental sustainability perspective |

Table 4 HCAI Guiding Principles

Human-centeredness is the fundamental principle of HCAI, ensuring that AI is designed, developed, deployed, and used with humans at the center throughout the AI lifecycle. It extends beyond system-level design to ensure that decisions at every stage reflect human values and priorities. Human-centeredness also calls for embedding AI within a broader sociotechnical ecosystem, recognizing that AI affects not only individual users but also institutions, communities, and society. By approaching human-centeredness as a multi-dimensional commitment, spanning design, development, deployment, governance, and sociotechnical integration, HCAI guiding principles ensure that AI technologies consistently align with and serve human needs across scales and contexts. The following sections provide further details on these guiding principles.

### 3.2.1 Human Augmentation

Human augmentation underscores the HCAI approach that AI should enhance and empower human capabilities, rather than replace or diminish them. This principle highlights that the true value of AI lies in complementing human strengths, such as judgment, creativity, empathy, and ethical reasoning, while compensating for human limitations in areas like memory, attention, data processing, and scalability. From a design and development perspective, augmentation refers to the creation of AI systems that support decision-making, expand human cognitive capabilities, and enable new forms of problem-solving and creativity. From a deployment and use perspective, it ensures that humans remain at the center of activity, with AI tools designed to enhance performance, productivity, user experience, and safety across various domains. At a broader sociotechnical level, human augmentation implies that organizations and societies should use AI to build human-centered AI ecosystems that amplify collective human potential and drive sustainable progress. In short, this principle reframes AI not as a substitute for human labor or intelligence, but as an enabler and a tool that powers human potential and advances societal well-being.

### 3.2.2 Human Controllability

Human controllability emphasizes that AI systems must remain under human authority and control, particularly in contexts involving safety, ethics, or high-stakes decisions. It requires humans to have the capacity to direct, oversee, intervene in, and override AI actions when necessary. This principle extends across the entire AI lifecycle: building systems with intuitive human-AI interaction mechanisms such as fallback modes or override options in design; embedding controls that are transparent, usable, and reliable under real-world conditions in development and deployment; ensuring operators have sufficient situation awareness, understandable user interfaces, and the authority to make final decisions in use. It also extends beyond individual human-AI interactions to encompass organizational and societal governance, including regulatory oversight, accountability mechanisms, and audit trails that ensure AI systems operate within established human boundaries. Human controllability balances AI performance with the enduring requirement that humans retain ultimate decision rights, preventing the delegation of responsibility or authority to machines alone.

### 3.2.3 Ethical Alignment

Ethical alignment ensures that AI systems are consistent with human values, rights, and societal expectations, going beyond technical optimization. It requires that AI be designed, developed, deployed, used, and governed in ways that uphold fairness, inclusivity, and respect for human dignity. This includes fairness, by preventing discriminatory outcomes and mitigating bias; inclusivity, by ensuring systems serve diverse populations across gender, age, culture, and ability, and by involving affected stakeholders; and privacy and data protection, by safeguarding individuals' controllability over their personal data. From a development perspective, ethical alignment calls for embedding value-sensitive design, bias audits, and impact assessments into the AI lifecycle. From a governance perspective, it requires accountability mechanisms, regulatory compliance, and culturally sensitive practices that adapt to different ethical contexts while aligning with universal principles such as human rights. At the organizational level, ethical alignment involves embedding fairness, inclusivity, privacy, and accountability into governance structures, development processes, and culture, ensuring that AI practices are transparent, auditable, and continuously monitored. Organizations must integrate ethical reviews, bias audits, and stakeholder engagement into the AI lifecycle to sustain responsible practices beyond individual systems. At the societal level, ethical alignment requires that AI systems reflect the diversity of ethical frameworks across cultures and disciplines and adhere to universally recognized values. Crucially, ethical alignment is not a one-time effort but an ongoing process of dialogue, monitoring, and recalibration, ensuring that as contexts evolve, AI continues to serve humans and society.

### 3.2.4 User Experience

User experience (UX) emphasizes that AI systems must be not only functional and reliable but also usable, accessible, understandable, and satisfying for the human users who interact with them. Good UX ensures that humans can understand, trust, and effectively engage with AI without confusion, frustration, or cognitive overload. UX extends beyond usability to encompass emotional, cognitive, and social dynamics of human-AI interaction. This includes multimodal user interfaces, emotional responsiveness, and adaptive personalization. From a design perspective, this means creating intuitive user interfaces, clear feedback mechanisms, and interaction flows that adapt to user needs and contexts. From a design and development perspective, UX requires embedding human-centered design methods, such as user research, participatory design, iterative prototyping, usability testing, and continuous feedback loops, throughout the AI lifecycle. This ensures that user interfaces, interaction flows, and explanations are grounded in real human needs rather than technical assumptions. It also ensures that AI systems support diverse users, provide accessibility accommodations for people with different abilities, and take into account cultural differences in interaction preferences. By treating UX as a core principle, HCAI ensures that technology fits seamlessly into human practices and workflows, enabling AI to empower and serve its users rather than impose additional burdens. This makes UX stand out as more than just user interface design, facilitating human-centered trust and adoption.

#### 3.2.5 Human-Led Collaboration

Human-led collaboration emphasizes that AI should be designed to act as a supportive collaborator, while retaining human leadership and decision authority in human-AI interaction, especially in high-stakes environments. Unlike automation-driven models that risk sidelining human judgment, this principle positions humans as leaders of shared tasks, with AI serving as an adaptive collaborator that brings complementary strengths, such as speed, scale, and pattern recognition. From a design and development perspective, this means creating interaction models that foster mutual adaptation, enabling AI systems to interpret human intent, context, and feedback while making their own actions understandable and predictable. From a governance and use perspective, it requires embedding safeguards to ensure that human operators retain the ability to set goals, make final decisions, and intervene when necessary. At a broader sociotechnical level, human-led collaboration necessitates cultivating trust, shared situation awareness, and enabling effective teamwork that extends across individuals, organizations, and ecosystems. By ensuring that AI augments rather than competes with humans, this principle promotes productive, safe, and ethical collaboration between humans and AI across diverse domains.

#### 3.2.6 Transparency and Explainability

Transparency and explainability are crucial for building trust, ensuring accountability, and the effective use of AI systems. AI systems should clearly communicate how decisions are made, enabling humans to understand system logic, foster trust, and support informed decision-making. User interfaces must visually represent reasoning and provide traceable outcomes. Transparency requires openness about how AI systems are designed, trained, validated, and deployed. Explainability complements transparency by focusing on the user experience: providing outputs and rationales that humans can understand, interpret, and act upon. This principle acknowledges that explanations must be tailored to the stakeholder (e.g., developers, regulators, end-users), who will each require distinct levels and forms of clarity. From a design perspective, explainability requires interpretable models, human-centered user interfaces for explanations, and iterative testing to ensure that explanations are truly usable. From a governance perspective, it requires policies that mandate the disclosure of capabilities, risks, and uncertainties. At a sociotechnical level, transparency and explainability ensure that AI does not become a "black box," thereby fostering informed oversight, trust, and responsible decision-making.

#### 3.2.7 Accountability and Responsibility

Accountability and responsibility ensure that humans, not AI systems, remain accountable for AI outcomes. Accountability must not be shifted to the AI itself; the ultimate responsibility lies with humans (e.g., developers, deployers, operators, or decision-makers), thereby reinforcing their ethical and legal responsibility for the consequences. Also, this principle requires that responsibility be clearly assigned across the AI lifecycle, from data collection and model development to deployment and monitoring. At the design and development stage, accountability entails documenting design choices, maintaining audit trails, anticipating potential harms, creating traceable interaction logs, establishing decision audit trails, and clearly attributing responsibility among human and AI agents. At the deployment and organizational levels, it requires clear lines of ownership, governance structures, and mechanisms for investigating and addressing failures or misuse. From a societal and regulatory perspective, this principle requires enforceable policies, liability frameworks, and ethical standards to ensure that when harm occurs, it can be attributed to responsible human actors. It also involves transparency toward affected stakeholders, allowing users and the public to understand who is accountable for AI behavior. It prevents liability displacement and reinforces trust that AI systems operate under human oversight, ethical integrity, and compliance with legal requirements**.** This positions accountability and responsibility as shared but clearly distributed duties across actors and levels, from developers to organizations to regulators.

### 3.2.8 Safety and Reliability

Safety and reliability prioritize human safety, maintain reliability in diverse scenarios to ensure resilience, and reduce potential risks to humans. It ensures that AI systems operate in ways that are trustworthy for humans, resilient in practice, and aligned with human needs. Safety means minimizing risks of harm not only in physical domains (e.g., automated vehicles) but also in psychological and social domains, such as avoiding user confusion, reducing cognitive overload, and preventing harm from bias or misinformation in content systems. Reliability emphasizes consistent, predictable performance that humans can depend on**,** reducing surprises and fostering confidence in daily use. From the HCAI perspective, it also requires error-tolerant systems, provides user alerts, enables emergency shutdowns, facilitates transparent recovery processes, and supports human situation awareness and decision-making. This allows humans to calibrate their trust appropriately and intervene effectively when needed. From a development and deployment perspective, it requires robust testing under realistic human use conditions and continuous monitoring. At the organizational and societal levels, safety and reliability require shared incident-reporting frameworks, standards that protect users, and accountability processes that prioritize human well-being over efficiency gains. From a sociotechnical lens**,** reliability also means long-term sustainability: AI systems must be maintainable, resilient to misuse, and designed to avoid cascading failures that could destabilize broader ecosystems.

### 3.2.9 Sustainability

Sustainability requires that AI systems be designed with consideration for technical, environmental, economic, and social sustainability. AI technologies are developed, deployed, used, and governed in ways that are responsible to both present and future generations, balancing environmental, social, and economic considerations. Achieving sustainability also involves adopting integrative approaches, such as data- and knowledge-dual-driven AI, as well as human-AI hybrid enhanced intelligence, to avoid a siloed technical approach to AI development. From the HCAI perspective, sustainability also emphasizes social and cultural dimensions, ensuring that AI does not exacerbate digital divides, displace vulnerable groups without support, or erode social cohesion, but instead contributes to equitable opportunities, inclusive participation, and resilient communities. From an environmental perspective, sustainability means reducing the ecological footprint of AI by lowering the energy and resource demands of large-scale training, designing efficient algorithms, and prioritizing green infrastructure. Economically, sustainability calls for systems that are maintainable and adaptable over time, avoiding short-lived solutions that create long-term risks or dependencies for humans and organizations. At the sociotechnical ecosystem level**,** it emphasizes embedding AI within broader sustainable development goals, such as responsible consumption, health, and education, ensuring that AI advances human flourishing. This perspective frames sustainability not only as an environmental issue but as a holistic human-centered concern: technical, ecological, social, cultural, and economic.

## 4. HCAI Methodological Framework (HCAI-MF)

While researchers have carefully investigated the concepts and frameworks behind HCAI, its practical implementation remains largely underdeveloped (Hartikainen et al., 2022; Friedrich et al., 2024; Schmager et al., 2025). A significant challenge is the lack of comprehensive methodologies that effectively guide HCAI adoption across diverse contexts (Bingley et al., 2023; Capel et al., 2023; Mazarakis *et al.*, 2023). In response, researchers have called for robust methodological frameworks and proposed various models addressing specific areas such as HCAI design and evaluation solutions (Desolda et al., 2025), interdisciplinary collaboration (Mazarakis *et al.*, 2023), requirement gathering (Ahmad *et al.*, 2023), organizational maturity (Hartikainen, 2023), design processes (Cerejo., 2021), and service science integration (Le et al., 2024). The lack of scalable, actionable methods impedes the operationalization of HCAI across the AI lifecycle (Desolda et al., 2025; Schmager et al., 2025). Collectively, the challenges indicate that HCAI remains at an early methodological stage, requiring structured frameworks, integrative methods, and cross-disciplinary collaboration to evolve into a mature socio-technical practice that bridges the gap between design philosophy and real-world implementation. Without structured methodological scaffolding that covers principles, processes, and interdisciplinary collaboration, HCAI risks remaining an aspirational design philosophy rather than a repeatable design practice. Developing a unified methodological framework is therefore essential to bridge the gap between HCAI's guiding ideals and real-world application.

A complete methodological framework should encompass design philosophy, guiding principles, requirements, approaches, methods, processes, and design paradigms. However, existing HCAI frameworks often lack these essential components, limiting their practical utility and hindering effective operationalization. To bridge these gaps and facilitate HCAI adoption in real-world practice, this chapter extends the Human-Centered AI Methodological Framework (HCAI-MF), as depicted in Figure 9, to include six key components: HCAI requirement hierarchy, HCAI method taxonomy, HCAI process, HCAI interdisciplinary collaboration approach, HCAI multi-level design paradigm, and HCAI maturity model (Figure 12). This HCAI-MF framework aims to provide comprehensive, structured, and actionable guidance for the design, development, and deployment of HCAI systems. This section elaborates on each of the six components of the HCAI-MF.

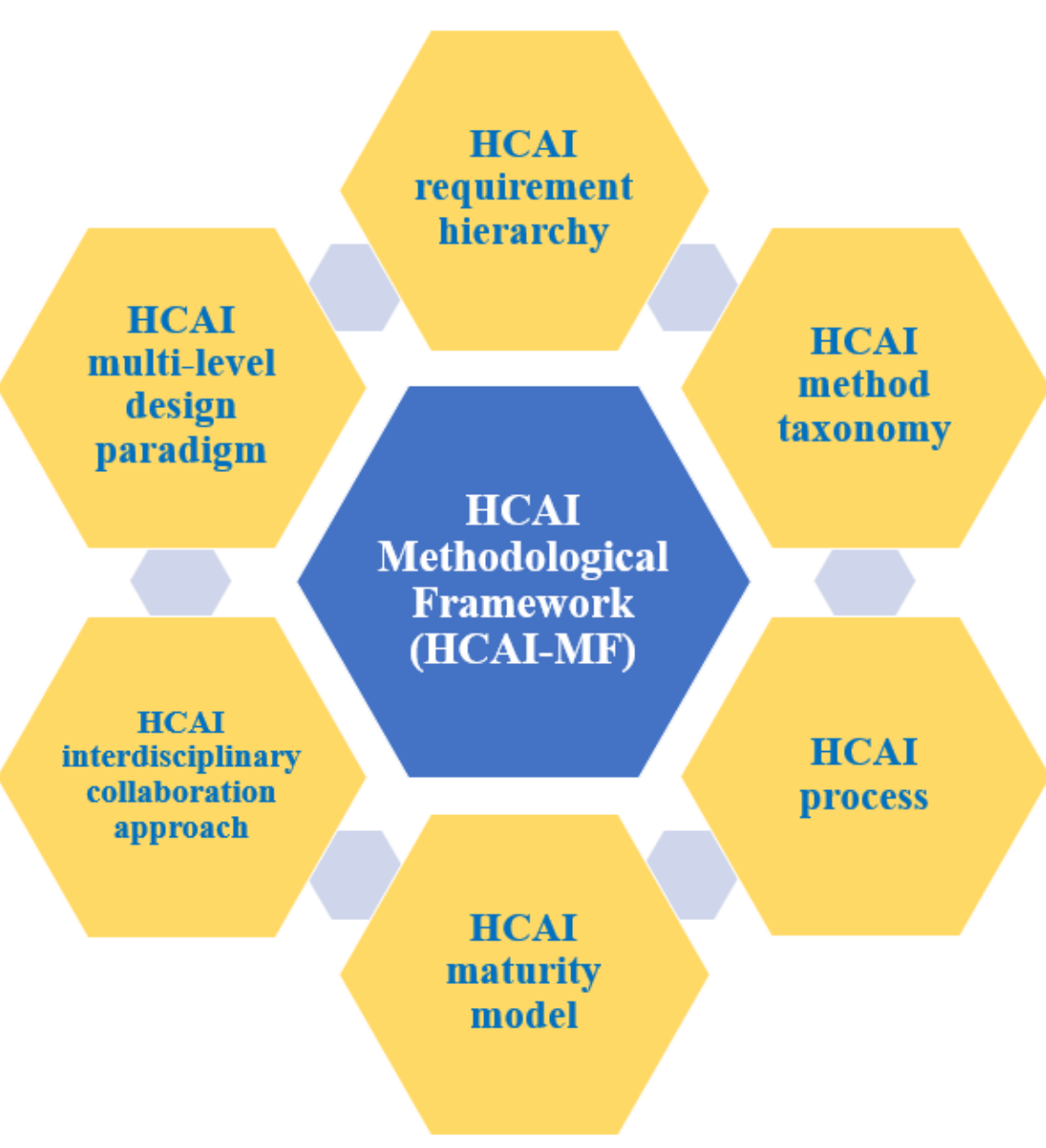

Figure 12 An extended HCAI methodological framework (HCAI-FM)

### 4.1 HCAI Requirement Hierarchy

Despite growing interest in HCAI, current practice continues to face challenges, particularly in the systematic elicitation of requirements for HCAI systems (Ahmad *et al.,* 2023). While HCAI guiding principles, as discussed previously, have been proposed, their conceptual boundaries often blur. For example, ethical frameworks such as fairness, transparency, and accountability frequently operate as high-level aspirations without clear operational criteria. Existing HCAI design guidelines further complicate the landscape by emphasizing broad notions such as human values, ethics, and privacy, which, although important, remain too abstract to provide actionable direction for system design and implementation (Yildirim et al., 2023). Compounding these issues is the lack of a clearly defined relationship among HCAI guiding principles, design guidelines, and product-level requirements. This lack of granularity, combined with unclear relationships among principles, design guidelines, and product-level requirements, makes it difficult to translate high-level HCAI aspirations into concrete, verifiable system specifications, thereby undermining the establishment of robust HCAI requirements as a foundation for development of HCAI solutions (Ahmad et al., 2023; Cerejo, 2021; Shneiderman, 2022; Xu, 2019). As a result, organizations struggle to derive actionable requirements that align AI system behavior with HCAI guiding principles and goals.

Xu, Gao, and Dainoff (2025) proposed a *Human-Centered AI Requirement Hierarchy (HCAI-RH)*, a structured framework that systematically links abstract values to concrete system requirements (Figure 13). Such a hierarchical structure linking principles to guidelines and requirements can help bridge this strategic-to-practical divide, making HCAI principles more actionable and measurable across projects. The hierarchy is organized into four interrelated levels:

- *HCAI Design Goals*: Broad, overarching aims that emphasize augmenting human capabilities, safeguarding well-being, and ensuring AI systems support, rather than replace or diminish human performance.

- *HCAI Guiding Principles*: High-level, strategic commitments that articulate industry-wide standards for aligning AI development with human-centered goals (refer to Section 3.2).
- *HCAI Design Guidelines*: Mid-level, domain-specific recommendations that translate HCAI guiding principles into actionable requirements. These guidelines may draw on industry standards or be tailored to organizational contexts, providing tactical direction for system design.
- *Product-Level Requirements*: Concrete, measurable specifications (or product detail requirements), both functional and non-functional (e.g., usability, safety, and human performance metrics), which operationalize HCAI design guidelines and ensure compliance with HCAI design goals and guiding principles in practice.

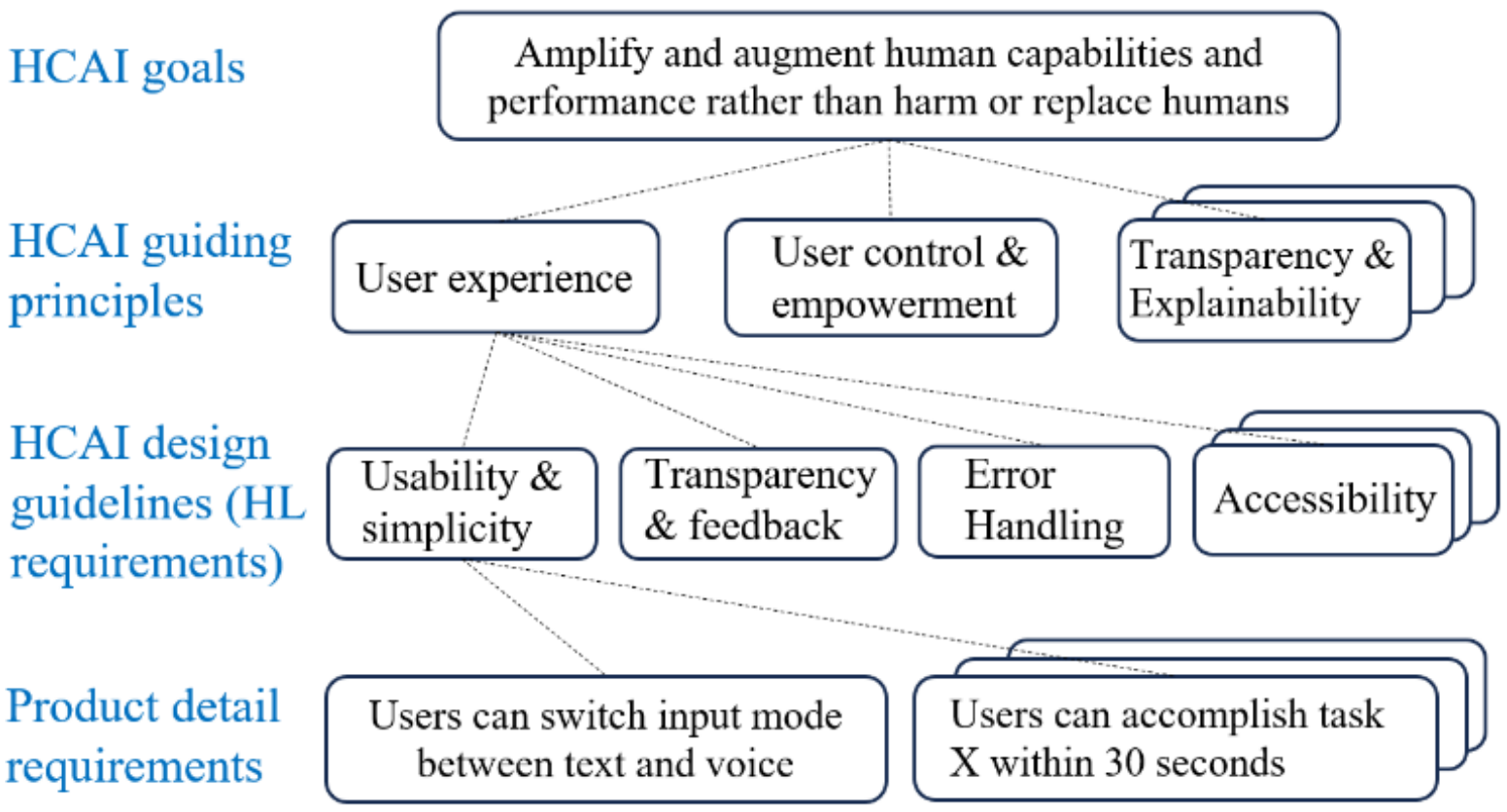

Figure 13. The Human-Centered AI Requirement Hierarchy (Xu, Gao, & Dainoff, 2025)

The HCAI requirement hierarchy provides a structured framework characterized by three defining features:

- *Means–end alignment*: Lower-level elements function as the "means" through which higher-level goals are realized, ensuring vertical consistency across the hierarchy.
- *One-to-many mapping:* Each higher-level element (e.g., guiding principle) is connected to multiple lower-level specifications, enabling comprehensive coverage and reducing the risk of gaps.
- *Goal-directed flow:* A top-down structure links HCAI strategic design goals and guiding principles to domain-specific guidelines and, ultimately, to actionable and verifiable product requirements.

By embedding these features, the HCAI requirement hierarchy grounds product-level specifications in overarching HCAI strategic goals and guiding principles. It is informed by human-centered requirement-gathering methods (e.g., user/stakeholder interviews, user participatory design). This integration bridges the gap between strategic aspirations and operational implementation, translating HCAI's strategic goals and guiding principles into concrete, testable product requirements. As a result, project teams can employ the hierarchy to derive HCAI-based product requirements and systematically deliver HCAI systems.

### 4.2 HCAI Method Taxonomy

Recent studies indicate that current HCAI practices remain limited by the absence of robust, actionable methods. Despite increasing interest in human-centered design and evaluation techniques for AI, HCAI still lacks comprehensive and actionable methods that can be applied consistently across the AI lifecycle (Bingley et al., 2020; Desolda et al., 2019; Xu et al., 2025). Most existing methods focus on isolated activities, such as explainability or usability, without a unified framework for efficiency and scalability. Practitioners from human-centered professional fields, such as human-computer interaction (HCI) and human factors, often encounter difficulties in integrating their methods into AI design and development processes, including generating interaction concepts, developing system conceptualizations, building prototypes, and conducting iterative user validation (van Allen., 2018; Stephanidis *et al.*, 2019). This fragmentation hinders cross-disciplinary learning and reduces the reproducibility of outcomes, highlighting a broader methodological gap in the field. To advance HCAI practice, scholars emphasize the importance of developing dedicated design and evaluation approaches and refining existing methods to enhance effectiveness and alignment with HCAI principles (Desolda et al., 2025; Xu, 2018; Ozmen Garibay et al., 2023; Umbrello & Natale, 2024).

#### 4.2.1 The HCAI method taxonomy

To address the gap, a structured *HCAI Method Taxonomy (HCAI-MT)* was introduced as one of the HCAI-MF components (Xu, Gao, & Dainoff, 2025). It aims to provide a comprehensive, scalable foundation for operationalizing HCAI (see Figure 14). The taxonomy systematically organizes HCAI methods into five categories: (1) human-centered strategy, (2) human-centered computing, (3) interaction technology and design, (4) human-centered controllability, and (5) human-centered AI risk management and governance. Table 5 further details 16 representative HCAI methods, including their definitions and key activities mapped across the AI lifecycle.

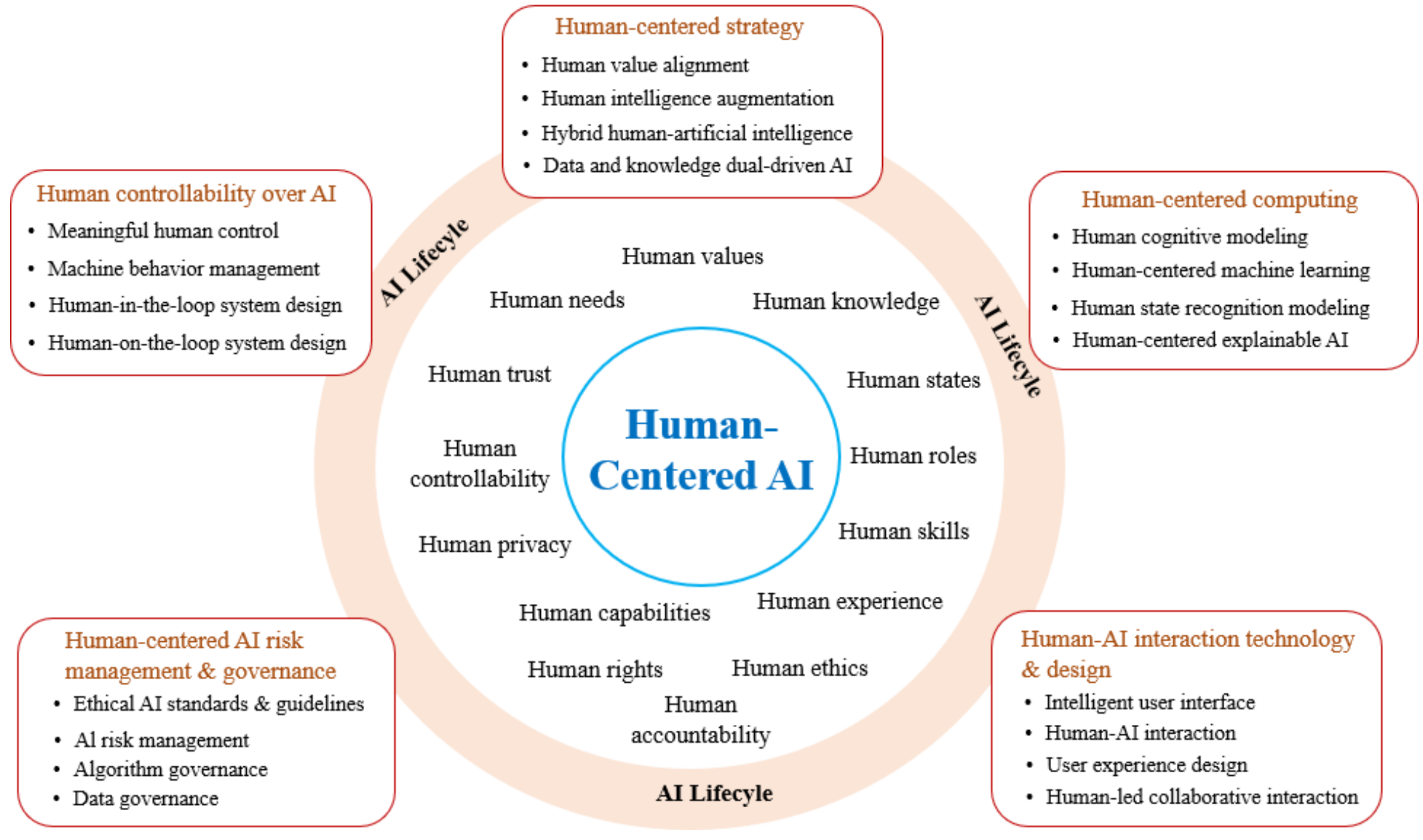

Figure 14 HCAI Method Taxonomy (Xu, Gao, & Dainoff, 2025)

Among these methods, some evolve from technology-centered approaches, enhanced by human-centered perspectives, such as human-centered explainable AI and human-centered machine learning. In contrast, others are adapted from human-centered disciplines (e.g., human factors, human–computer interaction) or from emerging AI practices (e.g., AI governance, hybrid human–artificial intelligence). Together, these interdisciplinary methods complement one another under the HCAI umbrella, collectively advancing shared HCAI goals. In addition, the HCAI method taxonomy is designed to be scalable and extensible, enabling the integration of new methods as the field continues to evolve.

Importantly, Figure 14 (labels within the big circle) and Table 5 (see the "human factors prioritized" column) emphasize that these methods explicitly address critical "human factors" throughout the AI lifecycle, such as human needs, values, experience, abilities, roles, knowledge, and controllability. These focal points align with the HCAI guiding principles (see the final column of Table 5), ensuring that the taxonomy not only categorizes methods but also reinforces their contributions to achieving HCAI goals. In this way, the HCAI method taxonomy embodies the essence of HCAI, consistently guiding AI projects to prioritize humans in HCAI practice by leveraging these methods throughout the AI lifecycle, as illustrated by the "AI lifecycle" circle in Figure 14.

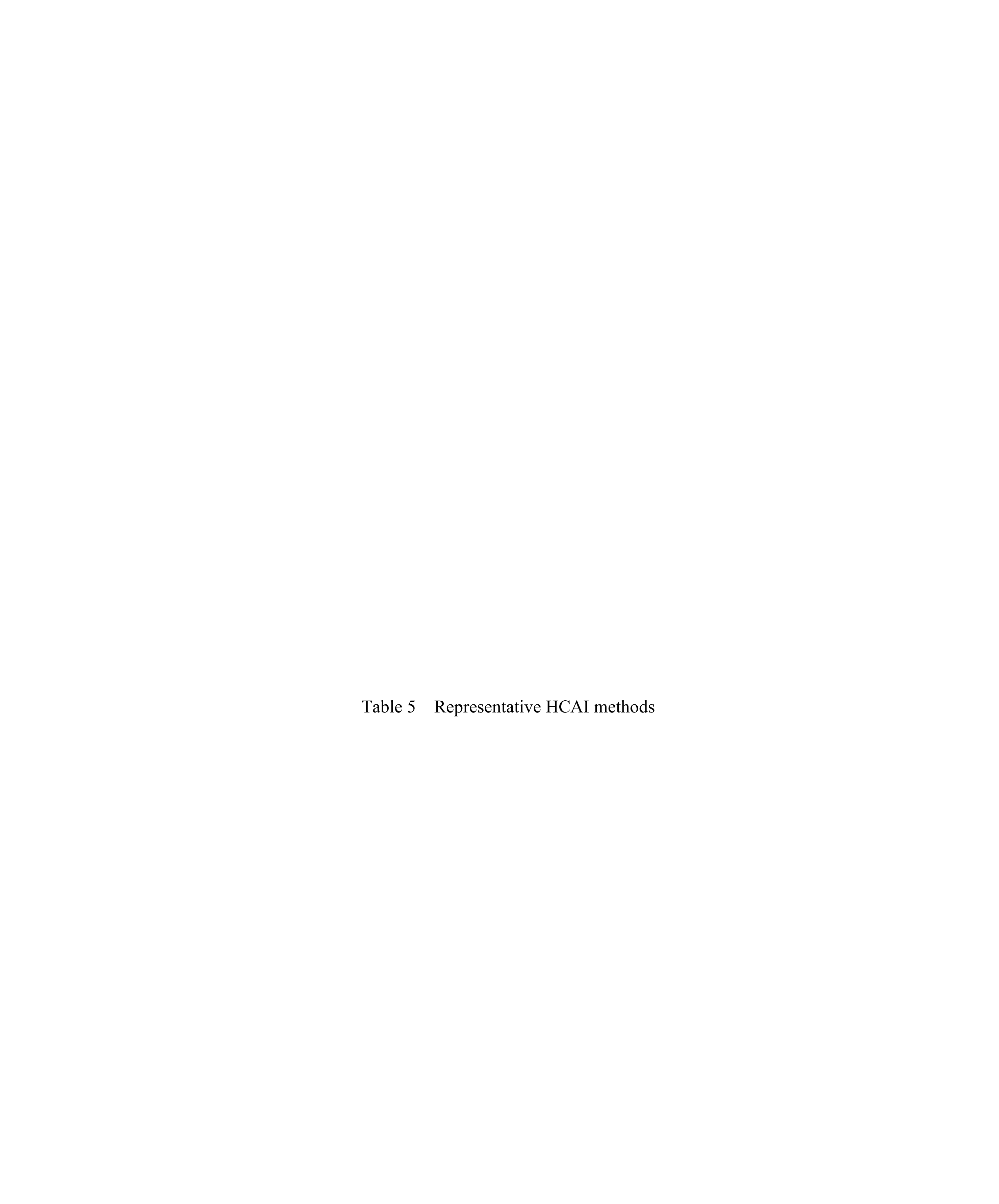

Table 5 Representative HCAI methods

| Category | HCAI methods | Description | "Human factors" focused | Key HCAI-driven activities across the AI lifecycle | HCAI Guiding Principles (see Table 3) |
|---|---|---|---|---|---|
| Human-centered strategy<br><br>(See Chapters 2, 4, 20, and 52) | Human value alignment with AI | Develops AI systems that align with human values, ethics, and social norms, ensuring that AI systems, like generative AI-based large language models (LLMs), make decisions and take actions that reflect these values and reduce the risk of unintended harm. | Human values | Define human values. Ensure fairness in data collection. Mitigate bias in modeling. Test with users to align with human values. Deploy with clear communication to users. Monitor users continuously and update as needed. | • Ethical Alignment<br>• Safety & Robustness |
| | Hybrid human-artificial intelligence | Integrates the unique strengths of both human and artificial intelligence to enhance decision-making, problem-solving, and overall performance. In hybrid AI, humans contribute qualities like creativity, intuition, and ethical judgment, while AI provides data processing, pattern recognition, and computational speed. | Human strength | Define roles and complementary strengths. Collect data on task-specific inputs with human insights. Choose models optimized for integrated human oversight. Train/fine-tune using HITL for hybrid interactions. Test collaborative performance. Deliver/monitor collaboration for evolving user needs. | • Human Controllability<br>• Human Augmentation<br>• Human-Led Collaboration |
| | Data and knowledge dual-driven | Integrates data-driven machine learning methods with knowledge-driven techniques (e.g., rules, expert systems). Integrates human expert knowledge into AI through technologies such as knowledge graphs, leveraging data and knowledge dual-driven AI to address issues in generative AI (e.g., hallucination, big data dependence, explainability). | Human knowledge | Define metrics and scenarios in which human knowledge complements data-driven insights. Acquire human knowledge (rules/expert insights). Select cognitive agent architectures for knowledge management. Train with human feedback. Audit for ethical metrics. Use tools for users to verify the knowledge base. Gather user feedback to refine the knowledge base. | • Transparency and Explainability<br>• Ethical Alignment<br>• Human Augmentation |
| Human-centered compu-ting<br><br>(See Chapters 19, 20, and 46) | Human cognitive modeling | Creates computational models that simulate human cognitive processes to enhance the design and functionality of AI systems. This approach ensures that AI systems are aligned with human thought processes, reasoning, and decision-making, fostering more intuitive and effective interactions between humans and AI. | Human cognitive capabilities | Identify goals and relevant cognitive tasks. Collect data on human behaviors & cognitive processes. Model cognitive processes. Train models using cognitive scenarios with user feedback. Assess alignment with users. Deploy UI for seamless cognitive interaction. Refine models to align with user needs. | • Human Augmentation<br>• Human-Led Collaboration<br>• User Experience |
| | Human-centered machine learning | Designs machine learning (ML) systems that prioritize human needs, values, and experiences throughout the AI system lifecycle, creating ML systems that are understandable, usable, fair, and aligned with users' expectations. It emphasizes transparency, interpretability, and human feedback loops, fostering trust and enhancing user experience. | Human values, needs, and feedback | User-centered problem definition, diverse data collection, explainable model selection, and user feedback integration. Usability testing with users and regular ethics audits uphold fairness and trust, while continuous monitoring and updates adapt models to evolving user needs and contexts. It is a user-participatory/interactive ML approach. | • Transparency and Explainability<br>• Ethical Alignment |
| | Human-centered explainable AI | Designs AI systems that prioritize interpretability. It aims to make AI decisions understandable and meaningful to users, enabling them to trust, interact with, and make informed decisions based on AI output by enhancing usability and aligning explanations with users' cognitive and emotional contexts. | Human knowledge, trust | Define user needs and domain knowledge levels. Select interpretable models. Validate explanations with users. Deploy transparent user interfaces with visualization. Continuously monitor and maintain relevance to ensure accountability, empowering users, and upholding ethical standards. | • Transparency and Explainability<br>• Ethical Alignment<br>• User Experience |
| Human-AI inter-action techno-logy and design<br><br>(See Chapters 11, 15, 10, and 16) | Intelligent user interface | Focuses on the design and functionality of UIs that leverage AI technologies to anticipate user needs and intentions, understand context and emotion, and respond dynamically to interactions (e.g., voice input). It centers around creating adaptive, responsive, personalized, intuitive, and supportive UI for users. | Human experience | Identify UX goals. Leverage diverse interaction data. Select multimodal technology, interaction model, and UI design paradigms to match user tasks and needs. Usability testing with user guides, refinements in evaluation, and deployment. Continuous monitoring/updates address user needs. | • User Experience<br>• Transparency and Explainability |
| | User experience design | Creates AI-driven interfaces and interactions that are intuitive, ethical, and aligned with users' needs. This approach considers not only usability but also transparency, interpretability, and trust, ensuring that AI systems are understandable and usable to users. | Human needs, experience | Conduct user research. Define UX requirements. Design explainable and intuitive UIs. Testing with users and iterative improvements improve UX, while personalization and accessibility make it adaptable and inclusive. Continuous monitoring with user inputs for improvements. | • Human Augmentation<br>• Human-Led Collaboration<br>• User Experience |
| | Human-AI interaction (e.g., robot, LLM) | Takes a broader view of how humans and AI systems interact. It considers all aspects of interaction, communication, collaboration, user control, and ethical implications using AI technology. Aims to create holistic interactions for humans to interact with AI, maintain oversight, trust, and human control. It | Human experience, trust, collaboration | Define goals and diverse interaction context. Develop multimodal interaction technology by leveraging AI-based approaches (e.g., intent recognition) and appropriate interaction design paradigms and models. Build explainability to foster transparency and trust. Personalization and adaptability are | • Human Augmentation<br>• Human-Led Collaboration<br>• User Experience |

| Category | HCAI methods | Description | "Human factors" focused | Key HCAI-driven activities across the AI lifecycle | HCAI Guiding Principles (see Table 3) |
|---|---|---|---|---|---|
| | | includes humans interacting with AI systems (e.g., robots, XR/metaverse, wearables, LLM). | | integrated to meet individual preferences, supported by rigorous testing for usability and performance. | |
| | Human-led collaboration with AI | Designs AI systems to ensure humans guide and oversee AI in a partnership, guaranteeing that AI augments rather than replaces human judgment. Humans leverage AI's computational strengths while retaining decision-making control, aligning AI outputs with human values, context, and goals | Human leadership role | Set human/AI roles and collaborative goals. Define human-AI collaboration models (e.g., shared tasks, situation awareness) with humans as leaders. Evaluate trust, human controllability, and usability. Deploy a seamless UI with monitoring/updates to refine trustworthy collaboration. | • Human Augmentation<br>• Human-Led Collaboration<br>• User Experience |
| Human-controllability over AI<br><br>(See Chapters 25, 26, and 28) | Human-in/on-the-loop system design | Designs AI systems to integrate humans into the AI decision-making, allowing for real-time input, intervention, and oversight, ensuring that AI systems can benefit from human judgment, adapt to nuanced contexts, and improve based on human feedback. | Human role, controllability | Define human intervention tasks and roles. Select models that can adapt to human input in real time. Test and refine models with human corrections and validate them with human-in-the-loop workflows. Deploy user controls and override options, while monitoring user feedback to improve workflows. | • Human Controllability<br>• Human Augmentation<br>• Ethical Alignment |
| | Machine behavior management | Monitors and regulates AI behaviors to align with human goals, ethical and safety requirements. Designs machine monitoring mechanisms, sets constraints, and enables human intervention when needed. It is essential for preventing unintended outcomes and promoting responsible AI. | Human ethics | Define AI machine behaviors for ethical and reliable actions. Use diverse training data and model-controlled responses. Testing, real-time monitoring, continuous feedback, and regular audits maintain alignment with ethical standards, while safe decommissioning ensures ethical accountability. | • Ethical Alignment<br>• Human Controllability |
| | Meaningful human control | Designs AI with humans' oversight, influence, and accountability. It ensures that humans can understand, guide, and, if needed, intervene in AI systems to prevent unintended harm for ethical standards and societal values. It ensures AI remains transparent/predictable to align with human intentions. | Human accountability | Define roles and metrics for intervention. Data collection focuses on scenarios requiring judgment. Design models for real-time alerts and accountability. Train/test with uses to refine control mechanisms and capabilities with human intentions. Deploy UI for human intervention, with continuous updates. | • Human Controllability<br>• Ethical Alignment<br>• Accountability and Responsibility |
| Human-centered AI risk management and governance<br><br>(See Chapters 31, 37, 38, and 39) | Ethical AI standards & governance | Develops standards, processes, and practices for ethical AI systems, ensuring they align with societal values, ethics, and legal standards. This focuses on accountability, transparency, fairness, and safety, addressing concerns like bias, privacy, and unintended harm. | Human values, accountability | Define and follow ethical AI standards with stakeholders. Data collection focuses on bias mitigation, while model development ensures ethical and transparency. Test and audit for ethical compliance. Deploy clear communication. Continuous monitoring and user feedback maintain ethical alignment. | • Ethical Alignment<br>• Safety & Robustness |
| | Algorithm governance | Establishes frameworks, policies, and practices to oversee algorithms, ensuring they operate transparently, ethically, and in alignment with human values. This sets guidelines for accountability, fairness, bias mitigation, and privacy in algorithmic decision-making processes for managing risk. | Human ethics | Define ethical and operational goals for algorithms. Collect quality data for privacy and bias mitigation. Select algorithms with built-in fairness and bias mitigation. Train with user input to align algorithms with ethical goals. Test compliance and societal impact. Deploy and monitor with user feedback. | • Ethical Alignment<br>• Safety & Robustness |
| | Data governance | Manages data used in AI systems, focusing on its quality, security, ethical use, and compliance with legal standards. Encompasses policies and practices to ensure the data is accurate, unbiased, and protected, facilitating responsible and transparent AI and mitigating risks, and promoting trust. | Human ethics, privacy | Define goals (e.g., quality, privacy). Collect data for ethical alignment (e.g., biases, fairness). Model design uses privacy-preserving techniques and usage logs. Train for policy compliance. Audit adherence to standards. Deploy safeguards and access controls. Monitor quality and security. | • Ethical Alignment<br>• Safety & Robustness |

#### 4.2.2 Implications for HCAI practice

The HCAI method taxonomy encompasses both strategic and tactical approaches, as well as hybrid approaches. Using these methods effectively throughout the AI lifecycle enhances their value and guides HCAI solutions. The following four practical scenarios illustrate the implications of the HCAI method taxonomy for HCAI practice.

**Scenario 1: Strategic planning**

At the start of the AI lifecycle, it is essential to set a strategy that guides progress toward HCAI solutions. For example, two complementary methods can be used to achieve the HCAI goal for generative AI. (1) *Hybrid Human–Artificial Intelligence* integrates the unique strengths of humans and machines to enhance decision-making, problem-solving, and overall system performance. In this approach, humans contribute creativity, intuition, contextual awareness, and ethical judgment, while AI provides computational efficiency, data processing, and pattern recognition capabilities (Pileggi, 2024). (2) *Data- and Knowledge-Dual-Driven AI* combines data-driven machine learning techniques with human knowledge-driven approaches such as rule-based reasoning and expert systems. By embedding human expertise through technologies such as knowledge graphs, this dual-driven paradigm mitigates challenges associated with large language models (e.g., hallucination, data dependence, and limited explainability) (Wang & Wang, 2024). Together, these methods help overcome siloed and unsustainable AI development practices that focus solely on machine intelligence or big data, enabling the creation of AI systems that are reliable, safe, and sustainable.

For further elaboration, refer to:

- (Chapter 3) Human value alignment in AI
- (Chapter 4) Hybrid Intelligence
- (Chapter 20) AI-Augmented Computational Modeling of Human Behavior
- (Chapter 52) A Sociotechnical Approach to Integrating Artificial Intelligence in Nuclear Power Plants

**Scenario 2: Modeling**

HCAI encourages a human-centered computational and modeling approach. For example, two methods can be used to achieve the goal. (1) *Human-Centered Machine Learning (HCML)* focuses on designing machine learning systems that prioritize human needs, values, and experiences across the entire AI lifecycle using methods such as user participatory design and human-in-the-loop approach. HCML aims to create models that are understandable, usable, fair, and aligned with user expectations by emphasizing transparency, interpretability, and continuous human-in-the-loop feedback, thereby fostering trust and improving user experience (Vaughan & Wallach, 2020). (2) *Human-Centered Explainable AI* develops AI systems, including generative AI-based large language models (LLMs), that emphasize interpretability and meaningful transparency using methods such as explainable algorithms and visualization technology. This approach seeks to make AI reasoning and outputs comprehensible to users, supporting informed human decision-making and promoting trust by aligning explanations with users' cognitive processes and emotional contexts (Rong et al., 2024). Together, these methods ensure that AI modeling not only achieves technical accuracy but also aligns with human understanding, trust, and ethical responsibility, bridging the gap between algorithmic performance and human-centered value.

For further elaboration, refer to:

- (Chapter 19) Human-centered machine learning
- (Chapter 22) Human-centered recommender systems
- (Chapter 23) Human-centered social computing
- (Chapter 46) LLM explainability

**Scenario 3: Interaction design**

Human-centered interaction design bridges technology and empathy, fostering seamless and trustworthy experiences between humans and AI systems. (1) *Intelligent User Interfaces* focus on designing and implementing user interfaces that leverage AI technologies to anticipate user needs and intentions, understand context and emotion, and respond dynamically to interactions such as voice or gesture input. The goal is to create adaptive, personalized, and intuitive interfaces that enhance usability and engagement (Brdnik et al., 2022). (2) *User Experience Design* emphasizes creating AI-driven interactions that are not only intuitive and efficient but also ethical, transparent, and trustworthy. This approach goes beyond usability to ensure that AI systems remain understandable, interpretable, and aligned with users' expectations and values using methods such as participatory design, iterative prototypes and user validation throughout the AI lifecycle (Asif, 2024).

For further elaboration, refer to:

- (Chapter 12) Intelligent adaptive systems

- (Chapter 15) Human-centered AI design principles for generative AI
- (Chapter 16) Designing human-centered AI experiences

**Scenario 4: Governance across multiple stages**

Human-centered AI risk management and governance methods should be applied at various stages of the AI lifecycle; two examples are provided below. (1) *Algorithmic Governance* establishes frameworks, policies, and oversight mechanisms to guide the development, deployment, and use of algorithms in ways that are transparent, ethical, and aligned with human values. It sets clear guidelines for accountability, fairness, bias mitigation, and privacy, thereby managing risks associated with algorithmic decision-making (Ebers & Gamito, 2021). (2) *Data Governance* focuses on managing the data that fuels AI systems, ensuring its quality, integrity, security, ethical use, and compliance with regulatory standards. It encompasses policies and practices that promote accuracy, reduce bias, and protect sensitive information, fostering trust and accountability in AI-driven decisions (OECD, 2024). Thus, these governance methods underpin human-centered oversight, ensuring that AI systems remain safe, fair, and socially responsible throughout their lifecycle.

For further elaboration, refer to:

- (Chapter 37) Ethical AI standards and governance: A perspective of human-centered AI
- (Chapter 38) Data Governance
- (Chapter 39) Algorithmic Governance

To sum up, the HCAI method taxonomy offers three benefits for HCAI solutions:

- *Goal-oriented design*: It specifies the critical "human factors" to be prioritized and systematically aligns the methods with HCAI guiding principles.
- *Comprehensive and scalable structure:* It integrates a broad range of methods that can be adapted to diverse HCAI practices, while remaining flexible enough to incorporate emerging developments as the field evolves.
- *Actionable guidance for practice*: It provides strategic and tactical approaches to address persistent HCAI challenges across the AI lifecycle, improving the effectiveness of implementation and fostering integration across disciplinary boundaries for HCAI solutions.

Collectively, these contributions position the HCAI method taxonomy as a practical, evolving framework that can guide HCAI practices while supporting ongoing refinement and innovation.

### 4.3 HCAI Design Paradigms

A design paradigm provides a distinct lens through which the scope, approaches, and methods of a field are defined and advanced. Throughout the computer era, different paradigms have been adopted to address evolving challenges. For example, human factors and engineering psychology have relied on the cognitive information-processing paradigm, examining how psychological processes such as human perception and attention interact with technical systems (Wickens et al., 2021). Similarly, fields such as HCI and UX have framed computers as supportive tools, emphasizing user models, interface concepts, and usability testing to optimize the user experience (Norman, 1986; Xu, 2003). As an emerging domain, HCAI requires equally robust design paradigms to guide practice and address novel challenges. Recent studies underscore both the influence of diverse factors on HCAI practice and the need for paradigmatic frameworks to overcome persistent limitations (NAS, 2021; Battistoni et al., 2023).

As discussed in Section 3.1.2, the *Hierarchical Human-Centered AI (hHCAI) framework* extends the scope of HCAI beyond human-AI interaction level practices of individual human-AI systems; it frames HCAI as a multi-level paradigmatic shift: individual → organizational → ecosystem → macrosocial, encompassing individual human-AI systems (keeping human-in/on/the loop), organizational environments (keeping organization-in-the-loop), cross-system ecosystems (keeping ecosystem-in-the-loop), and macrosocial contexts (keeping society-in-the-loop) (see Figure 8 in Section 3.1.2) (Xu & Gao, 2025). This section elaborates on the multi-level design paradigms of HCAI practice by applying the hHCAI framework.

#### 4.4.1 Keeping human-in/on-the-loop

Human-in-the-loop (HITL) and human-on-the-loop (HOTL) approaches have been adopted to provide real-time human oversight, allowing humans to monitor, control, and intervene when needed (Mosqueira-Rey et al., 2023; Yan et al., 2024). This ensures AI serves as a supportive tool rather than replacing human judgment. Both approaches represent complementary approaches to integrating human oversight and decision-making authority into AI systems. From the perspective of human controllability over AI, HITL ensures continuous, active human participation in critical decision-making processes, thereby enabling real-time intervention, ethical judgment, and contextual reasoning to mitigate automation bias and unintended consequences. In contrast, HOTL emphasizes supervisory control, in which humans monitor AI systems and intervene as needed, thereby promoting scalability and operational efficiency while maintaining accountability. From the perspective of hybrid human–machine intelligence, both paradigms advance synergistic intelligence by combining human cognitive flexibility, intuition, and ethical reasoning with AI's computational power, adaptability, and precision.

Significantly, the human-in/on-the-loop design model in HCAI should be recognized as more than just facilitating immediate operational interactions between individual humans and AI systems, as current HITL/HOTL approaches do; it also encompasses a broader human-centered perspective. As illustrated in Figure 15, the human-in/on-the-loop HCAI design paradigm encompasses the entire lifecycle of a human–AI system; the loop extends to the *loops* of requirements, design, development, deployment, use, operations, and governance. Humans are embedded throughout this cycle, as requirement definers, interaction designers, model trainers, evaluators, feedback providers, decision participants, human-AI collaboration leaders, operational controllers, and ultimate authorities with override capabilities. These responsibilities, which extend beyond the conventional HITL/HOTL approach, constitute a comprehensive, human-centered approach. Such an HCAI design paradigm ensures that AI systems remain aligned with human needs, values, and controllability throughout the entire AI lifecycle.

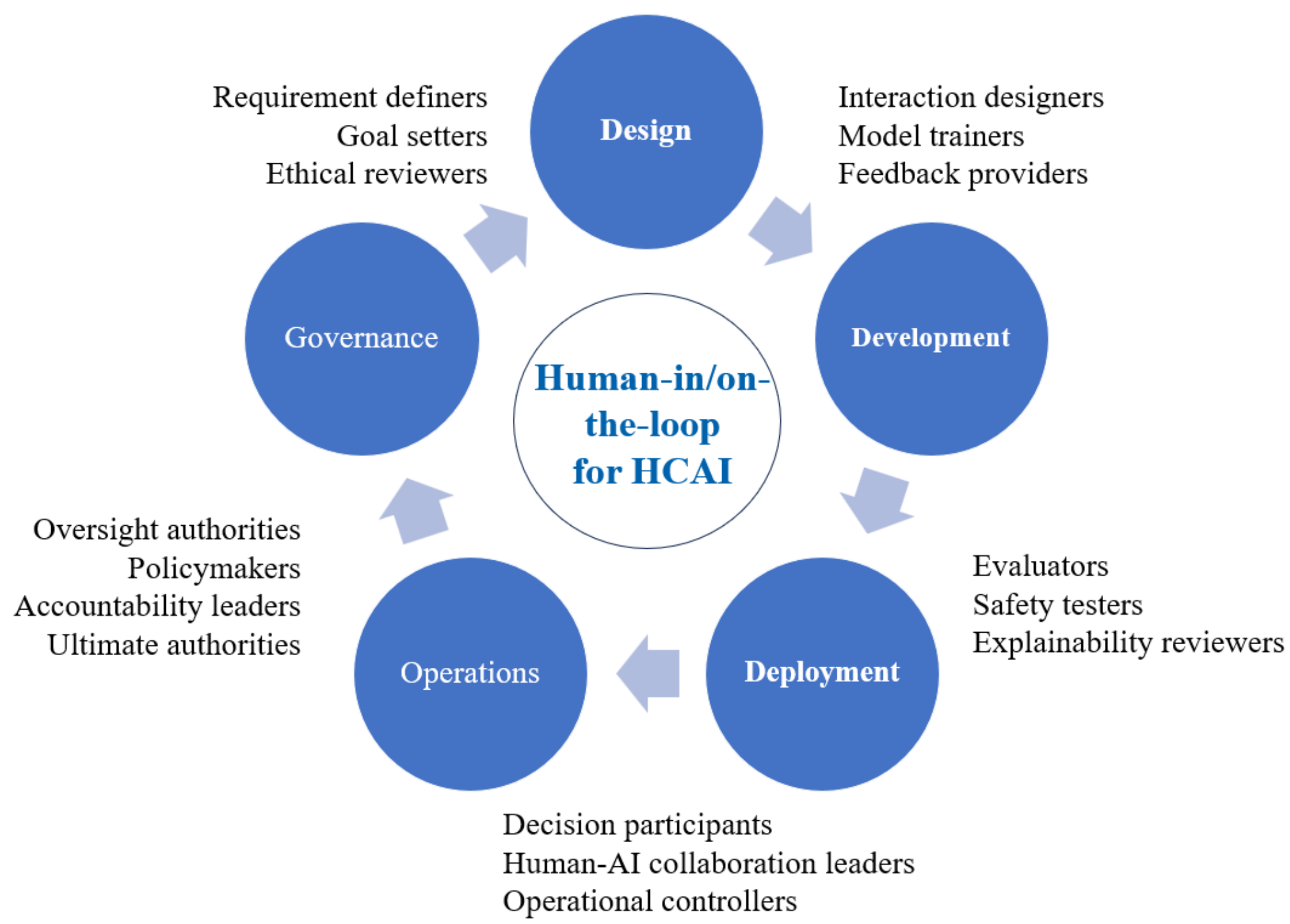

Figure 15 Illustration of the human-in/on-the-loop HCAI design paradigm

For further elaboration, refer to

- (Chapter 4) Hybrid intelligence
- (Chapter 19) Human-centered machine learning
- (Chapter 26) Human-in/on-the-loop design for human controllability
- (Chapter 28) Human-in/on-the-loop AI: Enabling human controllability and decision-making in spaceflight
- (Chapter 43) Human-centered and participatory AI auditing

### 4.4.2 Keeping organization-in-the-loop

While the HITL/HOLT design paradigm drives effective human-AI interaction, it has inherent limitations if considered in isolation. Current HCAI practices often concentrate narrowly on individual human-AI systems without systematically addressing broader organizational, ecosystem, or societal dynamics. As highlighted in the HCAI framework (see Section 3.1.2), future work must extend the HITL/HOTL paradigm toward multi-level integration, such as linking individual-level human oversight to organizational, ecosystem, and societal contexts.

HCAI practice cannot neglect the critical organizational context in which AI is developed, deployed, and used. Organizations are not only users of AI but also enablers, mediators, and regulators of its influence on people's work, values, and responsibilities (Pasmore et al., 2019). The "organization-in-the-loop (OITL)" HCAI paradigm emphasizes that HCAI practice must account for organizational structures, workflows, and cultures as essential design elements (Herrmann & Pfeiffer, 2023). By embedding HCAI within organizational contexts, such as decision-making flows, role allocation, governance structures, and learning mechanisms, organizations themselves become active participants in shaping AI outcomes. This view ensures that AI is not designed in isolation but co-evolves with organizational systems (Cascio & Montealegre, 2016; Ransbotham et al., 2021).

As Figure 16 illustrates, the OITL paradigm prioritizes systemic integration (human–technology–organization) through four dynamic loops—use, customization, task, and context. Governance and work design provide formal oversight, whereas organizational practices, collaboration, and culture provide informal scaffolding for the responsible evolution of AI. Continuous learning, adaptive management, and cross-organizational networks ensure that AI remains transparent, accountable, and human-centered as both technology and its contexts evolve (Herrmann & Pfeiffer, 2023).

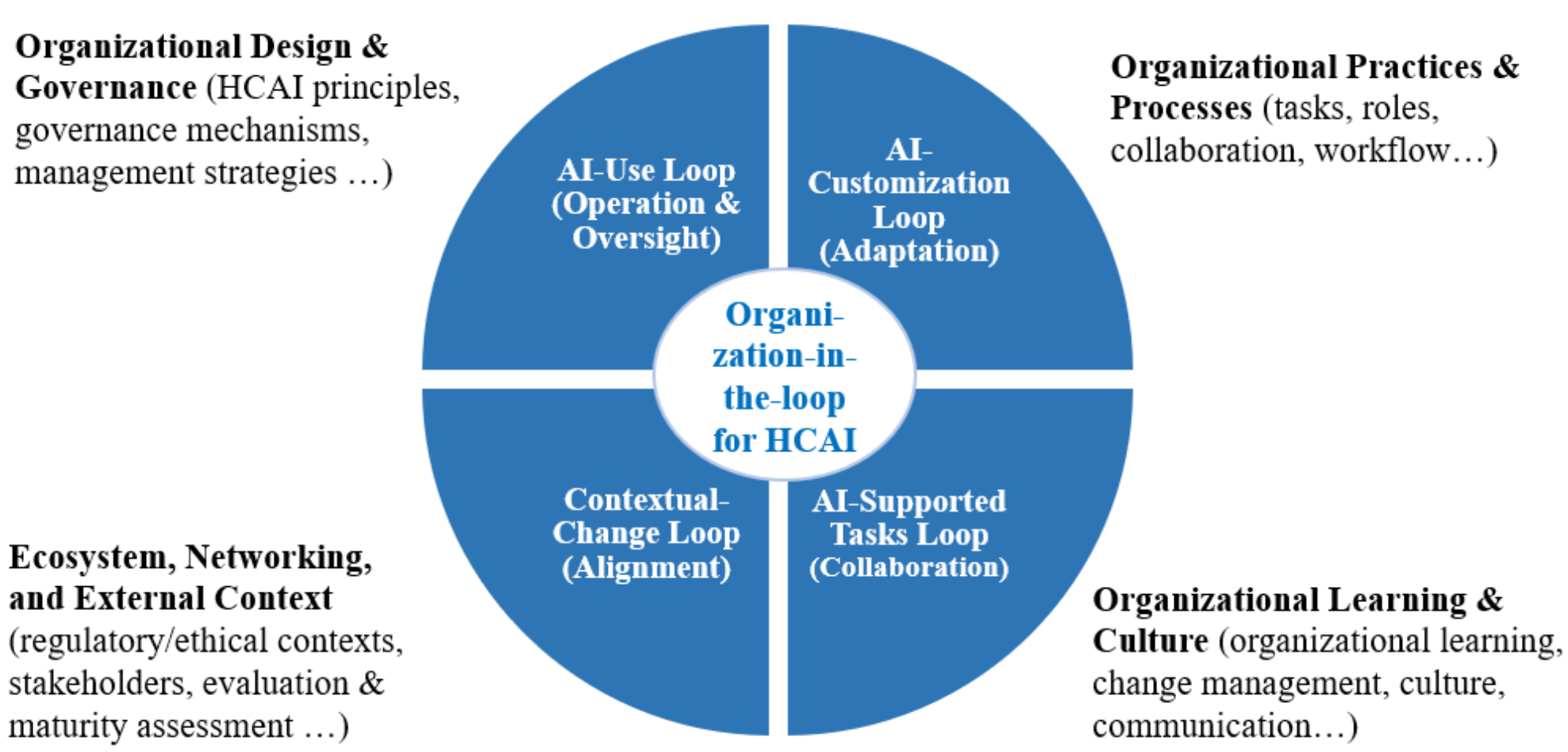

Figure 16 Illustration of keeping the organization-in-the-loop (based on Herrmann & Pfeiffer, 2023)

Adopting the OITL paradigm means treating organizational design as a continuous and integral component of AI development and deployment. Research in sociotechnical systems and organizational science suggests that AI can disrupt existing work systems, challenge established authority lines, and reshape job roles (Herrmann, 2025). Without deliberate organizational alignment, these disruptions risk undermining trust, fairness, and employee well-being. Conversely, by explicitly incorporating organizational adaptation into AI design, through transparent decision-making, participatory governance, and AI–human role complementarity, human-led human-AI collaboration, HCAI practice can enhance productivity, resilience, and accountability (Xu & Gao, 2025).

The OITL paradigm opens several pathways for future research and practice. First, it calls for methods to diagnose an organization's readiness for AI integration, including HCAI maturity models and frameworks to assess alignment between AI goals and organizational values. Second, it highlights the need for interdisciplinary design processes that involve human factors experts, ethicists, organizational leaders, and technical developers. Finally, it emphasizes the importance of ongoing organizational learning and co-adaptation, in which AI systems evolve alongside changing business processes, employee needs, and cultural expectations.

For further elaboration, refer to:

- (Chapter 56) Human-Centered AI Maturity Model (HCAI-MM): An Organizational Design Perspective
- (Chapter 57) Enterprise Strategy for Human-Centered AI
- (Chapter 58) Organizational Practices and Sociotechnical Design of Human-centered AI
- (Chapter 59) Human-centered AI in business modeling

#### 4.4.3 Keeping ecosystem-in-the-loop

Beyond organizational contexts, intelligent ecosystems are large-scale, interconnected environments in which multiple AI systems, humans, and organizations interact and adapt dynamically, such as intelligent transportation and healthcare systems (Tahaei et al., 2023; Yan et al., 2024). They extend beyond individual human–AI systems and the boundaries of individual organizations to encompass networks of systems that co-evolve within broader sociotechnical contexts. From a *human–AI joint cognitive ecosystem* perspective, these environments can be understood as distributed cognitive systems in which humans and AI collectively perceive, reason, learn, and make decisions across multiple levels and time scales (Xu, Gao, & Ge, 2024). In such ecosystems, cognition and control are not localized within a single agent but emerge from the coordinated activities and shared representations among human actors, AI agents, and organizational structures. Consequently, overall safety and performance depend not only on individual systems but also on their coordination and collaborative interactions across all subsystems, including human–human, human–AI, and AI–AI interactions, within an intelligent ecosystem (Stahl, 2021).

As Figure 16 illustrates, the "keeping ecosystem-in-the-loop" design paradigm reconceptualizes HCAI as a dynamic, multi-layered practice that emphasizes systemic solutions addressing HCAI at the ecosystem level, rather than limiting design to isolated systems. It integrates four foundational pillars: *(1) ecosystem-oriented system design*: focuses on designing AI systems as part of an interconnected, sociotechnical ecosystem and emphasizes coherence across subsystems and domains. This ensures that safety and performance are considered as emergent, system-wide outcomes across multiple AI systems rather than attributes of isolated components. *(2) human-centered governance and ethics:* grounds the framework in human values, emphasizing ethical oversight, accountability, and regulatory alignment in evolving AI ecosystems. This ensures AI development remains responsive to HCAI goals. *(3) dynamic collaboration and distributed intelligence:* encourages participatory governance where all stakeholders, such as users, institutions, and regulators, co-create and co-evolve the AI ecosystem. This enables human-AI collaboration by fostering shared objectives among multiple human and AI agents within an intelligent ecosystem. *(4) adaptation, learning, and co-evolution:* enables multiple AI systems to adapt in response to feedback from humans and society, and encourages continuous monitoring, learning, and governance across levels. This enables multiple AI systems and organizations to learn and adapt to evolving societal norms continuously.

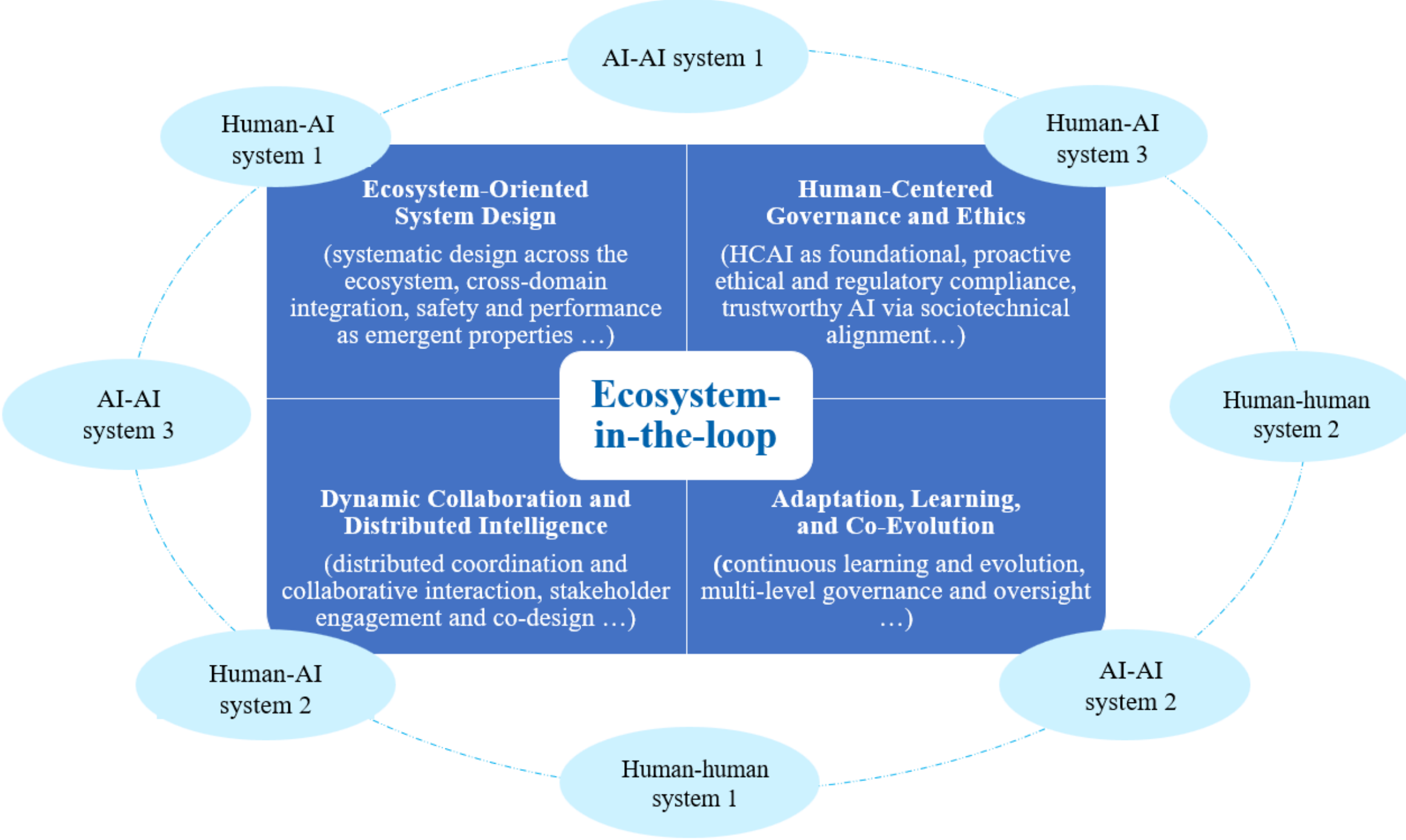

Figure 17 Illustration of the "keeping ecosystem-in-the-loop" paradigm

In practice, this entails establishing mechanisms for collaborative interaction, coordination, continuous learning, and adaptation across sub-systems within the ecosystems, enabling AI systems to evolve in response to human feedback and shifting societal norms. It also encompasses critical activities, including ensuring data privacy and security, fostering partnerships and stakeholder engagement, and maintaining regulatory compliance. The overarching goal is to optimize the interaction of multiple human-AI systems to build cohesive, human-centered AI ecosystems. This area is still in its early stages; developing theories and methods is critical.

For further elaboration, refer to:

- (Chapter 13) Human-centered digital twins
- (Chapter 29) Proactive Remote Operation of Automated Vehicles: Supporting human controllability
- (Chapter 49) Human-Centered AI in Healthcare

**4.4.4 Keeping society-in-the-loop**

Current HCAI efforts do recognize ethical issues such as fairness, privacy, and accountability; however, they often lack a broad sociotechnical perspective and a comprehensive framework to guide practice in real-world social settings. As a result, many HCAI practices remain narrowly focused, leaving unresolved the broader challenges of governance, societal impacts, and long-term sustainability. In sociotechnical systems (STS) theory, AI is not an isolated technical tool; it operates within a complex web of social, organizational, and institutional elements (Kudina et al., 2024). The sociotechnical view broadens the focus to encompass the joint optimization of AI technical and non-technical systems, ensuring that AI design accounts for human values, organizational dynamics, and collective well-being. In the absence of a broader perspective, efforts to address AI risks may focus exclusively on technical performance, thereby neglecting important societal considerations (Sartori & Theodorou, 2022; Michael et al., 2025).

The *society-in-the-loop* paradigm extends HCAI design paradigms by explicitly embedding societal values, concerns, and contexts into the full lifecycle of AI systems. The *Intelligent Sociotechnical Systems (iSTS)* framework provides the unifying conceptual foundation for this paradigm, framing AI-enabled systems as multi-level sociotechnical systems that require joint optimization across the individual, organizational, ecosystem, and societal levels (Xu & Gao, 2025). Within the iSTS perspective, intelligence is understood as an emergent property of dynamic interactions among humans, AI technologies, organizational processes, and social institutions, rather than as a capability residing solely in AI artifacts. This framing underscores the need to align technical design, human roles, governance mechanisms, and social norms throughout the AI lifecycle. Accordingly, the society-in-the-loop paradigm articulates six core approaches for HCAI practice: systematic design thinking, human-centered design, multi-level design approach, organizational adaptation and redesign, human–AI co-learning and co-adaptation, and an open ecosystem perspective (Xu & Gao, 2025).

As shown in Figure 18, the keeping society-in-the-loop design paradigm emphasizes that effective HCAI practice requires not only technical proficiency but also attention to joint optimization across technical and non-technical subsystems. Table 6 summarizes the key characteristics of major non-technical subsystems. For example, in healthcare, AI systems rely on technical components such as image recognition algorithms and cloud-based model deployment. At the same time, non-technical factors, including patient privacy laws, doctors' trust in AI, hospital workflows, and liability concerns, also play a critical role in their adoption. Thus, keeping society-in-the-loop means treating society as an active participant in shaping AI design, development, deployment, and use, rather than a passive recipient of outcomes. This involves systematically integrating non-technical subsystems. In doing so, AI systems become not only more technically capable but also socially responsible, reflecting the collective values, rights, and expectations of diverse communities.

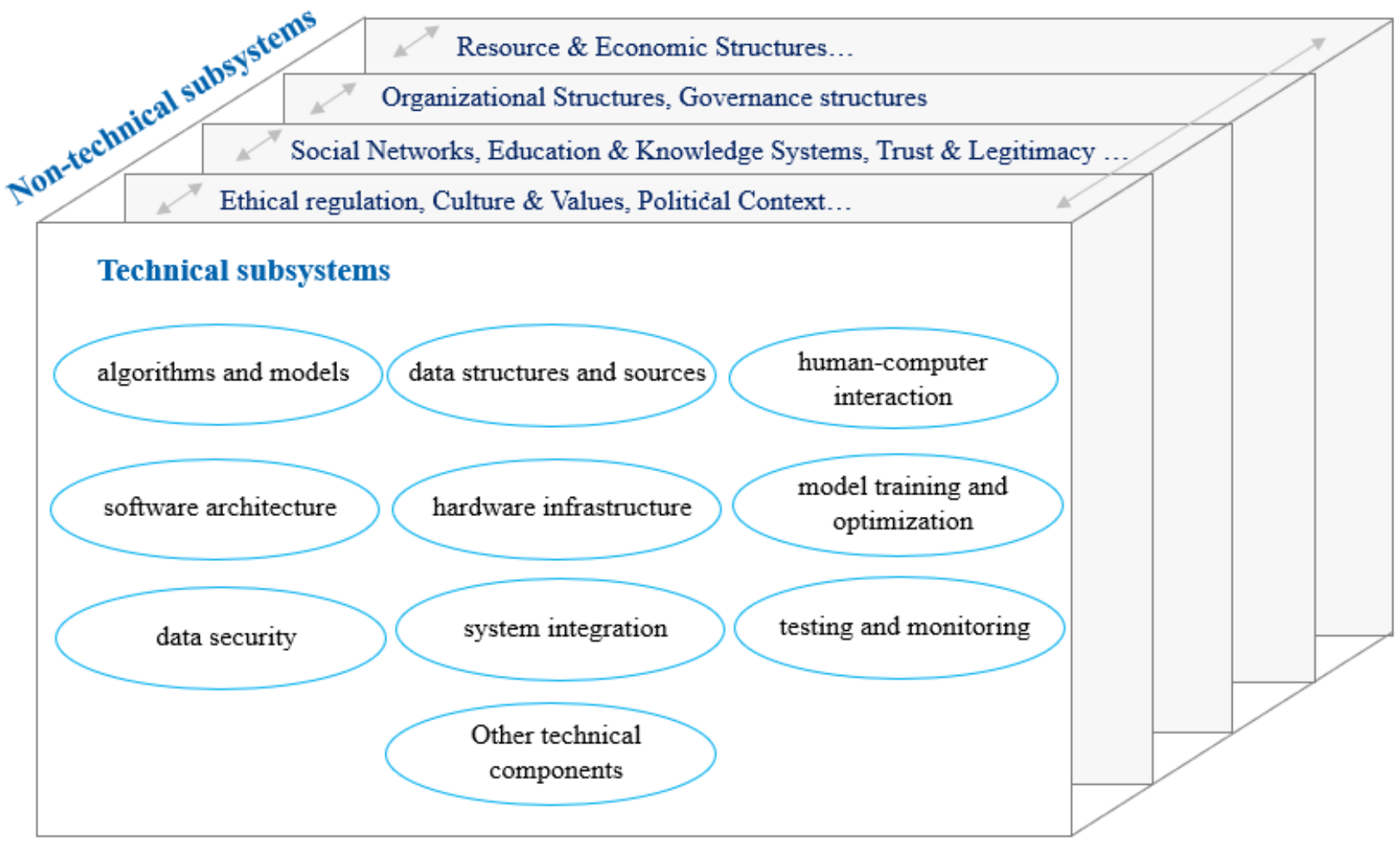

Figure 18 Illustration of the keeping society-in-the-loop design paradigm

Table 6 Key characteristics of major non-technical subsystems

| Key Dimensions | Key Non-Technical System Elements | Description |
|---|---|---|
| **Institutional Environment** | Laws, policies, and ethical norms | Regulatory frameworks, privacy laws, and ethical standards governing AI |
| **Organizational Structures** | Governance models, roles, internal workflows | How organizations deploy, regulate, and manage AI systems |
| **Culture & Values** | Societal expectations, ethical beliefs, and risk tolerance | Public acceptance, tolerance of bias, concepts like "automated justice" |
| **Human Factors** | User behavior, skills, and cognitive models | How users understand, interact with, and are trained to use AI systems |
| **Social Networks** | Relationships among stakeholders | Interactions between governments, companies, non-governmental organizations, and the public around AI |
| **Trust & Legitimacy** | Public trust and perceived legitimacy of AI | Whether AI decisions are seen as trustworthy, fair, and lawful |

| Key Dimensions | Key Non-Technical System Elements | Description |
| --- | --- | --- |
| **Resource & Economic Structures** | Ownership and control of data and AI infrastructure | Data monopolies, AI resource centralization, platform capitalism |
| **Education & Knowledge Systems** | Access to AI literacy, education equality | Who can understand and use AI, and the risks of the digital divide |
| **Political Context** | National strategies, political ideologies, and global competition | AI shaped by geopolitics, tech nationalism, and governance philosophies |

Future work in this area should focus on operationalizing this paradigm by developing interdisciplinary methods for systematically embedding societal input, optimizing governance and regulatory frameworks that adapt to evolving AI capabilities, and designing mechanisms for continuous feedback and adaptation as norms and expectations shift. Research is also needed to explore evaluation metrics that capture both technical performance and social impacts, ensuring that AI-based sociotechnical environments are sustainable, transparent, and aligned with the broader goals of human flourishing.

For further elaboration, refer to:

- (Chapter 4) Multi-level sociotechnical systems approach to human-centered AI
- (Chapter 52) A Sociotechnical Approach to Integrating Artificial Intelligence in Nuclear Power Plants
- (Chapter 55) AI transformation of organizations and society: A human-centered perspective

### 4.4.5 Implications of HCAI multi-design paradigms

Table 7 presents the major research topics across the four HCAI design paradigms to demonstrate their significance. The table shows that HCAI practice should extend beyond human-AI interaction in individual human-AI systems; full implementation is required at all levels by leveraging the multi-design paradigms. From a broad sociotechnical perspective, AI projects are no longer engineering projects; they are sociotechnical system projects that require interdisciplinary collaboration. The HCAI multi-design paradigms help overcome weaknesses in limited HCAI practice and drive the adoption of HCAI across all aspects, delivering comprehensive HCAI solutions.

Table 7 Key research topics across HCAI design paradigms

| **Research area** (selected) | **Keeping human-in/on-the-loop** | **Keeping organization-in-the-loop** | **Keeping ecosystem-in-the-loop** | **Keeping society-in-the-loop** |
| --- | --- | --- | --- | --- |
| AI-based machine behavior | AI agents act, learn, and adapt within continuous interaction cycles with humans | Alignment of AI agents' decisions, learning processes, and operational behaviors with organizational missions and standards | Machine learning and behavior evolution across multiple AI systems within an intelligent ecosystem | Effects of social factors on machine behavior and the ethical impacts of machine behavior |
| Ethical AI | HITL/HOTL approaches to detect, evaluate, and mitigate algorithmic bias through human participatory dataset curation and validation | Organizations' ethical responsibility throughout the AI lifecycle, from data collection to deployment and monitoring. | Human authority setting and compatibility across multiple AI systems from vendors with different cultures and norms | Ethical AI issues, norms, and governance in the social environment |
| Human-AI interaction | Shared tasks, roles, and decision-making between humans and AI across various contexts while retaining human authority | Organizations' role as a mediator in human-AI collaboration through governance, team configurations, and oversight mechanisms. | Interaction theory, modeling, technology, and design across multiple AI systems within an intelligent ecosystem | Impacts of social environments (cultural, ethical, etc.) on human-AI interaction design |
| Explainable AI (XAI) | Integration of humans in the loop to co-construct explanations through iterative interaction and feedback. | Organization's roles in traceability, accountability, and continuous monitoring for XAI | XAI issues across AI systems within an intelligent ecosystem (e.g., decision-making compatibility, conflict management across AI systems) | Public AI trust, public acceptance, and the relationship between XAI and culture, user knowledge, and ethics |

### 4.4 HCAI process

Effective processes are essential for advancing HCAI practice; however, research reveals challenges. Early phases of AI design still rely heavily on traditional software engineering practices (Hartikainen et al., 2022), and HCAI practitioners are often engaged only after requirements are set, limiting their influence (Silberg & Manyika, 2019). Studies highlight challenges, including prototyping and managing unpredictable AI behavior (Yang et al., 2020). While developers may consider ethics and security, broader HCAI perspectives often receive less attention (Bingley et al., 2023; Mazarakis et al., 2023). As a result, validation frequently overlooks bias, and HCAI principles are applied inconsistently (Silberg & Manyika, 2019).

To address these shortcomings, researchers have proposed process-oriented solutions (Ozmen Garibay et al., 2023). For example, Ahmad et al. (2023) applied HCAI principles to requirements gathering, Cerejo (2021) emphasized collaboration between AI and HCI professionals, and Battistoni et al. (2023) adapted the human-centered design (HCD) process for AI contexts. These contributions mark progress but remain fragmented. Current processes seldom integrate human feedback loops or ethical checkpoints throughout the AI lifecycle. Consequently, HCAI principles such as fairness, accountability, and transparency are inconsistently embedded across the design, testing, deployment, and monitoring stages. Like HCD, a comprehensive HCAI process should span the entire AI lifecycle and be guided by HCAI principles. Such a process must prioritize human needs, values, and roles not only during design and development but also in post-deployment activities, including ethical governance, monitoring AI behavior, and integrating user feedback, to ensure that AI systems are both technically proficient and socially responsible.

An HCAI process was proposed to address gaps in HCAI practice (Xu, Gao, and Dainoff, 2025) (Figure 19). The HCAI process integrates human-centered design (HCD) with a general AI system development lifecycle (ISO, 2019; ISO/IEC, 2023). This integration of the following two processes offers a comprehensive, integrated framework for developing HCAI systems, combining the strengths of established human-centered methods with the unique demands and challenges posed by AI technologies.

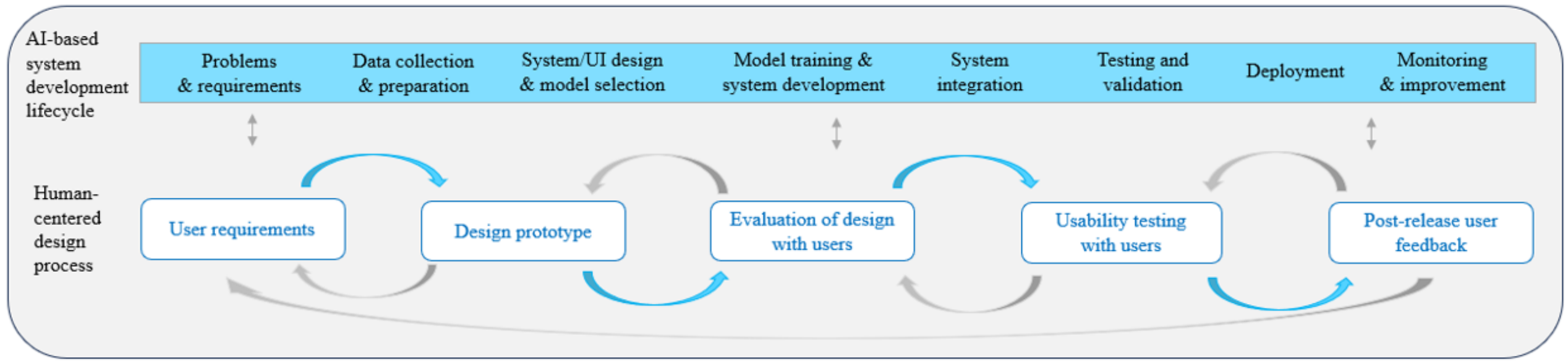

Figure 19 Illustration of the HCAI process

- *The HCD process:* a human-centered methodology that prioritizes human needs and experiences throughout system development (Xu, 2005; ISO, 2019). It emphasizes understanding problems from the user's perspective, incorporating user feedback at every stage. Interdisciplinary collaboration brings together diverse expertise, while iterative design, prototyping, and testing allow rapid evaluation and refinement of ideas. Thus, the process strikes a balance between user experience and organizational goals, delivering practical and impactful outcomes.
- *The AI system development lifecycle:* a generalized framework that spans all stages of the AI lifecycle, from problem identification and data preparation to deployment, monitoring, and continuous improvement (ISO/IEC, 2023).

Within the proposed HCAI process, these two perspectives are combined to ensure that HCAI guiding principles guide each stage of the lifecycle. Table 8 highlights key HCAI activities across the AI lifecycle, illustrated through four selected HCAI methods from the HCAI Method Taxonomy (Table 5). These illustrations demonstrate the four key features of the HCAI process:

- *Human-centered focus:* Keeps human needs, goals, and values at the center, with human involvement guided by HCAI principles.
- *Iterative improvement*: Supports ongoing prototyping, testing, reviews, and refinement based on human feedback.
- *Ethical development:* Ensures fairness, accountability, privacy, and bias are addressed throughout the lifecycle.
- *Interdisciplinary collaboration:* Brings together diverse expertise to develop HCAI solutions.

Table 8 Key HCAI activities across the AI lifecycle

| **HCAI methods** (Table 5) | **Problems & requirement** | **Data collection & preparation** | **Model selection & training, interaction design, system development** | **Test and validation** | **Deployment** | **Monitoring & refinement** | **Related chapters** |
|---|---|---|---|---|---|---|---|
| Human value alignment | Define human value | Ensure human value (e.g., fairness in data) | Mitigate issues (e.g., bias) in model selection, training, and development | Validate human value | Communicate with users clearly | Monitor user feedback, update the system | (Chapter 2) Human value alignment in AI |
| User experience design | Define user need/UX goals | Deploy user participatory method | Define interaction model and user interfaces | Test software & hardware with users | Provide user support | Track user feedback, update system | (Chapter 16) Designing human-centered AI experiences |
| Human-centered machine learning | Define human-centered problems and goals | Collect diverse data | Use diverse data, model behavior with controlled algorithms | Test for ethics | Gather and integrate user feedback, refine algorithms | | (Chapter 19) Human-centered machine learning |
| AI ethical governa-nce | Establish ethical policies | Mitigate ethical risks (e.g., data bias) | Choose/train/develop models for fairness/transparency | Perform ethical audits | Communicate users clearly about ethical AI | Monitor ethical issues, update system | (Chapter 37) Ethical AI standards and governance: A perspective of human-centered AI |

In summary, the HCAI process provides overarching guidance for embedding HCAI methods and activities throughout the AI lifecycle, while allowing flexibility for projects to adapt and optimize the process to achieve HCAI goals.

For further elaboration, refer to:

- (Chapter 14) Design thinking and AI: Facilitating HCAI solutions
- (Chapter 16) Designing human-centered AI experiences

### 4.5 HCAI interdisciplinary collaboration

Interdisciplinary collaboration is not merely supportive but foundational to the realization of HCAI. While AI has advanced through contributions from fields such as cognitive neuroscience, the unprecedented challenges it poses cannot be addressed by technical expertise alone (Yang et al., 2020; Mazarakis et al., 2023; Ahmad et al., 2023). Research demonstrates that while computer science and AI disciplines deliver computational power to build AI systems, they lack sufficient grounding in human needs, practices, and values, insights that fields such as HCI and human factors are uniquely positioned to provide (Xu & Dainoff, 2023). Neither domain on its own can ensure responsible and effective AI systems; it is their integration that makes HCAI viable (Mazarakis et al., 2023). The history of Human-Centered Design (HCD) demonstrates that interdisciplinary collaboration is essential for the success of new technologies, and this lesson is particularly relevant to AI. As Desolda et al. (2025) emphasize, HCAI requires the convergence of diverse fields, such as HCI, AI, software engineering, and ethics, not as an option but as a necessity for developing rigorous design and evaluation solutions. Without such integration, HCAI risks remaining aspirational rhetoric rather than actionable practice.

As discussed above, all components of HCAI-MF, including requirements, methods, processes, and design paradigms, require interdisciplinary collaboration. Table 9 highlights representative interdisciplinary approaches and methods presented throughout this book's chapters. It is apparent that HCAI cannot be successful without interdisciplinary collaboration. Collectively, these disciplines and approaches complement one another by maximizing their respective strengths and mitigating individual weaknesses, enabling the joint delivery of robust HCAI solutions.

Table 9 Illustration of interdisciplinary collaboration for HCAI

| **HCAI Areas** | **Area Focus** | **Primary Participating Disciplines/Approaches** | **Relevant Chapters in This Handbook** (selected) |
|---|---|---|---|

| Human-AI interaction | Employ technical and design approaches to enable humans to engage, communicate, interact, and collaborate effectively | Human-Computer Interaction (HCI), AI, Psychology, Computer Science, Human Factors, and Interaction Design | (Chapter 5) Human-Centered Artificial Social Intelligence<br>(Chapter 8) Theory of mind in human-AI interaction and AI<br>(Chapter 10) Human-AI interaction in LLM<br>(Chapter 11) Human-robot interaction<br>(Chapter 12) Intelligent adaptive systems |
|---|---|---|---|
| Human-centered AI design | Employ approaches such as design thinking, interaction design, and user experience design to create HCAI solutions | Design Thinking, Interaction Design, User Experience (UX) Design, HCI, AI, CS, Visual and Communication Design, Ethics and Responsible AI | (Chapter 13) Design thinking and AI: Facilitating HCAI solutions<br>(Chapter 15) Human-centered AI design principles for generative AI<br>(Chapter 16) Designing human-centered AI experiences<br>(Chapter 18) Intercultural design for human-centered AI solutions |
| HCAI in computing | Employ approaches such as algorithmic design, computational modeling, and data-driven methods to develop HCAI solutions | AI, Simulation and Modeling, Machine Learning, CS, HCI, Cognitive Science, Psychology, Data Science, Ethics and Responsible AI | (Chapter 19) Human-centered machine learning<br>(Chapter 20) AI-Augmented Computational Modeling of Human Behavior<br>(Chapter 22) Human-centered recommender systems<br>(Chapter 23) human-centered social computing<br>(Chapter 46) LLM explainability |
| Human controllability over AI | Employ approaches such as human-in/on-the-loop design, meaningful human control, accountability, and traceability design to ensure human controllability | AI, HCI, Control Systems Engineering, CS, Ethics and Responsible AI, Robotics, Systems Engineering, Law and Policy, Cybersecurity, Safety-Critical System Design, Human Factors | (Chapter 25) Human-controllable AI: Meaningful Human Control<br>(Chapter 26) Human-in/on-the-loop design for human controllability<br>(Chapter 27) Testing Obedience and Control in AGI: Exploring Irrational Commands and the AI Control Problem<br>(Chapter 28) Human-in/on-the-loop AI: Enabling human controllability and decision-making in spaceflight<br>(Chapter 29) Proactive Remote Operation of Automated Vehicles: Supporting human controllability |
| Human-centered AI risk management | Employ approaches such as risk reporting, risk measurements and assessment, and risk management to avoid and mitigate risk for HCAI solutions | Risk Management, AI, Ethics and Responsible AI, CS, Cybersecurity, Law and Public Policy, Sociology, Data Science, Systems Engineering | (Chapter 31) AI risk management frameworks<br>(Chapter 32) AI risk measurement<br>(Chapter 33) A Human-Centered Privacy Approach to AI<br>(Chapter 34) AI risk and trust<br>(Chapter 35) Addressing overreliance on AI<br>(Chapter 36) Enabling diversity and gender equity in HCAI |
| Human-centered AI governance | Employ approaches such as HCAI standards, algorithmic governance, and data governance for HCAI solutions | Public Policy, Law and Regulation, Ethics and Responsible AI, AI, Sociology, Data Governance, CS, Risk Management, Organizational Studies | (Chapter 38) Data governance<br>(Chapter 39) Algorithmic governance<br>(Chapter 40) International governance of AI<br>(Chapter 41) Governance of AI-based algorithms<br>(Chapter 43) Human-centered and participatory AI auditing |

Effective interdisciplinary collaboration remains a barrier to HCAI practice. While AI researchers, human factors experts, and ethicists recognize the need for cooperation, real-world projects often suffer from communication breakdowns, delayed participation by human-centered specialists, and divergent priorities across disciplines (Mazarakis et al., 2022; Xu & Dainoff, 2023; Desolda et al., 2025). To overcome inefficiencies in interdisciplinary collaboration for HCAI, organizations must adopt structured strategies that institutionalize collaboration rather than rely on ad hoc efforts, following an integrated approach. The following structured approach aims to shift interdisciplinary collaboration from a loosely coordinated effort to a systematic organizational practice, enabling HCAI implementation at scale with measurable accountability.

- *Establish shared goals and vision:* Define a unified strategy for HCAI that aligns technical innovation with HCAI principles, ensuring all participants work toward common objectives.
- *Assemble interdisciplinary teams*: Assemble project teams that intentionally combine expertise from AI, human-centered design, ethics, and relevant application domains to capture the full range of perspectives.
- *Co-design with stakeholders:* Involve diverse stakeholders, including developers, domain experts, users, ethicists, and business owners, throughout the design cycle to ensure inclusive and comprehensive input.
- *Address ethical conflicts systematically:* Create ethical review boards and establish baseline principles that reconcile disciplinary and cultural differences while respecting contextual diversity.

- *Institutionalize HCAI processes and methods:* Integrate the HCAI process and methodological toolkit into organizational workflows to guide collaboration and ensure consistency.
- *Define transparent decision-making frameworks:* Develop clear governance structures and accountability mechanisms that distribute decision-making authority fairly and openly.
- *Foster continuous communication:* Provide training, collaboration tools, and open channels for dialogue to bridge disciplinary language and conceptual gaps.
- *Measure success holistically:* Evaluate outcomes not only by technical performance but also by HCAI success metrics such as user satisfaction, productivity, ethical soundness, and societal impact.

### 4.6 HCAI maturity model

An HCAI maturity model provides organizations with a structured, progressive roadmap for operationalizing HCAI as an integral component of the HCAI-MF (Winby & Xu, 2025; Hartikainen et al., 2023). Because organizations vary widely in their readiness, capabilities, culture, governance, and human-factor integration, a maturity model helps benchmark current states, identify capability gaps, and prioritize incremental improvements across the AI lifecycle. As AI systems scale and risks grow, the maturity model translates HCAI principles such as safety, transparency, explainability, oversight, and ethical alignment into actionable processes, artifacts, and interdisciplinary practices. By guiding the development of cross-domain collaboration, lifecycle documentation, post-deployment monitoring, and organizational governance, the maturity model supports continuous improvement, alignment with compliance requirements, and long-term trustworthiness. Ultimately, the HCAI maturity model operationalizes the goals of HCAI-MF by helping organizations institutionalize human-centered methods, maximize strengths across disciplines, mitigate emergent risks, and deliver responsible, reliable, and value-aligned AI solutions.

Whereas Figure 10 in Section 3.1.2 depicts the conceptual structure of the HCAI-MF, Table 10 offers a concise summary of its stages across the process. Assessment under the HCAI-MM spans five dimensions. *Design and engineering practices* ensure HCAI principles, such as user experience, trust, and explainability, are embedded in AI systems. *Risk and incident management* requires proactive anticipation, tracking, and organizational learning from failures. *Organizational governance* formalizes policies, structures, and cross-disciplinary collaboration to sustain HCAI efforts. *Metrics and evaluation* focus on both qualitative and quantitative measures of system performance, accountability, and human impact. Finally, *culture and ecosystem engagement* emphasize embedding HCAI values into organizational culture while aligning practices with industry standards, regulatory expectations, and broader societal goals. For further elaboration, refer to Chapter 57 ("*Human-Centered AI Maturity Model (HCAI-MM): An Organizational Design Perspective*") of this handbook.

Table 10 Stages in HCAI-MM [adapted from (Winby & Xu, 2025)]

| Maturity Level | Description | Characteristics | Assessment Criteria | Objectives |
|---|---|---|---|---|
| **Level 1: Initial**<br>Entry, sanction, and start-up design phase in progress<br>The organization is primarily focused on establishing a basic understanding and some ad-hoc HCAI practices | • Building sanctions and resource preparation for the transformation process<br>• Establishing ad-hoc practices and increasing awareness of human-centered principles.<br>• Isolated efforts to address human factors in AI. | • Aligning sanctions and resources for HCAI transformation<br>• AI projects initiated without user input.<br>• Limited understanding of HCAI principles. AI initiatives are generally unstructured and reactive. | • Readiness assessment in progress<br>• Entry, sanction, start-up phase in progress<br>• No formal processes for user engagement<br>• Minimal or no stakeholder feedback mechanisms<br>• Low awareness of HCAI<br>• Isolated efforts to address human factors in AI. | • Complete readiness assessment<br>• Complete entry, sanction, and startup<br>• Begin building awareness of HCAI<br>• Communicate the importance of HCAI |
| **Level 2: Developing**<br>Research and analysis providing awareness of opportunities and constraints<br>Developing understanding and structuring HCAI | • Growing understanding human-centered design.<br>• A structured HCAI methodology and program are in place | • User interviews or surveys conducted for select projects<br>• Basic usability testing implemented.<br>• Early collaboration between disciplines<br>• Ad-hoc training and education programs on HCAI | • Technical and social systems analysis is in progress.<br>• Some processes to gather user feedback<br>• Limited consideration of ethical implications<br>• Early integration of HCAI design in AI projects.<br>• Implementation of usability testing and basic feedback | • Research and analysis design phase completed<br>• Foster a culture of learning around HCAI<br>• Begin documenting AI projects with a focus on understanding user needs and ethical implications.<br>• Starting the adoption |

| practices<br><br>Begin to implement HCAI frameworks and processes | | • External HCAI design standards are being adopted. | loops. | of use cases. |
|---|---|---|---|---|
| **Level 3: Defined**<br>Design phase prototyping in progress.<br>Formalization of HCAI practices.<br>Actively employ and continuously refine HCAI practice | • Implementing, testing, and adjusting HCAI approaches and methods<br>• Established HCAI design processes | • HCAI governance established<br>• Organizational HCAI design guidelines published<br>• Proactive HCAI-related training launched<br>• Structured deliberations in the design lab<br>• Structured initiatives to incorporate user input in processes | • Structured feedback mechanisms in place<br>• Standardized processes for ethical review/governance.<br>• Formal processes for HCAI integrated into the AI lifecycle<br>• Standardized practices for ensuring ethical alignment<br>• Work system team units have all structured properties in place | • Design Phase completed: prototyping, testing, and ongoing HCAI iterations.<br>• Create consistent processes for involving stakeholders and users in AI system design.<br>• Develop policies and guidelines to address ethical considerations in AI. |
| **Level 4: Managed**<br>HCAI implementation established with widespread diffusion in the organization.<br>Quantitative HCAI metrics and institutionalizing HCAI.<br>Organizations are recognized for HCAI competence, focusing on performance, accountability, and societal impact. | • Implementation with examples of HCAI strategies across the organization<br>• HCAI metrics in place. | • HCAI is defined in the organization's strategy<br>• Continuous user involvement throughout the lifecycle.<br>• Rigor HCAI governance established<br>• Data-driven decision-making based on user feedback<br>• Design lab deliberations yielding changes in work design and innovation.<br>• Comprehensive training programs on HCAI | • KPIs established for HCAI and ethical compliance.<br>• Metrics collected and acted on.<br>• Use of data-driven metrics to assess HCAI quality.<br>• Continuous monitoring of user feedback and trust in AI systems. | • Continuous HCAI implementation and iterations.<br>• Ensure all AI projects align with established HCAI policies and ethical standards.<br>• Evaluate the impact of AI systems on users and society regularly. |
| **Level 5: Optimizing**<br>Continuous improvement and innovation. Leading change/reconfigurability at all org levels through HCAI.<br>Becoming an HCAI organization.<br>Continuous performance improvement and positive experience. | • Advanced HCAI practices with a culture of continuous improvement and innovation.<br>• Evidence of continuous adaptive reconfigurability. | • HCAI is an integral part of organizational and business strategy.<br>• Active user community engaged for feedback and co-design.<br>• HCAI becomes part of the organizational culture and best practices<br>• The organization is recognized as a leader in HCAI practices within the industry. | • Regularly updated AI systems based on user insights.<br>• Formalized training programs on human-centered design.<br>• Advanced HCAI practices are embedded across the organization.<br>• HCAI becomes part of the performance dashboard (monitoring & tracking).<br>• Proactive risk management | • Identity as an HCAI organization.<br>• Proactively advocate for ethical AI development within the industry and regulatory landscape.<br>• Serve as best practice for other organizations in HCAI practice |

## 5. Strategies for HCAI Practice

HCAI concepts, frameworks, and methodologies provide the foundational basis for HCAI practice; however, realizing HCAI in real-world systems requires effective, sustained implementation. As with any emerging field, HCAI practice faces notable challenges.

As highlighted in previous sections, HCAI practice continues to face multifaceted challenges across the technical, ethical, and organizational dimensions throughout the AI lifecycle. For instance, siloed structures, limited interdisciplinary collaboration, and the absence of formal governance models hinder the institutionalization of HCAI principles and the consistent oversight of responsible AI deployment within organizations. Ethical and value alignment remains another critical area, as current frameworks frequently lack mature operational mechanisms to translate high-level principles, such as fairness, accountability, and inclusivity, into actionable design practices. Beyond the organizational sphere, challenges also emerge in the domains of data, algorithms, and societal ecosystems. On a broader scale, regulatory frameworks, cross-sector coordination, and public engagement have not kept pace with rapid technological advancements, leaving gaps in addressing the societal impacts of AI on sustainability, equity, and human control over AI. Altogether, these challenges underscore that HCAI is inherently a socio-technical endeavor, requiring not only advanced technical solutions but also robust collaboration at the ethical, organizational, and ecosystem levels. This section offers strategies to guide the practical adoption and organizational integration of HCAI.

### 5.1 Integrative Multi-Level and Multi-Layered Approaches

Integrative multi-level and multi-layered approaches integrate two complementary frameworks that together bridge conceptual hierarchy and practical implementation in HCAI practice, which avoid siloed solutions in HCAI practice: (1) the Hierarchical HCAI (hHCAI) approach (see Figure 8 in Section 3.1.2); (2) the three-layer HCAI implementation approach (Figure 19 below). Both provide a unified strategic view of HCAI practice, connecting theoretical foundations with actionable implementation strategies.

**The Hierarchical Strategy for HCAI Practice**

As discussed in Section 4, current HCAI practice focuses more on human-AI interaction within individual human-AI systems. The hierarchical HCAI (iHCAI) framework expands and promotes a strategy for HCAI practice from "keeping human-in/on-the-loop" to "keeping organization-in-the-loop", "keeping ecosystem-in-the-loop", and " keeping society-in-the-loop" system level through extending HCAI practice to the levels of organizations, ecosystems, and social systems (i.e., individual systems → organization → ecosystem → society), enabling comprehensive HCAI solutions.

**A three-layer HCAI implementation strategy**

The three-layer HCAI Implementation strategy emphasizes a three-layer implementation structure and implementation contexts (i.e., project → organization → society) across the AI lifecycle (see Figure 20) (Xu, Gao, & Dainoff, 2025). Table 11 details the implementation strategy, with key activities defined and illustrated throughout this book.

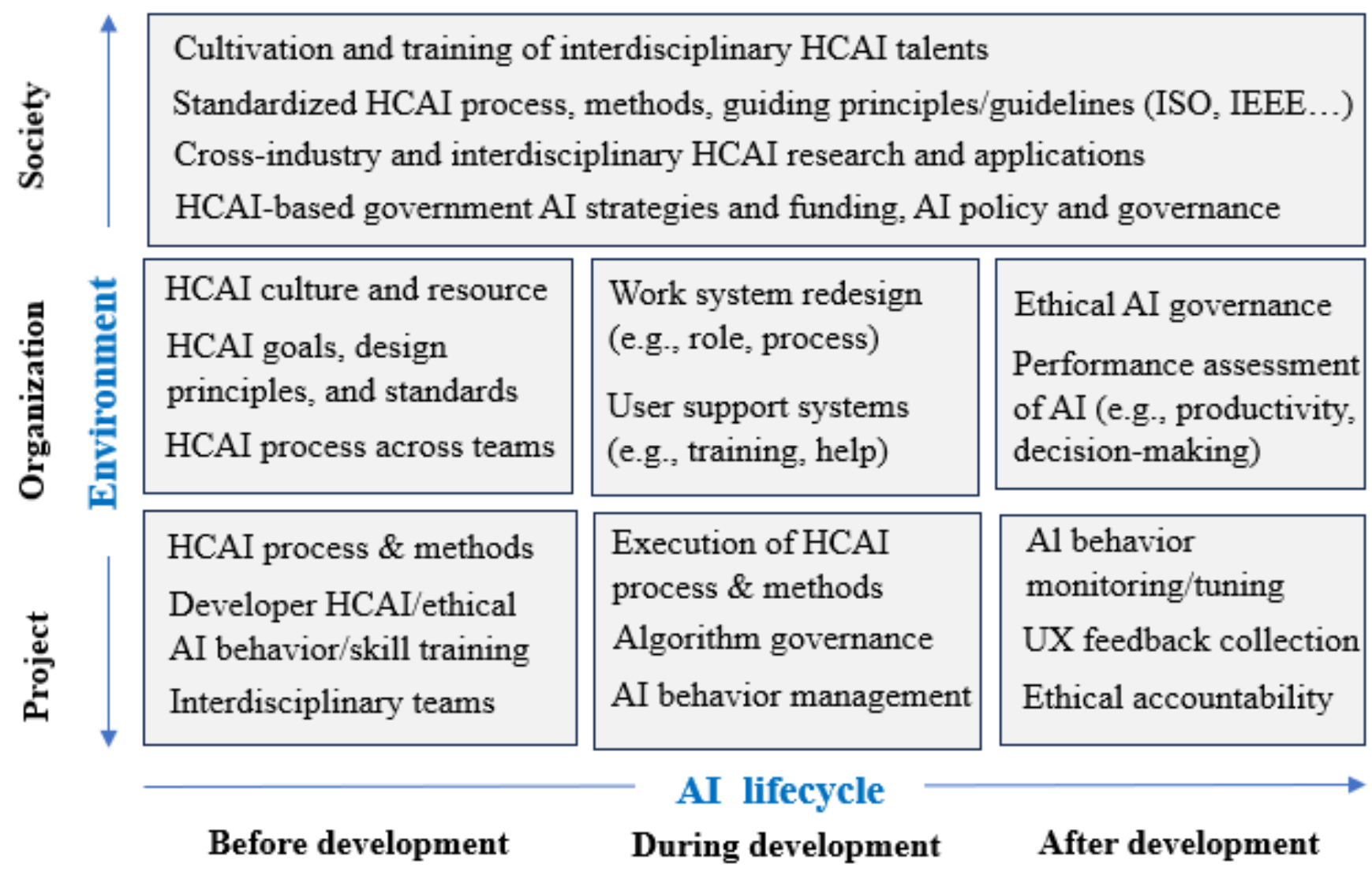

Figure 20 The three-layer (project → organization → society) HCAI implementation strategy across the AI lifecycle

Table 11 Key activities across the three-layer implementation strategy

| Implementation Environment | Key HCAI Actions | Related Chapters in This Handbook (selected) |
|---|---|---|
| Projects | **Before Development**<br>• Define HCAI strategy and plan<br>• Set up a cross-disciplinary development team<br>• Establish HCAI process<br>• Select interdisciplinary and HCAI methods<br>• Provide HCAI/ethical AI skill training for project teams<br>**During Development**<br>• Execute HCAI process and methods<br>• Manage AI behavior (e.g., user-participatory algorithm fine-tuning/training)<br>• Execute ethical AI/algorithmic/data governance<br>**During Development**<br>• Monitor and optimize AI system behavior<br>• Track user design feedback<br>• Track human accountability (e.g., developers) for ethical AI | (Chapter 1) Human-centered AI (HCAI): Foundations and Approaches<br>(Chapter 15) Human-centered AI design principles for generative AI<br>(Chapter 16) Designing human-centered AI experiences<br>(Chapter 19) Human-centered machine learning<br>(Chapter 25) Human-Controllable AI: Meaningful Human Control<br>(Chapter 26) Human-in/on-the-loop design for human controllability<br>(Chapter 37) Ethical AI standards and governance: A perspective of human-centered AI<br>(Chapter 38) Data governance<br>(Chapter 39) Algorithmic governance |
| Organizations | **Before Development**<br>• Define organizational HCAI goals and guiding principles<br>• Establish organization-level HCAI/ethical AI design guidelines<br>• Set up organizational AI governance<br>• Foster an HCAI organizational culture<br>• Build HCAI R&D resources (e.g., personnel, equipment)<br>**During Development**<br>• Design new work systems (e.g., roles, processes, human-AI function allocation, decision-making processes)<br>• Develop user support systems<br>**During Development**<br>• Implement HCAI/ethical AI governance<br>• Evaluate organizational performance (e.g., productivity, decision efficiency)<br>• Manage technological change and impacts (e.g., organization, users) | (Chapter 1) Human-centered AI (HCAI): Foundations and Approaches<br>(Chapter 14) Design thinking and AI: Facilitating HCAI solutions<br>(Chapter 37) Ethical AI standards and governance: A perspective of human-centered AI<br>(Chapter 55) AI transformation of organizations and society: A human-centered perspective<br>(Chapter 56) Human-Centered AI Maturity Model (HCAI-MM): An Organizational Design Perspective<br>(Chapter 57) Enterprise Strategy for Human-Centered AI<br>(Chapter 58) Organizational Practices and Sociotechnical Design of Human-centered AI<br>(Chapter 59) Human-centered AI in business modeling |
| Society | • Establish government and industry policy for ethical AI<br>• Establish HCAI/ethical AI design standards at international, national, and industry levels<br>• Implement effective ethical AI governance<br>• Organize cross-industry, cross-disciplinary HCAI-related research and applications<br>• Cultivate interdisciplinary HCAI talent (e.g., college programs) | (Chapter 3) Multi-level sociotechnical systems approach to human-centered AI<br>(Chapter 31) AI risk management frameworks<br>(Chapter 37) Ethical AI standards and governance: A perspective of human-centered AI<br>(Chapter 40) International governance of AI<br>(Chapter 55) AI transformation of organizations and society: A human-centered perspective |

In summary, implementing HCAI practice requires coordinated actions across three interrelated environments (projects, organizations, and society) throughout the AI Lifecycle. At the *project level*, successful implementation begins before AI development by forming interdisciplinary teams, adopting human-centered processes, and training developers in ethical AI. During development, teams execute HCAI processes, manage AI behavior through a user-participatory approach during model training and tuning, and ensure algorithmic governance. After deployment, they continuously monitor system performance, gather user feedback, and maintain accountability for ethical outcomes. At the *organizational level*, HCAI practice is enabled by setting organizational HCAI goals, establishing HCAI/ethical AI principles and design standards, fostering responsible AI culture, and building infrastructure to support HCAI research and applications. Organizations must also redesign work systems, provide user support, and manage technological change while monitoring fairness, privacy, and productivity. At the *societal level*, implementation depends on establishing standards and governance frameworks for ethical AI, setting international and industry-wide HCAI standards, promoting interdisciplinary research collaborations, and cultivating HCAI talent. Together, these multi-level strategies operationalize the HCAI vision throughout the AI lifecycle.

### 5.2 Enterprise Strategy

HCAI practice is an enterprise transformation that begins with rethinking how organizations align strategy, structure, and culture around human-centered values. Rather than treating AI as a purely technical or efficiency-driven enabler, enterprises must integrate HCAI principles into their strategies, including strategic planning, business process reengineering, and work-system redesign (Dong, 2024; Mazarakis et al., 2023). This integration embeds HCAI and ethics considerations at every stage of product and service development, fostering cross-functional collaboration among engineers, designers, ethicists, and domain experts. Global frameworks such as the OECD AI Principles (OECD, 2024), the EU AI Act (EU AI Act, 2024), and the NIST AI Risk Management Framework (NIST, 2023) provide actionable guidance for aligning AI initiatives with organizational missions and measurable social impact. Successful enterprises demonstrate that adopting HCAI as a guiding philosophy can become a differentiating force when supported by leadership commitment, ethical design culture, and enterprise-wide readiness for responsible AI integration (Mayer et al., 2025).

HCAI practice also includes HCAI-driven business reengineering and workforce transformation, which requires redesigning organizational workflows and decision-making systems to emphasize HCAI principles, such as transparency, participatory design, and accountability. Work systems should promote human-led collaboration with AI by clarifying roles, responsibilities, and oversight mechanisms to ensure meaningful human control (Interaction Design Foundation, 2024). Enterprises must invest in employee reskilling and incentive systems to prepare teams for hybrid intelligence environments, where human creativity and machine intelligence complement one another (Beamery, 2025; IBM, 2024). Decision-making processes should evolve from efficiency-driven automation toward human-centered deliberation that values explainability, fairness, and stakeholder inclusion. Moreover, by adopting multidimensional performance metrics that integrate human, social, and environmental outcomes alongside financial indicators, organizations can balance innovation with responsibility. In doing so, HCAI becomes a cornerstone of enterprise resilience, enabling sustainable innovation, trust-based stakeholder relationships, and long-term societal legitimacy in the AI era (Dong, 2024; Mazarakis et al., 2023).

For further elaboration, refer to

- (Chapter 56) Human-Centered AI Maturity Model (HCAI-MM): An Organizational Design Perspective
- (Chapter 57) Enterprise Strategy for Human-Centered AI
- (Chapter 58) Organizational Practices and Sociotechnical Design of Human-centered AI
- (Chapter 59) Human-centered AI in business modeling

### 5.3 Human-centered AI risk management and governance

From an HCAI practice strategy perspective, effective AI risk management begins by embedding human and ethical considerations into a continuous lifecycle program that the entire organization actively implements (Shneiderman, 2020b). It requires collaboratively mapping sociotechnical risks with diverse stakeholders (e.g., developers, domain experts, users, and affected communities) to identify potential harm and misuse. Each risk should be prioritized by impact and likelihood, with clearly defined controls observable across all lifecycle stages. To establish organizational consistency and resilience, HCAI practice should integrate recognized governance frameworks such as the NIST AI Risk Management Framework (Govern–Map–Measure–Manage) for process discipline, ISO/IEC 23894:2023 for structured AI risk processes, and ISO/IEC 42001:2023 for developing an AI management system that formalizes roles, metrics, and continuous improvement mechanisms (NIST, 2023; ISO/IEC, 2023a; ISO/IEC, 2023b). Together, these scaffolds institutionalize human-centered accountability, ensuring responsible AI practices. Evaluation plans should include performance metrics, failure analyses, and abuse testing. Finally, continuous post-deployment monitoring, structured user feedback channels, and incident reporting help close the feedback loop, enabling adaptive learning and ongoing risk mitigation. Through this lifecycle approach, HCAI risk management evolves into a dynamic organizational capability that continuously aligns AI behavior with HCAI principles.

In HCAI practice, human-centered AI governance serves as the structural backbone that translates ethical and risk-management principles into accountable, measurable organizational practices. Governance ensures that policies are both principled and actionable by aligning internal frameworks with established global standards, such as the OECD AI Principles for values alignment, the EU AI Act for risk-based obligations in high-risk systems, and sector-specific guidance, such as the FDA's Good Machine Learning Practice (GMLP) (OECD, 2019; EU, 2024; FDA, 2021). Effective implementation requires traceability from policy to control to evidence, supported by a structured framework that defines accountability across leadership, ethics committees, product lines, and research teams. Organizations should maintain a centralized AI governance system that documents purpose, data lineage, evaluation results, and post-deployment performance plans. Human-centered AI governance further mandates user agency, transparency, and contestability. This should be tied to measurable key performance indicators (KPIs), such as safety, efficacy, fairness, robustness, and human factors metrics. These KPIs must be regularly reviewed through the organization's AI management system (ISO/IEC 42001:2023) and risk management program (ISO/IEC, 2023c;

NIST, 2023), creating a continuous feedback loop that sustains trust, accountability, and alignment between AI behavior and HCAI principles throughout the AI lifecycle.

For further elaboration, refer to:

- Section 6 Human-centered AI risk management
- Section 7: Human-centered AI governance

**5.4 Design strategy**

In HCAI practice, enterprise design strategy must evolve beyond technical optimization to embrace approaches such as design thinking, user experience (UX) design, and human-in-the-loop (HITL) / human-on-the-loop (HOTL) design as integral components of the AI lifecycle. Traditional AI pipelines, which focused on modeling and data accuracy, often neglect user experiences, expectations, and contextual constraints of the humans and systems affected (Norman, 2019; Shneiderman, 2022). Specifically, integrating *Design Thinking* into AI development adds a strategic design layer that uncovers latent user needs, frames problems from human and ethical perspectives, and validates solutions iteratively before large-scale deployment (Brown & Katz, 2019).

To operationalize the vision of design strategy for HCAI practice, organizations should establish AI readiness assessments that evaluate design maturity, user trust, and HCAI/ethical preparedness; deploy interdisciplinary collaboration frameworks that unite professionals like UX designers, AI engineers, data scientists, ethicists, and policy leaders; and integrate AI governance checkpoints into the design workflow to ensure alignment with HCAI/ethical principles. For example, practical implementation includes creating design playbooks for HITL/HOTL systems, embedding human-centered evaluation metrics (e.g., usability, explainability, perceived control), and maintaining post-deployment user feedback channels to refine AI behavior. Ultimately, design strategies such as design thinking in HCAI practice enable organizations to integrate empathy into engineering, transparency into governance, and human purpose into AI systems, thereby transforming AI from a technology-driven process into an experience-driven innovation journey aligned with real-world human and organizational goals.

For further elaboration, refer to:

- (Chapter 14) Design thinking and AI: Facilitating HCAI solutions
- (Chapter 16) Designing human-centered AI experience
- (Chapter 26) Human-in/on-the-loop design for human controllability
- (Chapter 59) Human-Centered AI in Business Modeling

**5.5 Methodological strategy**

Overall, organizations should define a methodological strategy for HCAI practice by tailoring the HCAI methodology framework (HCAI-MF) to their organization's size, maturity, and goals. The overall strategy includes aligning HCAI initiatives with business objectives, demonstrating value through pilot projects, promoting leadership advocacy, building awareness through training, defining KPIs for trust and ethical compliance, standardizing toolkits and processes, and prioritizing high-impact projects. More specifically,

- *HCAI requirements.* Adopt a stepwise approach: establish organizational HCAI guiding principles, translate them into design guidelines, and define measurable product requirements. Provide staff training and develop case examples to illustrate HCAI principles, and tailor design guidelines for specific applications. Showcase HCAI benefits through initiatives to secure early leadership support.
- *HCAI Method Adoption.* Identify key HCAI design goals and choose suitable HCAI methods. Strengthen expertise through training and hiring to promote interdisciplinary teamwork. Create a phased plan, integrate HCAI into current workflows, and standardize methods.
- *HCAI Process.* Embed HCAI activities across the AI lifecycle, before, during, and after development. Define milestones, integrate processes into workflows, and monitor progress. Use standardized tools, start with small pilot projects, secure leadership support, and motivate participation with clear roles and recognition.
- *HCAI Interdisciplinary Collaboration.* Form interdisciplinary teams with clear roles, communication channels, and decision-making frameworks. Build shared vocabulary through joint training, encourage open communication, and organize teams around defined goals. Establish ethical review boards to align values and promote collaboration across disciplines and cultures.

For further elaboration, refer to

- (Chapter 14) Design thinking and AI: Facilitating HCAI solutions
- (Chapter 16) Designing human-centered AI experience

- (Chapter 56) Human-Centered AI Maturity Model (HCAI-MM): An Organizational Design Perspective
- (Chapter 57) Enterprise Strategy for Human-Centered AI
- (Chapter 59) Human-Centered AI in Business Modeling

### 5.6 Progressive HCAI practice

Establishing effective HCAI practice is an evolutionary process that progresses over time rather than occurring overnight. HCAI practice within organizations needs a structured framework to evaluate, guide, and continuously improve their ability to design, develop, deploy, and use HCAI systems. Establishing an HCAI maturity framework for the organization can be used to diagnose gaps and needs, and to define a roadmap to achieve organizational HCAI goals. The HCAI-MM, as discussed in Section 4.6, defines a five-level progression: *Initial, Developing, Defined, Managed,* and *Optimizing* (Winby & Xu, 2025). HCAI maturity is demonstrated through increasing stakeholder engagement, structured governance, ethical oversight, interdisciplinary collaboration, continuous learning, and measurable human-AI performance outcomes. These stages reflect an organization's evolution from ad hoc awareness of human-centered principles to the full institutionalization of HCAI practices as part of its culture and strategy.

By guiding organizations through progressive stages of maturity, an organizational HCAI maturity model fosters continuous learning, ethical alignment, and socio-technical adaptability, ensuring that AI serves as a tool to enhance human judgment, creativity, and decision-making rather than replace it. Ultimately, an HCAI maturity model positions human-centered AI as both a strategic capability and a governance system, linking ethical AI design with adaptive organization structures that enhance trust, creativity, decision quality, and long-term competitiveness.

For further elaboration, refer to:

- (Chapter 14) Design thinking and AI: Facilitating HCAI solutions
- (Chapter 56) Human-Centered AI Maturity Model (HCAI-MM): An Organizational Design Perspective
- (Chapter 57) Enterprise Strategy for Human-Centered AI

## 6. Structure of this handbook

This handbook centers on HCAI, providing a comprehensive overview of the latest research and practical applications. Each chapter presents findings on emerging HCAI approaches, discusses current implementation status, and concludes with an analysis of key challenges and unresolved issues. Moreover, these chapters collectively identify promising directions for future inquiry and innovation. In doing so, the handbook not only reflects the current HCAI landscape but also anticipates its future evolution toward more mature, interdisciplinary, and sustainable research and applications. Like the earlier emergence of fields such as human-centered design and human-computer interaction, HCAI represents a new yet rapidly evolving paradigm, one that continues to refine and expand its theories, methodologies, and practices to ensure that AI technologies meaningfully serve and empower humanity. The handbook is organized into the following ten sections.

*Overview, Concepts, and Foundations.* This opening section establishes the conceptual and theoretical foundations of Human-Centered AI (HCAI). It includes an opening chapter, *Human-Centered AI: Foundations and Approaches*, which frames the handbook's purpose, scope, and overall structure. This chapter outlines major HCAI concepts, frameworks, methodologies, challenges, and practice strategies. It establishes the philosophical underpinnings, core principles, and methodological frameworks of HCAI, situating the field within contemporary AI developments. The chapter also presents an integrated map of the handbook's structure and thematic focus, illustrating how its parts interconnect. The section also includes six other chapters. In addition to foundational constructs such as human values, hybrid intelligence, and sociotechnical systems, this section highlights emerging forms of artificial social intelligence, artificial emotional intelligence, and embodied intelligence as essential dimensions of next-generation AI systems. This section provides a unifying framework for understanding AI as a deeply human-centered technology, laying the foundation for the handbook's subsequent sections.

*Human-AI Interaction.* This section focuses on the interactional dynamics between humans and AI systems, examining how people perceive, interpret, collaborate with, and adapt to AI technologies. Topics include theory of mind in human–AI interaction, trust calibration, explainability, transparency, and user experience considerations across diverse AI applications. The chapters emphasize that effective human–AI interaction requires more than user interface usability; it depends on mutual intelligibility, appropriate role allocation, and the alignment of system behavior with human cognitive, emotional, and social expectations. This section unpacks how humans and AI systems interact in practice. These chapters highlight both technical design challenges and more profound questions about human–AI relationships and their social dynamics.

*Human-Centered AI Design.* This section comprises five chapters aimed at practitioners and designers. It showcases methodologies for embedding HCAI principles into the design of AI-enabled systems, marrying theoretical rigor with hands-on practice. These chapters address how human-centered principles can be systematically embedded into AI design processes. Drawing on design thinking, participatory design, and interdisciplinary collaboration, the chapters explore methods for translating human needs, values, and constraints into concrete requirements for AI systems. This section highlights design frameworks, processes, and tools that support iterative development, stakeholder engagement, and value-sensitive innovation, demonstrating how HCAI can translate abstract principles into actionable design practices.

*Human-Centered AI in Computing.* This section includes five chapters that examine the integration of human-centered AI principles within core computing and modeling technologies. Topics span human-centered recommender systems, social computing, human-AI hybrid intelligence architectures, and system-level design considerations. The chapters emphasize that human-centeredness must be addressed not only at the user interface level but also within algorithms, data pipelines, and system architectures, ensuring that AI computing systems meaningfully support human goals and ethical, explainable, and human-centered AI, thereby tying technical system design to HCAI principles.

*Human Controllability over AI.* This section, with six chapters, focuses on the critical issue of human control, oversight, and meaningful intervention in AI systems. The chapters explore concepts such as human-in-the-loop and human-on-the-loop systems, levels of automation, shared decision-making, and responsibility allocation. This section underscores that controllability is a foundational requirement for trustworthy, human-centered AI systems, particularly in high-stakes domains, and that effective control must be designed as a sociotechnical property rather than treated as a purely technical safeguard. This section outlines how to design AI systems that are controllable, interpretable, and safe for humans, providing a comprehensive guide to ensuring alignment with human control and societal norms.

*Human-Centered AI Risk Management.* This section includes seven chapters that address the identification, assessment, and mitigation of AI-related risks from an HCAI perspective. This section navigates the landscape of AI risk, from technical safety and incident reporting to accountability frameworks and measurement systems. The chapters also examine operational risk metrics, scenario-based planning, and organizational readiness for managing AI-related failures or harm, offering a detailed toolkit for practitioners to anticipate, measure, and mitigate risks in real-world AI systems. Specifically, the chapters examine technical, ethical, social, organizational, and environmental risks, including bias, safety, reliability, privacy, and unintended consequences. By integrating a human-centered AI approach, governance frameworks, and lifecycle-based risk management approaches, this section demonstrates how HCAI can proactively reduce harm and support the deployment of resilient, accountable AI.

*Human-Centered AI Governance.* The seven chapters in this section examine AI governance structures, policies, and organizational practices necessary to ensure that AI systems remain aligned with human values and societal goals. The chapters analyze AI governance at multiple levels across organizational, institutional, and societal, covering topics such as accountability, regulatory frameworks, standards, and stakeholder participation. This section emphasizes that effective AI governance is inherently human-centered, requiring transparency, inclusiveness, and adaptive mechanisms that reflect evolving social norms and technological capabilities. Spanning seven chapters, this governance-focused section provides a comprehensive overview of AI governance in the HCAI landscape.

*Human-Centered Large Language Models (LLMs).* As one of the most widely used generative AI technologies, LLMs power applications like chatbots, content creation tools, and coding assistants. However, LLMs should be developed to ensure that their outputs are trustworthy, transparent, and aligned with user intent, particularly given their tendency to generate convincing yet potentially misleading or biased content. With five chapters specifically focused on LLMs, this section examines how to build and deploy LLMS that are transparent, trustworthy, and human-centered. Contributions include strategies for explainability, bias mitigation, responsible deployment, interaction design, risk mitigation, and interaction design optimization, showing how LLMs can be aligned with HCAI principles. This section highlights how HCAI principles can guide the development and deployment of LLMs as human-centered AI solutions.

*Sectoral Applications for Human-Centered AI.* This application-driven section comprises six chapters that explore HCAI applications across real-world domains, including healthcare, the judiciary, nuclear power plants, cybersecurity, environmental sustainability, and the arts, showcasing HCAI in action. These chapters examine contextual challenges, impact measurement, stakeholder engagement, and socio-technical dynamics. Through cross-domain analysis, this section demonstrates how HCAI principles can be tailored, applied, and evaluated in diverse global contexts. This section highlights the adaptability of HCAI and demonstrates how it can enhance effectiveness, safety, and acceptance across application domains.

*Human-Centered AI Practice*. The final section synthesizes theory and application by focusing on practical implementation, organizational adoption, HCAI maturity, human-centered business transformation, and social impact of AI adoption across diverse workforces. In this section, the focus shifts to organizational transformation and real-world integration of HCAI, drawing on case studies to illustrate how institutions operationalize HCAI principles at scale. These chapters address toolkits, processes, methods, and maturity models that support sustained HCAI practice within real organizations. The section emphasizes that HCAI is not a one-time design choice but an ongoing organizational capability that requires leadership commitment, interdisciplinary collaboration, and continuous learning to ensure AI systems remain aligned with human needs over time.

## 7. Conclusion

Human-Centered Artificial Intelligence (HCAI) represents an integrative, multi-level evolution of how humans design, develop, interact with, and govern AI systems. The contents in this chapter span a wide range of HCAI concepts, frameworks, methodologies, and practical strategies, revealing a converging trajectory: AI must advance in ways that are reliable, safe, trustworthy, and fundamentally human-aligned. HCAI reframes AI not as an isolated technological entity but as part of an interconnected web of human–AI systems, organizational contexts, ecosystems, and sociotechnical structures in which human values and needs are preserved at every level of design, development, deployment, use, and governance.

AI is growing fast in many exciting ways, such as, generative AI, which can create images, videos, and sounds, turn data into new content; multimodal AI, which can work with different types of input at once, like combining text, images, and audio, to understand the world more like humans do; agentic AI, where systems can make decisions, use tools, and carry out some tasks automatically; embodied AI, where intelligence is placed into robots or virtual avatars that can move, see, and interact with their surroundings. These trends are helping AI become more powerful, interactive, and human-aware, demanding more cross-disciplinary collaboration and continuous advancements of HCAI methodology and practice throughout the AI lifecycle. The evolution from conceptual principles to operational maturity, guided by frameworks such as the HCAI Methodological Framework (HCAI-MF) and the HCAI Maturity Model (HCAI-MM), provides a structured roadmap for embedding HCAI in practice across individuals, organizations, ecosystems, and societies. As the *Handbook of Human-Centered AI* unfolds, the subsequent chapters build on this foundation, presenting methods, strategies, toolkits, and case studies that translate human-centered values into actionable design and governance practices. Ultimately, the success of AI will be measured not solely by its technical sophistication but by its enduring capacity to enhance human performance, safeguard human dignity, and strengthen societal benefits and trust.

## References

AIAAIC (2025). *AI, Algorithmic, and Automation Incident and Controversy*. Retrieved November 24, 2025, from https://www.aiaaic.org/

AIID (2025). *The AI Incident Database*. Retrieved November 24, 2025, from https://incidentdatabase.ai,

Akbarighatar, P., et al. (2024). Operationalizing responsible AI principles through design and development. *AI and Ethics*. Springer. https://doi.org/10.1007/s43681-024-00524-4

Ahmad, K., et al. (2023). Requirements practices and gaps when engineering human-centered AI systems. *Applied Soft Computing, 143*, 110421. https://doi.org/10.1016/j.asoc.2023.110421

Amershi, S., Weld, D., Vorvoreanu, M., Fourney, A., Nushi, B., Collisson, P., ... & Horvitz, E. (2019). Guidelines for human-AI interaction. In Proceedings of the 2019 CHI conference on human factors in computing systems (pp. 1-13).

Asif, M. (2024). AI and human interaction: Enhancing user experience through intelligent systems. *Frontiers in Artificial Intelligence Research, 1*(2), 209–249.

Auernhammer, J. (2020). Human-centered AI: The role of Human-centered Design Research in the development of AI, in Boess, S., Cheung, M. and Cain, R. (eds.), *Synergy - DRS International Conference 2020*, 11-14 August, Held online. https://doi.org/10.21606/drs.2020.282

Battistoni, P., et al. (2023). Can AI-oriented requirements enhance human-centered design of intelligent interactive systems? Results from a workshop with young HCI designers. *Multimodal Technologies and Interaction, 7*(3), 24. https://doi.org/10.3390/mti7030024

Beamery (2025). *Upskilling & reskilling with AI: A strategic view.* https://beamery.com/resources/blogs/upskilling-reskilling-with-ai-a-strategic-view

Bingley, W. J., Curtis, C., Lockey, S., Bialkowski, A., Gillespie, N., Haslam, S. A., ... & Worthy, P. (2023). Where is the human in human-centered AI? Insights from developer priorities and user experiences. Computers in Human Behavior, 141, 107617.

Brill, J.C., Cummings, M. L., Evans, A. W. *III*., Hancock, P. A., Lyons, J. B., & Oden, K. (2018). Navigating the advent of human-machine teaming. *Proceedings of the Human Factors and Ergonomics Society 2018 Annual Meeting, 455-459*.

Brdnik, S., Heričko, T., et al. (2022). Intelligent user interfaces and their evaluation: A systematic mapping study. *Sensors, 22*(15), 5830. https://doi.org/10.3390/s22155830

Brown, T., & Katz, B. (2019). *Change by design: How design thinking transforms organizations and inspires innovation* (Revised ed.). Harper Business.

Capel, T., & Brereton, M. (2023, April). What is human-centered about human-centered AI? A map of the research landscape. In *Proceedings of the 2023 CHI conference on human factors in computing systems* (pp. 1-23).

Cascio, W. F., & Montealegre, R. (2016). How technology is changing work and organizations. *Annual Review of Organizational Psychology and Organizational Behavior, 3*(1), 349–375. https://doi.org/10.1146/annurev-orgpsych-041015-062352

Cerejo, J. (2021, April 15). The design process of human-centered AI — Part 2: *Empathize & Hypothesis Phase in the development of AI-driven services. Medium.* https://bootcamp.uxdesign.cc/human-centered-ai-design-process-part-2-empathize-hypothesis-6065db967716

Chatila, R., et al. (2019). The IEEE global initiative on ethics of autonomous and intelligent systems. In *Robotics and Well-Being* (pp. 11–16). Springer.

Clegg, C. W. (2000). Sociotechnical principles for system design. *Applied Ergonomics, 31*(5), 463–477. https://doi.org/10.1016/S0003-6870(00)00009-0

Coskun, H. (2026) Human-Centered AI Applications for Medical Informatics. IGI Global.

Costea, S. & Kamaldeen, A. (2023). *Localizing AI ethics principles: Adapting global AI ethics guidelines to diverse cultural and community contexts*. ResearchGate. https://www.researchgate.net/publication/394494458

Coursaris, C. K., Beringer, J., Léger, P. M., & Öz, B. (Eds.). (2025). *The Design of Human-Centered Artificial Intelligence for the Workplace*. Springer Nature.

DAIL. (2025). *Database of AI Litigation*. Retrieved November 24, 2025, from https://oecd.ai/en/incidents.

de Sio, S. F., & den Hoven, V. J. (2018). Meaningful human control over autonomous systems: A philosophical account. Frontiers in Robotics and AI, 5, 15. doi: 10.3389/frobt.2018.00015.

Dignum, F., & Dignum, V. (2020). *How to center AI on humans*. In *Proceedings of the 1st International Workshop on New Foundations for Human-Centered AI (NeHuAI 2020),* Santiago de Compostela, Spain, September 4, 2020.

Di Fede, G., Alrabie, L., Andolina, S. (2026). Human-Centered LLM. In: Xu, W. (eds) *Handbook of Human-Centered Artificial Intelligence*. Springer, Singapore. https://doi.org/10.1007/978-981-97-8440-0_49-2

Desolda, G., Esposito, A., Lanzilotti, R., et al. (2025). From human-centered to symbiotic artificial intelligence: A focus on medical applications. *Multimedia Tools and Applications, 84*, 32109–32150. https://doi.org/10.1007/s11042-024-20414-5

Dong, M., Bonnefon, J. F., & Rahwan, I. (2024). Toward human-centered AI management: Methodological challenges and future directions. *Technovation*, *131*, 102953.

Ehsan, U., & Riedl, M. O. (2020). Human-centered explainable AI: Towards a reflective sociotechnical approach. arXiv preprint arXiv: 2002.01092.

Ebers, M., & Cantero Gamito, M. (2021). Algorithmic governance and governance of algorithms: Legal and ethical challenges. In *Data Science, Machine Intelligence, and Law* (Vol. 1). Springer.

European Parliament and the Council of the European Union (EU AI Act). (2024). *Regulation (EU) 2024/1689 of the European Parliament and of the Council on laying down harmonised rules on Artificial Intelligence (Artificial Intelligence Act).* EUR-Lex. https://artificialintelligenceact.eu/the-act/

Endsley, M. R. (1995). Toward a theory of situation awareness in dynamic systems. *Human Factors, 37*(1), 32–64.

Endsley, M. R. (2023). Ironies of artificial intelligence. *Ergonomics*, 66(11), 1–13.

European Union. (2024). *Regulation (EU) 2024/1689 of the European Parliament and of the Council of 13 June 2024 laying down harmonised rules on artificial intelligence (Artificial Intelligence Act).* Official Journal of the European Union. https://eur-lex.europa.eu/eli/reg/2024/1689/oj

Fields, M., & Liu, Y. (2025). College Student Opinions on Generative Artificial Intelligence for Educational Experiences. *Proceedings of the Human Factors and Ergonomics Society Annual Meeting*, *69*(1), 267-271.

Food and Drug Administration. (2021). *Good Machine Learning Practice for medical device development: Guiding principles.* https://www.fda.gov/medical-devices/software-medical-device-samd/good-machine-learning-practice-medical-device-development-guiding-principles U.S. Food and Drug Administration

Floridi, L., Cowls, J., Beltrametti, M., Chatila, R., Chazerand, P., Dignum, V., … Vayena, E. (2018). AI4People—An ethical framework for a good AI society: Opportunities, risks, principles, and recommendations. *Minds and Machines, 28*(4), 689–707. https://doi.org/10.1007/s11023-018-9482-5

Fortinet. (2025). *Global threat landscape report (as summarized in TechRadar).* https://www.techradar.com/pro/security/ai-powering-a-dramatic-surge-in-cyberthreats-as-automated-scans-hit-36-000-per-second

Friedrich, J., et al. (2024). Human-centered AI development in practice—Insights from a multidisciplinary approach. *Zeitschrift für Arbeitswissenschaft, 78*(3), 359–376.

Gao, Q., Xu, W., Pan, H., Shen, M., & Gao, Z. (2025). Human-Centered Human-AI Collaboration (HCHAC). *arXiv preprint arXiv:2505.22477*.

Google PAIR (2019). People + AI Guidebook: Designing human-centered AI products. pair.withgoogle.com/

Germanakos, P., Juhasz, M., Kongot, A., Marathe, D., & Sacharidis, D. (Eds.). (2025). *Human-Centered AI: An Illustrated Scientific Quest*. Springer Nature.

Gunning, D., Stefik, M., Choi, J., Miller, T., Stumpf, S., & Yang, G. Z. (2019). XAI—Explainable artificial intelligence. *Science robotics*, *4*(37), eaay7120.

Hartikainen, M., et al. (2022). Human-centered AI design in reality: A study of developer companies' practices. In *Proceedings of the Nordic Human-Computer Interaction Conference* (pp. 1–11). ACM.

Hartikainen, M., Väänänen, K., & Olsson, T. (2023). Towards a human-centred artificial intelligence maturity model. In *Extended Abstracts of the 2023 CHI Conference on Human Factors in Computing Systems* (pp. 1-7).

Herrmann T. & Pfeiffer, S. (2023). Keeping the organization in the loop: A socio-technical extension of human-centered artificial intelligence. AI & SOCIETY, 38(4), 1523-1542.

Herrmann T. (2025). Organizational Practices and Socio-Technical Design of Human-Centered AI. In W. Xu (ed.), *Handbook of Human-Centered Artificial Intelligence*, Springer Nature Singapore. https://link.springer.com/referencework/10.1007/978-981-97-8440-0

AI-HLEG (High-Level Expert Group on Artificial Intelligence). (2019). Ethics Guidelines for Trustworthy AI. https://digital-strategy.ec.europa.eu/en/library/ethics-guidelines-trustworthy-ai#:~:text=On%208%20April%202019%2C%20the,and%20human%2Din%2Dcommand%20approaches

IBM. (2024). *Upskilling and reskilling for talent transformation in the era of AI.* IBM Think Insights. https://www.ibm.com/think/insights/ai-upskilling

Interaction Design Foundation. (2024). *What is human-centered AI (HCAI)?* https://www.interaction-design.org/literature/topics/human-centered-ai

ISO (International Organization for Standardization). (2019). ISO 9241-220: Ergonomics of human-system interaction Part 220: Processes for enabling, executing, and assessing human-centered design within organizations.

ISO/IEC. (2023a). ISO/IEC 5338:2023, *Information Technology—AI System Life Cycle Processes*, International Organization for Standardization, Dec. 2023.

ISO/IEC. (2023b). *ISO/IEC 23894:2023—Information technology—Artificial intelligence—Guidance on risk management.* https://www.iso.org/standard/77304.html ISO

ISO/IEC. (2023c). *ISO/IEC 42001:2023—Artificial intelligence management system (AIMS)—Requirements.* https://www.iso.org/standard/42001 ISO

Jobin, A., Ienca, M., & Vayena, E. (2019). The global landscape of AI ethics guidelines. *Nature Machine Intelligence, 1*(9), 389–399.

Kaber, D. B. (2018). A conceptual framework of autonomous and automated agents. *Theoretical Issues in Ergonomics Science*, 19(4), 406-430.

Kaluarachchi, T., Reis, A., & Nanayakkara, S. (2021). A Review of Recent Deep Learning Approaches in Human-Centered Machine Learning. Sensors, 21(7), 2514.

Kang, H., Lee, S. C., Jeong, J., Kim, H., Cha, M., & Jeon, M. (2025, November). Together or Apart: Designing Boundaries for Personal Intelligent Agents. In *Proceedings of the 13th International Conference on Human-Agent Interaction* (pp. 572-574).

Kudina, O., & van de Poel, I. (2024). A sociotechnical system perspective on AI. *Minds and Machines, 34*, 21.

Le Dinh, T., et al. (2024). A methodological framework for designing human-centered artificial intelligence services. In *Proceedings of the International Conference on Conceptual Modeling* (pp. 21–39). Springer Nature.

Li, F.F. (2018). How to make A.I. that's good for people. The New York Times. https://www.nytimes.com/2018/03/07/opinion/artificial-intelligence-human.html. March, 7, 20

Li, F. F. (2023). *The worlds I see: Curiosity, exploration, and discovery at the dawn of AI*. Flatiron books: a moment of lift book.

Liu, T., Yang, Q., & Tao, D. (2017). Understanding How Feature Structure Transfers in Transfer Learning. In IJCAI (pp. 2365-2371).

Mazarakis *et al.*, "What is critical for human-centered AI at work? Towards an interdisciplinary theory," *Frontiers in AI*, vol. 6, p. 1257057, 2023.

Mayer, H., Yee, L., Chui, M., & Roberts, R. (2025). *Superagency in the workplace: Empowering people to unlock AI's full potential.* McKinsey & Company. https://www.mckinsey.com/capabilities/mckinsey-digital/our-insights/superagency-in-the-workplace-empowering-people-to-unlock-ais-full-potential-at-work

Mosqueira-Rey, E., Hernández-Pereira, E., Alonso-Ríos, D., Bobes-Bascarán, J., & Fernández-Leal, Á. (2023). Human-in-the-loop machine learning: A state of the art. *Artificial Intelligence Review, 56*(4), 3005–3054.

McGregor, S. (2025). AI Incident Database. https://incidentdatabase.ai/ accessed Nov. 5, 2025.

Michael, K., Vogel, K. M., Pitt, J., & Zafeirakopoulos, M. (2025). Artificial intelligence in cybersecurity: A socio-technical framing. *IEEE Transactions on Technology and Society, 6*(1), 15–30.

Naikar, N., Hoffman, R., Roth, E. M., Klein, G., Militello, L. G., & Dominguez, C. (2025). Should we Make AI More Tool-like or Teammate-Like? *Journal of Cognitive Engineering and Decision Making*, *0*(0). https://doi.org/10.1177/15553434251346904

NAS (National Academies of Sciences, Engineering, and Medicine). (2021). Human-AI teaming: State-of-the-art and research needs. https://www.nationalacademies.org/read/26355/chapter/1

National Institute of Standards and Technology (NIST). (2023). *Artificial Intelligence Risk Management Framework (AI RMF 1.0).* U.S. Department of Commerce. https://nvlpubs.nist.gov/nistpubs/ai/NIST.AI.100-1.pdf

Nie, J., et al. (2023). Data and domain knowledge dual-driven artificial intelligence: Survey, applications, and challenges. *Expert Systems*, e13425.

Norman, D. A. (1986). *User Centered System Design: New Perspectives on Human-Computer Interaction.* Hillsdale, NJ: Lawrence Erlbaum Associates.

Norman, D. (2013). *The design of everyday things: Revised and expanded edition*. Basic books.

Organisation for Economic Co-operation and Development (OECD). (2019). *Recommendation of the Council on Artificial Intelligence.* https://legalinstruments.oecd.org/en/instruments/oecd-legal-0449 OECD Legal Instruments

(OECD) (2024). AI, Data Governance and Privacy: Synergies and Areas of International Co-operation. *OECD iLibrary*, Jun. 2024. https://www.oecd.org/en/publications/ai-data-governance-and-privacy_2476b1a4-en.html

OECD. (2025). OECD AI Incidents Monitor, an evidence base for trustworthy AI. Retrieved November 24, 2025, from https://oecd.ai/en/incidents

Ozmen Garibay, O., Winslow, B., Andolina, S., Antona, M., Bodenschatz, A., Coursaris, C., ... & Xu, W. (2023). Six human-centered artificial intelligence grand challenges. *International Journal of Human-Computer Interaction*, 39(3), 391-437.

Paeth, K., & McGregor, S. (2025). AI Risk, Safety, and Incident Reporting. In W Xu (Ed.) *Handbook of Human-Centered Artificial Intelligence*. Singapore: Springer Nature Singapore.

Pasmore, W., Winby, S., Mohrman, S. A., & Vanasse, R. (2019). Reflections: Sociotechnical systems design and organization change. *Journal of Change Management, 19*(2), 67–85.

Pan, H., Xu, W., Shen, M., & Gao, Z. (2025). Human-Centered Artificial Social Intelligence (HC-ASI). In W. Xu (ed.), *Handbook of Human-Centered Artificial Intelligence*, Springer Nature Singapore. https://link.springer.com/referencework/10.1007/978-981-97-8440-0

Parasuraman, R., Sheridan, T. B., & Wickens, C. D. (2000). A model for types and levels of human interaction with automation. IEEE Transactions on Systems, Man and Cybernetics-Part A: Systems and Humans, 30: 286–297.

Pileggi, S. F. (2024). Ontology in hybrid intelligence: A concise literature review. *Future Internet, 16*(8), 268. https://doi.org/10.3390/fi16080268

Rahwan, I., Cebrian, M., Obradovich, N., Bongard, J., Bonnefon, J.-F., Breazeal, C., ... & Wellman, M. (2019). *Machine behaviour*. Nature, 568(7753), 477-486.

Raunak, M. S., & Kuhn, D. R. (2024). AI failures. *Computer, 57*(11), 14–16.

Raunak, M. S., & Kuhn, D. R. (2025). AI failures—Part II. *Computer, 58*(11), 28–30.

Ransbotham, S., Khodabandeh, S., Fehling, R., LaFountain, B., & Kiron, D. (2021). *Expanding AI's impact with organizational learning*. MIT Sloan Management Review and Boston Consulting Group.

Régis, C., Denis, J. L., Axente, M. L., & Kishimoto, A. (2024). *Human-centered AI: A multidisciplinary perspective for policy-makers, auditors, and users*. Taylor & Francis.

Richards, D., Vythilingam, R., & Formosa, P. (2023). A principlist-based study of the ethical design and acceptability of artificial social agents. *International Journal of Human–Computer Studies, 172*, 102980.

Ridley, M. (2025). Human-centered explainable artificial intelligence: An Annual Review of Information Science and Technology (ARIST) paper. *Journal of the Association for Information Science and Technology*, *76*(1), 98-120.

Riedl, M. O. 2019. "Human-Centered Artificial Intelligence and Machine Learning." *Human Behavior and Emerging Technologies* 1 (1): 33–36.

Rong, Y., Leemann, T., Nguyen, T. T., Fiedler, L., Qian, P., Unhelkar, V., ... & Kasneci, E. (2023). Towards human-centered explainable ai: A survey of user studies for model explanations. *IEEE transactions on pattern analysis and machine intelligence*, *46*(4), 2104-2122.

Sanderson, C., Douglas, D., & Lu, Q. (2023). *Implementing responsible AI: Tensions and trade-offs between ethics aspects*. In *2023 International Joint Conference on Neural Networks (IJCNN)* (pp. 1–7). IEEE.

Sartori, L., & Theodorou, A. (2022). A sociotechnical perspective for the future of AI: Narratives, inequalities, and human control. *Ethics and Information Technology, 24*(1), 4.

Schmager, S., Pappas, I. O., & Vassilakopoulou, P. (2025). Understanding Human-Centred AI: a review of its defining elements and a research agenda. *Behaviour & Information Technology*, 44(15), 1-40.

Shneiderman, B., (1987). *Designing the User Interface: Strategies for Effective Human-Computer Interaction,* Addison-Wesley Publ. Co., Reading, MA

Shneiderman, B. (2020a). Human-centered artificial intelligence: Reliable, safe & trustworthy. *International Journal of Human-Computer Interaction*, 36(6), 495-504.

Shneiderman, B. (2020b). Bridging the gap between ethics and practice: Guidelines for reliable, safe, and trustworthy human-centered AI systems. *ACM Transactions on Interactive Intelligent Systems, 10*(4), Article 26.

Shneiderman, B. (2020c). Design lessons from AI's two grand goals: Human emulation and useful applications. *IEEE Transactions on Technology and Society, 1*(2), 73–82.

Shneiderman, B. (2021) *19th Note: Human-Centered AI Google Group.* In Human-Centered AI (Sept.12, 2021). Available: https://groups.google.com/g/human-centered-ai/c/syqiC1juHO.c

Shneiderman, B. (2022). *Human-Centered AI*. Oxford University Press.

Sison, A. J. G., Daza, M. T., Gozalo-Brizuela, R., & Garrido-Merchán, E. C. (2024). ChatGPT: More than a "weapon of mass deception" ethical challenges and responses from the human-centered artificial intelligence (HCAI) perspective. *International Journal of Human–Computer Interaction*, *40*(17), 4853-4872.

Shively, R.J., Lachter, J., Brandt, S.L., Matessa, M., Battiste, V., Johnson, W.W. (2018). Why Human-Autonomy Teaming?. In: Baldwin, C. (eds) Advances in Neuroergonomics and Cognitive Engineering. AHFE 2017. Advances in Intelligent Systems and Computing, vol 586. Springer, Cham.

Silberg, J., & Manyika, J. (2019). *Notes from the AI frontier: Tackling bias in AI (and in humans).* McKinsey Global Institute.

Stahl, B. C. (2021). *Artificial intelligence for a better future: An ecosystem perspective on the ethics of AI and emerging digital technologies* (p. 124). Cham, Switzerland: Springer Nature.

Stefan, F., Dzbor, M., & Simperl, G. (2020). Fair AI. *Business & Information Systems Engineering, 62*(4), 379–384.

Stephanidis, C., et al. (2019). Seven HCI grand challenges. *International Journal of Human–Computer Interaction, 35*(14), 1229–1269.

Stanford Institute HAI (Institute for Human-Centered Artificial Intelligence (HAI). (2025). AI *Index Report 2025 (incl. Responsible AI section).* https://hai.stanford.edu/ai-index/2025-ai-index-report

Subramonyam, H., Im, J., Seifert, C., & Adar, E. (2022). Human-AI Guidelines in Practice: Leaky Abstractions as an Enabler in Collaborative Software Teams. arXiv preprint arXiv: 2207.01749.

Sun, L., Xu, W., & Gao, Z. (2025). A Human-centered privacy (HCP) to AI. ). In W. Xu (ed.), *Handbook of Human-Centered Artificial Intelligence*, Springer Nature Singapore. https://link.springer.com/referencework/10.1007/978-981-97-8440-0.

Tahaei, M., et al. (2023). Human-centered responsible artificial intelligence: Current and future trends. *Extended Abstracts of the 2023 CHI Conference on Human Factors in Computing Systems* (pp. 1–4). ACM.

Trist, E. L., & Bamforth, K. W. (1951). Some social and psychological consequences of the longwall method of coal-getting: An examination of the psychological situation and defences of a work group in relation to the social structure and technological content of the work system. *Human Relations, 4*(1), 3–38.

Umbrello, S., & Natale, S. (2024). Reframing deception for human-centered AI. *International Journal of Social Robotics, 16*, 2223–2241.

van Allen, P. (2018). Prototyping ways of prototyping AI. *Interactions, 25*(6), 46–51.

Vaughan, J. W., & Wallach, H. (2020). A human-centered agenda for intelligible machine learning. In T. Miller (Ed.), *Machines we trust: Getting along with artificial intelligence* (pp. 33–47). MIT Press.

Walker, C. P., Schiff, D. S., & Schiff, K. J. (2024). Merging AI incidents research with political misinformation research: introducing the political Deepfakes incidents database. In *Proceedings of the AAAI Conference on Artificial Intelligence.* 38(21), pp. 23053-23058.

Wang, H., & Wang, M. (2024). Data and Knowledge-Driven Artificial Intelligence. In W. Xu. (Ed). *Human-AI Interaction: Enabling Human-Centered AI Through Interdisciplinary Collaboration.* Beijing: Tsinghua University Press.

Wickramasinghe, C. S., et al. (2020). Trustworthy AI development guidelines for human-system interaction. In *2020 13th International Conference on Human System Interaction (HSI)* (pp. 130–136). IEEE.

Winby, S., & Xu, W. (2025). Human-Centered AI Maturity Model (HCAI-MM): An Organizational Design Perspective. In W. Xu (ed.), *Handbook of Human-Centered Artificial Intelligence*, Springer Nature Singapore. https://link.springer.com/referencework/10.1007/978-981-97-8440-0

Wickens, C. D., Helton, W. S., et al. (2021). *Engineering psychology and human performance* (5th ed.). Routledge.

Wu, C. J., Raghavendra, R., Gupta, U., Acun, B., Ardalani, N., Maeng, K., ... & Hazelwood, K. (2022). Sustainable ai: Environmental implications, challenges and opportunities. *Proceedings of machine learning and systems*, *4*, 795-813.

Xu, W. (2003a). User-centered design: Challenges and opportunities for human factors. *Journal of Ergonomics, 9*(4), 8–11.

Xu, W. (2003b). Psychological research of human-automation interaction in automated flight deck. *Psychological Sciences*, *26*(3), 523-524.

Xu, W. (2005). Recent trends of research and applications on human-computer interaction. *Chinese Journal of Ergonomics, 11*(4), 37–40.

Xu, W. (2018). User-centered design (III): Methods for user experience and innovative design in the intelligent era. *Chinese Journal of Applied Psychology, 25*(1), 3–17.

Xu, W. (2019). Toward human-centered AI: A perspective from human-computer interaction. Interactions, 26(4), 42-46.

Xu, W. (2020). User-centered design (V): From automation to the autonomy and autonomous vehicles in the intelligence era. *Chinese Journal of Applied Psychology*, *26*(2), 108-129.

Xu, W. (2021). From automation to autonomy and autonomous vehicles: Challenges and opportunities for human-computer interaction. *Interactions*, *28*(1), 48-53.

Xu, W. (2025). Human-Centered Human-AI Interaction (HC-HAII): A Human-Centered AI Perspective. *arXiv preprint arXiv:2508.03969*, https://arxiv.org/abs/2508.03969

Xu, W. & Ge, L. (2020). Engineering psychology in the era of artificial intelligence. *Advances in Psychological Science*, 28(9), 1409-1425.

Xu, W. & Gao, Z. (2024). Applying human-centered AI in developing effective human-AI teaming: A perspective of human-AI joint cognitive systems. *Interactions*, 31 (1), 32-37.

Xu, W., & Gao, Z. (2025). An intelligent sociotechnical systems (iSTS) framework: Enabling a hierarchical human-centered AI (hHCAI) approach. *IEEE Transactions on Technology and Society, 6*(1), 31–46.

Xu, W., Gao, Z., & Dainoff, M. (2025). An HCAI methodological framework: Putting it into action to enable human-centered AI. *arXiv preprint arXiv:2311.16027*. https://arxiv.org/abs/2311.16027

Xu, W., Dainoff, M. J., Ge, L., & Gao, Z. (2023). Transitioning to human interaction with AI systems: New challenges and opportunities for HCI professionals to enable human-centered AI. *International Journal of Human-Computer Interaction,* 39(3), 494-518.

Xu, W. & Dainoff, M. J. (2023). Enabling Human-Centered AI: A New Junction and Shared Journey Between AI and HCI Communities. *Interactions,* 30(1), 42-47

Xu, W., Gao, Z., & Ge, L. (2024). New research paradigms and agenda of human factors science in the intelligence era. *Acta Psychologica Sinica.* 56(3), 363-382.

Yampolskiy, R.V. (2019). Predicting future AI failures from historic examples. foresight, 21,138-152.

Yan, J., Gao, T., et al. (2024). Teleoperation control of autonomous underwater vehicles toward human-on-the-loop: Needs, analyses, and solutions. *IEEE Systems, Man, and Cybernetics Magazine, 10*(3), 2–13.

Yang, Q., et al. (2020). Re-examining whether, why, and how human-AI interaction is uniquely difficult to design. In *Proceedings of the 2020 CHI Conference on Human Factors in Computing Systems* (pp. 1–12). ACM.

Ye, P. & Wang, F. (2026). Human-Centered Social Computing. In W. Xu (ed.), *Handbook of Human-Centered Artificial Intelligence*, Springer Nature Singapore. https://link.springer.com/referencework/10.1007/978-981-97-8440-0

Yildirim, N., et al. (2023). Investigating how practitioners use human-AI guidelines: A case study on the People+AI Guidebook. In *Proceedings of the 2023 CHI Conference on Human Factors in Computing Systems* (pp. 1–13).

Zheng, N. N., Liu, Z. Y., Ren, P. J., Ma, Y. Q., Chen, S. T., Yu, S. Y., ...& Wang, F. Y. (2017). Hybrid-augmented intelligence: collaboration and cognition. *Frontiers of Information Technology & Electronic Engineering*, 18(2), 153-179.

Zhou, J. & Chen, F. (2023). AI ethics: From principles to practice. AI & SOCIETY, 38, 2693–2703.